\documentclass[twocolumn]{aastex62}

\usepackage{amssymb,amsmath}
\usepackage{bm}
\usepackage{graphicx}
\usepackage[utf8]{inputenc}
\DeclareUnicodeCharacter{2009}{~}
\usepackage{natbib}
\usepackage[caption=false]{subfig}

\accepted{January 15, 2021}
\submitjournal{ApJ}

\begin{document}


\title{The Cosmic-Ray Composition between 2 PeV and 2 EeV Observed 
  with the TALE Detector in Monocular Mode}

\author[0000-0001-6141-4205]{R.U. Abbasi}
\affiliation{Department of Physics, Loyola University Chicago, Chicago, Illinois, USA}

\author{M. Abe}
\affiliation{The Graduate School of Science and Engineering, Saitama University, Saitama, Saitama, Japan}

\author[0000-0001-5206-4223]{T. Abu-Zayyad}
\affiliation{High Energy Astrophysics Institute and Department of Physics and Astronomy, University of Utah, Salt Lake City, Utah, USA}
\affiliation{Department of Physics, Loyola University Chicago, Chicago, Illinois, USA}

\author{M. Allen}
\affiliation{High Energy Astrophysics Institute and Department of Physics and Astronomy, University of Utah, Salt Lake City, Utah, USA}

\author{Y. Arai}
\affiliation{Graduate School of Science, Osaka City University, Osaka, Osaka, Japan}

\author{E. Barcikowski}
\affiliation{High Energy Astrophysics Institute and Department of Physics and Astronomy, University of Utah, Salt Lake City, Utah, USA}

\author{J.W. Belz}
\affiliation{High Energy Astrophysics Institute and Department of Physics and Astronomy, University of Utah, Salt Lake City, Utah, USA}

\author{D.R. Bergman}
\affiliation{High Energy Astrophysics Institute and Department of Physics and Astronomy, University of Utah, Salt Lake City, Utah, USA}

\author{S.A. Blake}
\affiliation{High Energy Astrophysics Institute and Department of Physics and Astronomy, University of Utah, Salt Lake City, Utah, USA}

\author{R. Cady}
\affiliation{High Energy Astrophysics Institute and Department of Physics and Astronomy, University of Utah, Salt Lake City, Utah, USA}

% ...
  
\collaboration{(Telescope Array Collaboration)}

\correspondingauthor{Tareq Abu-Zayyad}
\email{tareq@cosmic.utah.edu}

\begin{abstract}
  We report on a measurement of the cosmic ray composition by the 
  {\bf T}elescope {\bf A}rray {\bf L}ow-{\bf E}nergy Extension (TALE) air
  fluorescence detector (FD).
  By making use of the Cherenkov light signal in addition to air        
  fluorescence light from cosmic ray (CR) induced extensive air showers, 
  the TALE FD can measure the properties of the cosmic rays with energies 
  as low as $\sim 2$~PeV and exceeding 1~EeV.  
  In this paper, we present results on the measurement of $X_{\rm max}$
  distributions of showers observed over this energy range.
  Data collected over a period of $\sim 4$ years was analyzed for this study.
  The resulting $X_{\rm max}$ distributions are compared to the Monte Carlo (MC) 
  simulated data distributions for primary cosmic rays  with 
  varying composition and a 4-component fit is performed. 
  The comparison and fit are performed for energy bins, of width 0.1 or 0.2
  in $\log_{10} (E/{\rm eV})$, spanning the full range of the measured
  energies.  
  We also examine the mean $X_{\rm max}$ value as a function of
  energy for cosmic rays with energies greater than $10^{15.8}$ eV.  
  Below $10^{17.3}$ eV, the slope of the mean $X_{\rm max}$ as a 
  function of energy (the elongation rate) for the data is significantly 
  smaller than that of all elements in the models, indicating that the
  composition is becoming heavier with energy in this energy range. 
  This is consistent with a rigidity-dependent cutoff of events from galactic
  sources.  Finally, an increase in the $X_{\rm max}$ elongation rate is
  observed at energies just above $10^{17}$~eV indicating another change in the
  cosmic rays composition.
\end{abstract}


\section{Introduction}
\label{sec:introduction}

The TALE detector was designed to look for structure in the
energy spectrum and associated change in composition of cosmic rays
below the ``ankle'' structure at $10^{18.6}$~eV. A measurement of the 
cosmic ray energy spectrum using TALE observations in the energy range 
between $10^{15.3}$~eV and $10^{18.3}$~eV was published in a recent 
article~\citep{Abbasi:2018xsn}.  Here we present our results on the cosmic
ray composition from $10^{15.3}$~eV to $10^{18.3}$~eV.  Only the high elevation
telescopes of TALE observing $31^{\circ}$ to $55^{\circ}$, are used in this
analysis.  See section~\ref{sec:tale_detector} for the experimental setup.

Previous observations of cosmic rays composition for energies greater
than $10^{18}$~eV, such as those reported by HiRes~\citep{Abbasi:2004nz}, 
the Telescope Array (TA)~\citep{Abbasi:2014sfa}, 
and Auger~\citep{Aloisio:2013hya}, all suggest that the
transition from galactic to extragalactic sources occurs at an energy
below that of the ankle.  This transition is expected to be observable
in the form of a composition getting heavier up to a ``transition energy''
and then becoming lighter at higher energies.
Below the transition energy, galactic sources dominate the observed flux, 
while above the transition the sources of cosmic rays are mostly
extragalactic.

Several observations of the cosmic ray energy spectrum, including the one 
using TALE data, indicate the presence of a ``knee''-like structure in the 
$10^{17}$ decade, {a \it second knee}.  A change in the spectral index of the 
cosmic ray flux is also expected in the case 
of transition from galactic to extragalactic sources.  It is therefore logical 
to expect to see a correlated change in the flux and composition in the 
transition region.  This paper reports on the observation of just such a
correlated change.

We describe the detector and data collection 
in section~\ref{sec:tale_detector}.  We then briefly discuss event selection 
and event reconstruction procedures in section~\ref{sec:data_process}.
In section~\ref{sec:MC} we describe the Monte Carlo (MC) simulation,
and present the results of MC studies of the event reconstruction performance.
Section~\ref{sec:comp_analysis} presents an overview of the composition 
measurement procedures.  A discussion of the systematic uncertainties is 
presented in section~\ref{sec:systematics}.  
The measured composition is shown in section~\ref{sec:results_and_discussion},
along with a brief discussion of the measured results.  The paper concludes
with a summary in section~\ref{sec:summary}.

As the second paper on TALE data analysis, it is unavoidable that some of the 
material presented in this paper reproduces already published material in 
\citep{Abbasi:2018xsn}.  Furthermore, we refer the reader to that publication 
for a more detailed description of the TALE detector and data analysis.

\section{TALE Detector and Operation}
\label{sec:tale_detector}

The Telescope Array is an international collaboration with members from Japan,
U.S., South Korea, Russia, and Belgium.  The observatory is located in the
West Desert of Utah, about 150 miles southwest of Salt Lake City, and is the
largest cosmic ray detector in the northern hemisphere.  In operation
since 2008, TA consists of 507 scintillator surface detectors (SD), arranged
in a square grid of 1.2~km spacing~\citep{AbuZayyad:2012kk}.  A total of 38
telescopes are distributed among three FD stations located on the periphery
of the SD array~\citep{AbuZayyad:2012qk,Tokuno:2012mi}. The FD telescopes
observe the airspace above the SD array.  TA is the direct successor to both 
the Akeno Giant Air Shower Array (AGASA) and the High Resolution 
Fly’s Eye (HiRes) experiments~\citep{Teshima:1985vs,Sokolsky:2011zz}.
Telescope Array incorporates both the scintillation counter technique of
AGASA and the air fluorescence measurements of HiRes.  The goal of the
Telescope Array is to clarify the origin and nature of ultra-high energy cosmic 
rays~(UHECR) and the related extremely high energy phenomena in the universe.
The previous measurements of the energy spectrum, composition, 
and anisotropy in the arrival direction distribution for energies
above $10^{18.2}$~eV have been 
published~\citep{AbuZayyad:2012ru,Abbasi:2014sfa,Abbasi:2014lda}

A TA Low-Energy extension (TALE) fluorescence detector \citep{tale_icrc2011}
began operation in 2013 at the northern FD station (Middle Drum).  
Ten new TALE telescopes were added to the 14 telescopes which made up  the
TA FD at the site.  All 24 telescopes were refurbished from components
previously used by HiRes, and updated with new communications hardware.
The original 14 TA FD telescopes came from HiRes-I and were distributed in
two ``rings'' viewing $3^{\circ}$ to $31^{\circ}$ in elevation.
They are instrumented with Sample-and-Hold electronics.
The TALE FD telescopes, added to TA in 2013, came from HiRes-II and 
view $31^{\circ}$ to $59^{\circ}$ in elevation, directly above the field of view
of the main Telescope Array Telescopes.  The TALE telescopes are instrumented
with FADC electronics.

\begin{figure}[htb!]
  \centering
  \epsscale{1.2}
  \plotone{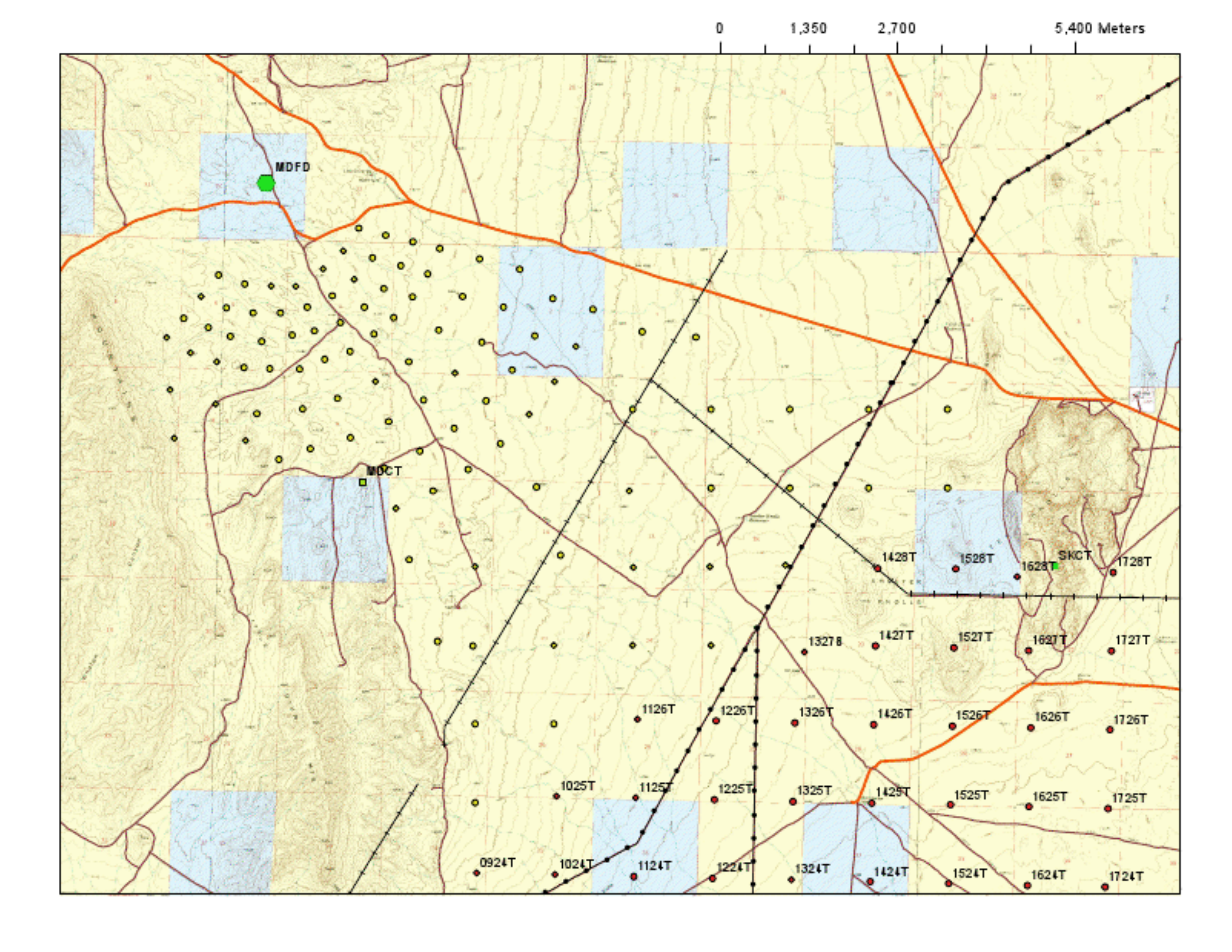}
  \caption{Map of the TALE detector.  
    The locations of the TALE surface detectors is indicated by the yellow
    circles.  The TALE detectors merge with the main Telescope Array surface
    array at the lower right of the figure.  
    Telescope Array surface detectors are shown by red circles with detector
    numbers.  The location of the TALE FD site can be seen near the top left
    of the figure indicated on the map by the green hexagon labeled MDFD}
  \label{fig:tale_sd_config}
\end{figure}

\begin{figure}[htb!]
\centering
\epsscale{1.0}
\plotone{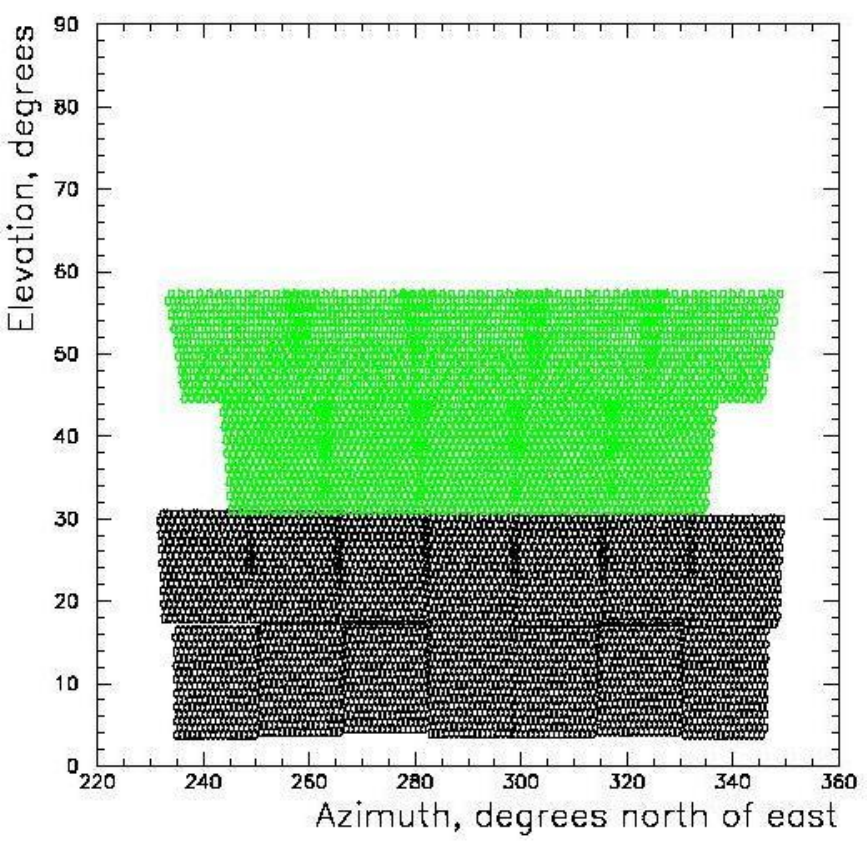}
\caption{Schematic of TALE/MD telescopes field of view, showing azimuthal
  and elevation coverage.
  Each small hexagon represents the field of view of a single PMT, 
  while the broad box outlines show the field of view of a 256 PMT telescope
  camera.  The view of the 14 TA telescopes is shown in black, 
  while that of the 10 TALE telescopes is shown in green.}
\label{fig:tale_fov}
\end{figure}

In addition to the ten new, high-elevation angle, FD telescopes, TALE also
added 103 new SD counters arranged in a graded spacing array.
See Figure~\ref{fig:tale_sd_config}.  Both the TA and TALE telescopes view
approximately southeast, over the Telescope Array and TALE SD arrays.
This arrangement is illustrated in Figure~\ref{fig:tale_fov}.

The TALE scintillator detectors were added after the telescopes in 2017.
Since the telescopes were taking data before the scintillators were deployed, 
the analysis described in this paper is based only
upon observations made by the FD component of TALE. 

The TALE FD telescopes were assembled from refurbished HiRes-II
telescopes~\citep{Boyer:2002zz}.  
The telescope mirrors are each made from four truncated circular segments 
which were assembled into a clover-leaf pattern.
The unobscured viewing area of each spherical mirror is approximately
$3.7$~m$^{2}$.  The focal plane of the telescope camera consists of 
256 (16x16) hexagonal photo-multiplier tubes (PMTs).
The Philips/Photonis XP3062 PMTs are 40-mm (flat-to-flat) where 
each PMT/pixel views a $1^{\circ}$  cone in the sky.  
The field-of-view (FOV) of each camera is about $16^{\circ}$ 
in azimuth by $14^{\circ}$ in elevation.

The TALE telescope electronics consist of a 
10 MHz FADC readout system with 8-bit resolution. 
Analog sums over the rows and columns of pixels, also
sampled at 8-bits, allow recovery of saturated PMTs in most cases.  
The trigger logic of the telescope also uses the digitized summed signals.
Systems for telescope GPS timing, inter-telescope triggers 
and communication to a central data acquisition (DAQ) computer 
use new hardware that resulted in a significant improvement 
in throughput over the old HiRes-II system, and 
are documented in ~\citep{Zundel_thesis}.

Useful events recorded by the TALE detector appear as tracks for which
the observed signal is comprised of a combination of direct Cherenkov
light (CL) and fluorescence light (FL), with some contribution from
scattered CL.  Contributions of light generated by these mechanisms are
both proportional to the number of charged particles in the extensive
air shower (EAS) at any point along its development.  Thus, CL signals
can be analyzed in a manner analogous to that for FL to determine the
energy of the cosmic ray, as well as to determine the depth of shower maximum
($X_{\rm max}$) which is related to the composition of the primary particle.

There are some important differences between the CL and FL measurements.
First, FL is emitted isotropically along the shower from the particles. 
In contrast, the CL is strongly peaked in the forward direction along the
shower axis.  As a result, CL falls off rapidly as the incident angle of
the shower to the detector increases.
In addition, the CL also accumulates along the shower track and
therefore increases in overall intensity as the shower develops.
Both types of light also undergo scattering in the atmosphere,
from both air molecules (Rayleigh scattering) as well as from particulate
aerosols.

At the lowest energies observable by TALE, events are dominated by CL.  
At higher energies, however, the FD becomes more sensitive to the 
isotropically emitted FL.
In the energy region between $10^{16.5}$eV and $10^{17.5}$eV, 
the shower events are typically recorded with a mix of both CL and FL.  Based 
upon our experience with the TALE energy spectrum calculation, we concluded
that a composition analysis, which requires accurate reconstruction of the
shower geometry, should be restricted to use only those events with a
significant contribution of direct CL.  Therefore, we restrict our analysis
to events with direct CL~$>35\%$ of the total recorded signal; see
Section 3 and Table 2 of~\citep{Abbasi:2018xsn}.

TALE FD data collected between June 2014 and November 2018 is included in
this analysis.  This data set includes, as a subset, the data set used for 
the energy spectrum paper~\citep{Abbasi:2018xsn}.  However, this data set
is more than double the size of the original spectrum data set.  The criteria
used for good-weather determination is the same as in the original analysis.
The total, good-weather, detector on-time in this period is $\sim$~2700 hours.

\section{Event Processing and Reconstruction}
\label{sec:data_process}

Most TALE FD data events are the result of noise triggers or very low-energy
air showers that can not be reliably used for physics analysis. An event
processing chain is used that filters out low quality events.  After filtering, 
the remaining events are subjected to full shower reconstruction, which
includes the determination of shower geometry, energy, and the depth of
shower maximum.

The event reconstruction procedure consists of the following main steps:

First, the shower-detector plane (SDP) is reconstructed from the pattern and 
pointing direction of the triggered PMT pixels.

Next, the arrival time of light at the detector (in each pixel) 
is fit as a function of the viewing angle of the pixel in the SDP:
\begin{equation}
  \label{eq:time_fit_flyseye}
  t_{i} = t_{0} + \frac{R_{p}}{c}\tan\left(\frac{\pi - \psi - \chi_{i}}{2}\right),
\end{equation}
where $R_{p}$ is the impact parameter or distance of closest approach from the
detector to the shower track, $\psi$ is the incline angle of the track within
the SDP, $t_{0}$ is a time offset, and $\chi_{i}$ is the viewing angle of the
{\it i}-th pixel.

The PMT signal is then fit to the light profile expected for a given energy 
and shower $X_{\rm max}$ according to the Gaisser-Hillas parameterization:
\begin{equation}
  \label{eq:time_gaisser_hillas}
  \resizebox{.85 \columnwidth}{!} {
    $N(x) = N_{\rm max} \left(\frac{x -
      X_{0}}{X_{\rm max}-X_{0}}\right)^{(X_{\rm max}-X_{0})/\lambda}
    \exp\left(\frac{X_{\rm max}-x}{\lambda}\right)$.
  }
\end{equation}
The parametrization gives the number of charged particles, $N$, at
atmospheric depth, $x$, along the shower track. $N_{\rm max}$, $X_{\rm max}$,
$X_{0}$, and $\lambda$ are parameters.  Here, $N_{\rm max}$ is the number of 
shower particles at the point of maximum shower development, $X_{\rm max}$.  
$X_{0}$ is a fit parameter roughly indicating the starting depth of the shower
and $\lambda = 70$~g/cm$^{2}$.  In combination with $X_{0}$, $\lambda$ sets
the width of the shower profile curve.
The fit produces two numbers of interest: $X_{\rm max}$, the depth of shower
maximum development and the shower's calorimetric energy. The calculation
of the total shower energy follows from the fit results, as explained below.

\begin{deluxetable}{l|c}[htb!]
  \tablecaption{Quality cuts applied to events used for the 
    composition measurement.
    Events meeting these conditions remain in the analysis. 
    \label{table:quality_cuts}}
  \tablecolumns{2}
  \tablehead{
    \colhead{{\bf Variable}} & \colhead{{\bf CL}}
  }
  \startdata
  Angular Track-length [deg]                        & $trk > 6.1^{\circ}$                \\ 
  Inverse Angular Speed [$\mu$s~deg$^{-1}$]          & $0.014<1/\omega<0.1$              \\
  Shower Impact Parameter [km]                      & $0.4<R_{p}<5.0$                    \\
  Shower Zenith Angle [deg]                         & $28^{\circ} < \theta < 65^{\circ}$   \\
  Shower $X_{\rm max}$ [g~cm$^{-2}$]                   & $435<X_{\rm max}<920$               \\ \hline
  Estimated Fit Error on Energy                    & $\delta E/E < 0.6$                \\ 
  Estimated Fit Error on $X_{\rm max}$ [g~cm$^{-2}$]  & $\delta X_{\rm max} < 200$          \\ 
  Timing Fit $\chi^{2} / dof$                        & $\chi^{2}_{tim}<4.5$                \\
  Profile Fit $\chi^{2} / dof$                       & $\chi^{2}_{pfl}<12$                 \\
  \enddata
\end{deluxetable}

The profile constrained geometry fit (PCGF)~\citep{AbuZayyad:2000vea} was
used to reconstruct this TALE data.  When applied to TALE events with
a significant CL signal, we found that the PCGF results in very good geometry
resolutions~\citep{Abbasi:2018xsn}.  Based upon MC studies, we determined
that a direct CL fraction of
at least 35\% was optimal for maintaining good geometrical
reconstruction and at the same time increasing event statistics at 
higher energies, approaching $10^{18}$~eV.

The PCGF reconstruction produces an estimate for the shower calorimetric 
energy.  To obtain the total shower energy, i.e. the primary CR particle
energy, a missing energy correction is applied.
This correction is composition dependent, and is therefore applied after
the best fit composition parameters have been determined.
We refer to the primary mixture obtained 
from fitting TALE data as ``TXF'', for {\bf T}ALE $\bm{X}_{max}$ 
distributions {\bf F}its, see Section~\ref{sec:comp_analysis}.  
Post-reconstruction, event selection criteria (quality cuts) are summarized in 
Table~\ref{table:quality_cuts}.

\section {Simulation}
\label{sec:MC}

We use Monte Carlo simulations to study the detector efficiency and
reconstruction resolution.  Two sets of simulations were generated for this 
analysis using different hadronic interaction models.  The first set of
simulations was based upon QGSJetII-03~\citep{Ostapchenko:2007qb} 
and a second set was based upon EPOS-LHC~\citep{Pierog:2013ria}.
QGSjetII-03 is the model that was previously used for the TALE energy
spectrum measurement~\citep{Abbasi:2018xsn}, while the EPOS-LHC model
is a ``post-LHC'' model, i.e. a hadronic interaction model that has been
updated with LHC data.

\begin{figure}[htb!]
\centering
\epsscale{1.2}
\plotone{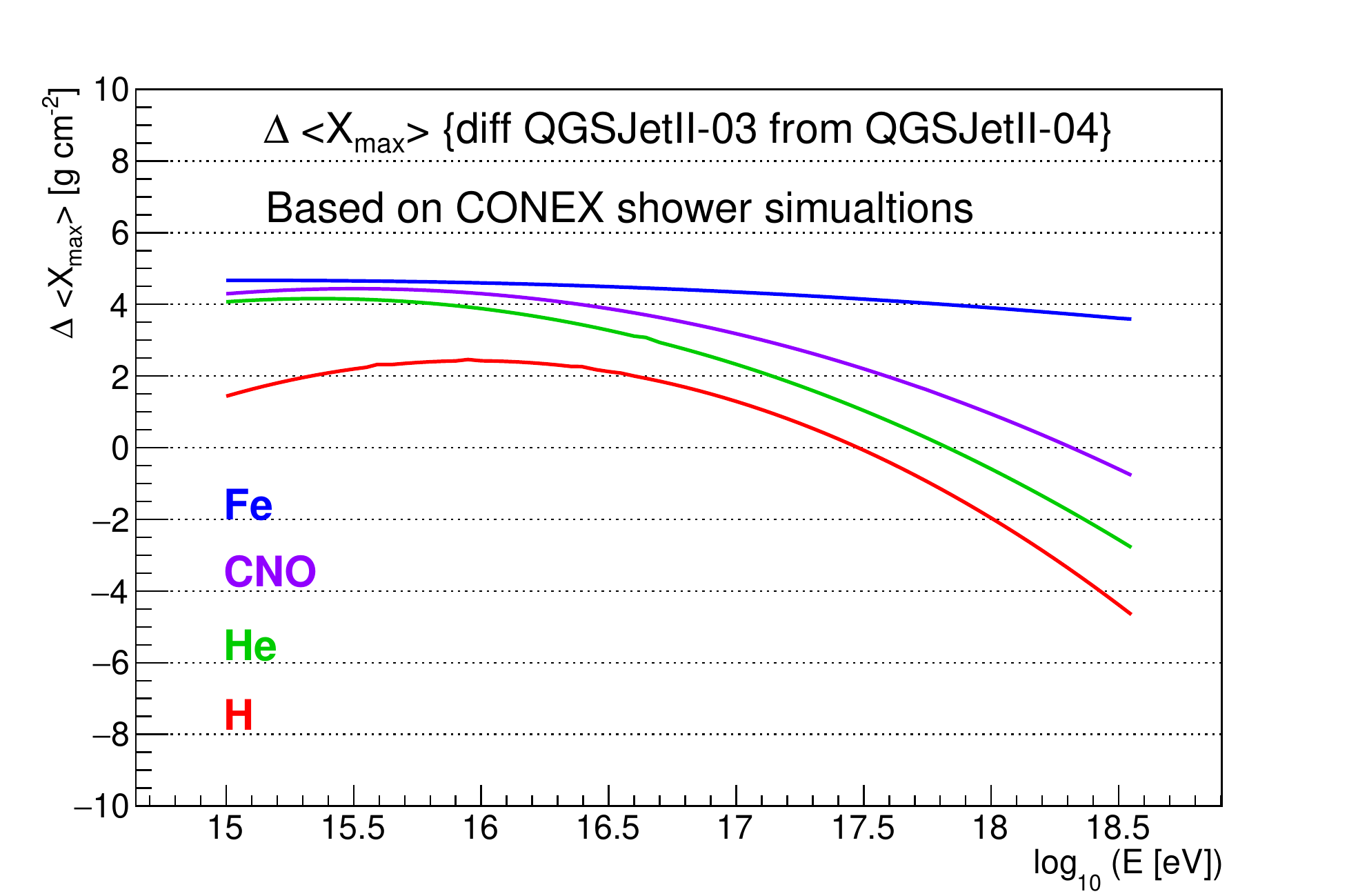}
\caption{Difference in mean $X_{\rm max}$ predictions for the
         QGSjetII-03 and QGSjetII-04 hadronic interaction models. 
         The difference between the mean shower maximum in showers generated 
         using QGSjetII-03 versus those using QGSJetII-04 is shown as a
         function of energy for four different cosmic ray primaries:
         H, He, CNO, and Fe. 
         The shower simulations were generated using CONEX.}
\label{fig:diff_qgs0203_qgs0204}
\end{figure}

The full processing of a set of simulations is a time consuming process.
This made it unfeasible for us to perform the full analysis using other,
post-LHC, hadronic interaction models, such as
QGSJetII-04~\citep{Ostapchenko:2010vb}.
We do note, however, that a comparison of CONEX~\citep{Bergmann:2006yz}
simulations in the energy range of interest for this publication, shows
that the air shower's $\langle X_{\rm max} \rangle$ predictions of 
QGSJetII-03 are within 5.0~g/cm$^{2}$ of the QGSJetII-04 model for all
of the four primaries used in this analysis, as demonstrated in
Figure~\ref{fig:diff_qgs0203_qgs0204}.

For both sets of simulations, with the QGSjetII-03 and EPOS-LHC hadronic
interaction models, a uniform mixture of four primaries \{H, He, N, Fe\}
was simulated for the energy range of 10$^{15}$~-~10$^{18.5}$~eV. 
The showers were thrown following a power law flux with a spectral index
of -2.92.  The MC shower events were then re-weighted to fit a broken power
law spectrum consistent with the TALE energy spectrum measurement.

A detailed simulation of the TALE detector response to cosmic ray generated 
air showers is performed.  For each hadronic interaction model, a library
of air showers generated using the CONEX package is used as input to the
detector MC.  Light production by shower particles and light propagation
to the detector, including accurate photon arrival time determination,
is performed.  This is followed by a detailed simulation of the detector
optics and electronics signal development, and finally by detector
trigger and event forming logic simulation.  \added{A more complete discussion
  of the simulation can be found in section 4 of~\citet{Abbasi:2018xsn}.}
 
The MC is generated for each data collection time interval that the TALE
telescope station was operated.  Nightly atmospheric conditions (three hour
interval GDAS~\citep{GDAS} database), and nightly detector calibration
information is incorporated into the simulation.  Each MC data set is about
twice size of the actual data set. 

\begin{figure*}[htb!]
  \centering
  \gridline{
    \fig{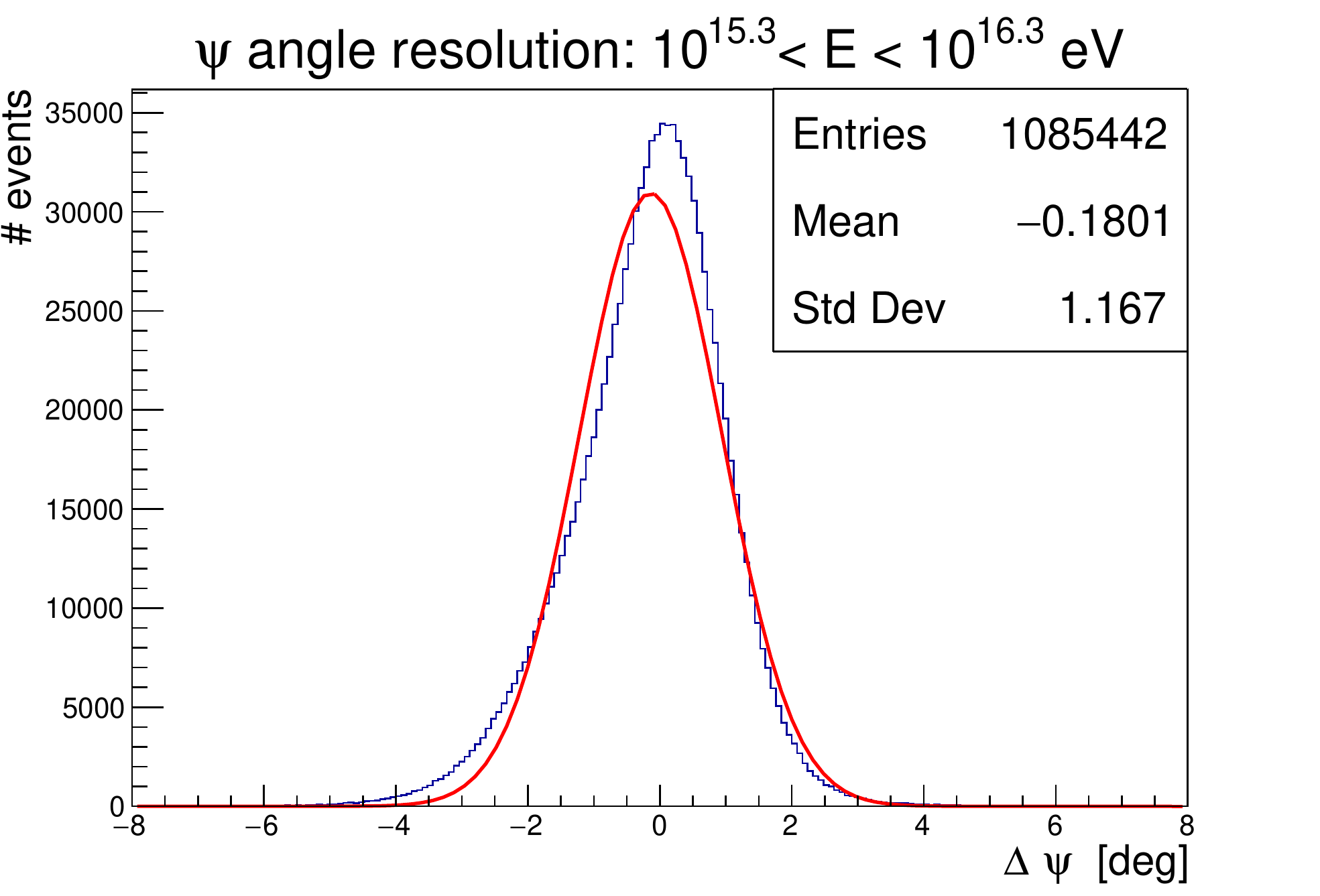}{0.33\textwidth}{(a)}
    \fig{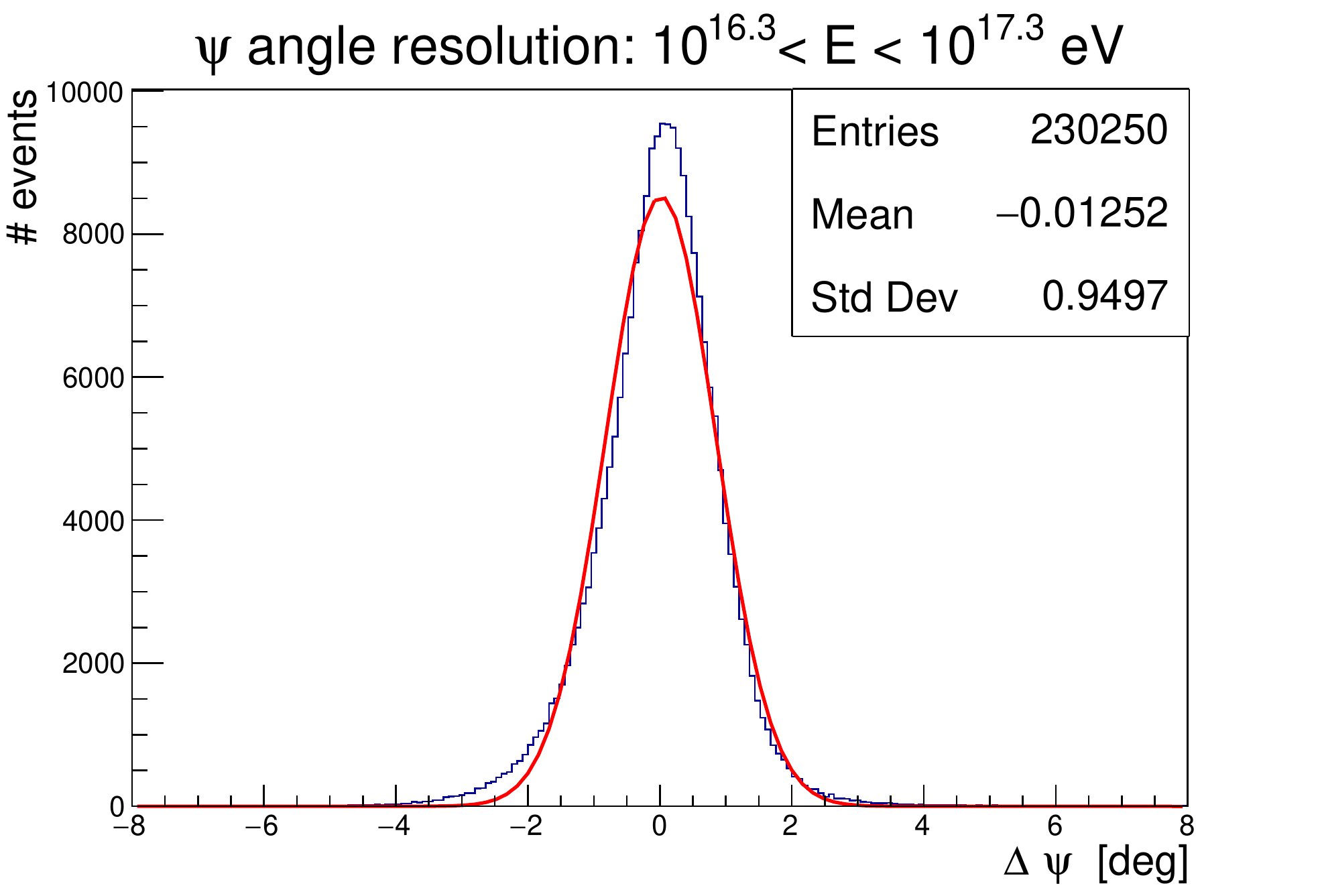}{0.33\textwidth}{(b)}
    \fig{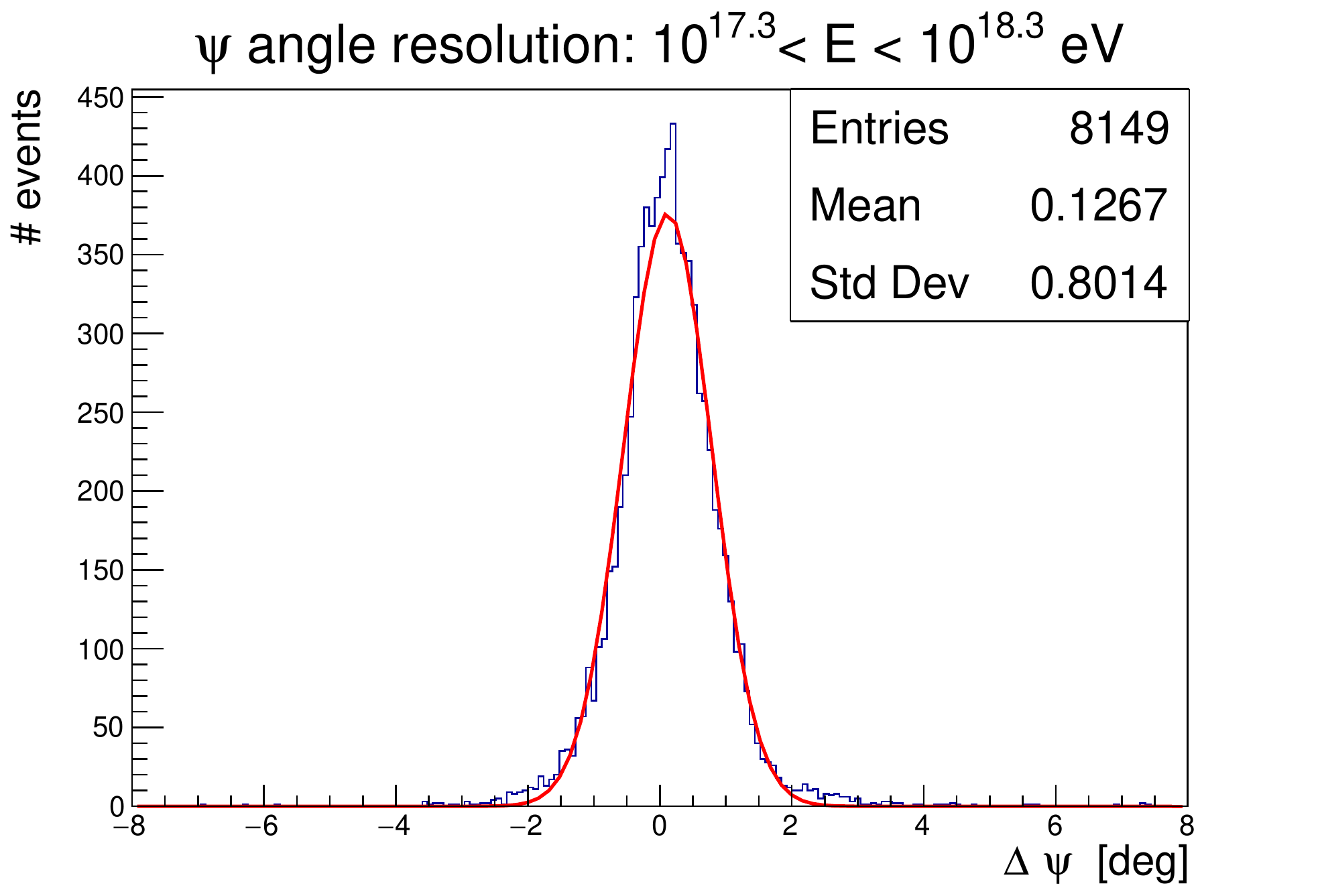}{0.33\textwidth}{(c)}
  }
  \caption{Reconstruction resolution of the shower angle in the shower-detector
    plane, $\psi$.
    The red curve is a Gaussian fit to the distribution of the uncertainty,
    $\Delta \psi$~[deg].  Distribution statistics are displayed in top-right
    box.
    The MC uses the EPOS-LHC hadronic generator and a mixed composition,
    matching the TXF results.
    From left to right, the histograms show the distribution of the uncertainty,
    $\Delta \psi$~[deg], for events reconstructed in three energy ranges:
    $10^{15.3}$ - $10^{16.3}$~eV, $10^{16.3}$ - $10^{17.3}$~eV,
    and $10^{17.3}$ - $10^{18.3}$~eV.
    The resolution of the shower angle, $\psi$, along with the resolution of the
    impact parameter, $R_{p}$, determines the accuracy of the shower track
    reconstruction.}
  \label{fig:res_plots_spectral_set_1}
\end{figure*}

\begin{figure*}[htb!]
  \centering
  \gridline{
    \fig{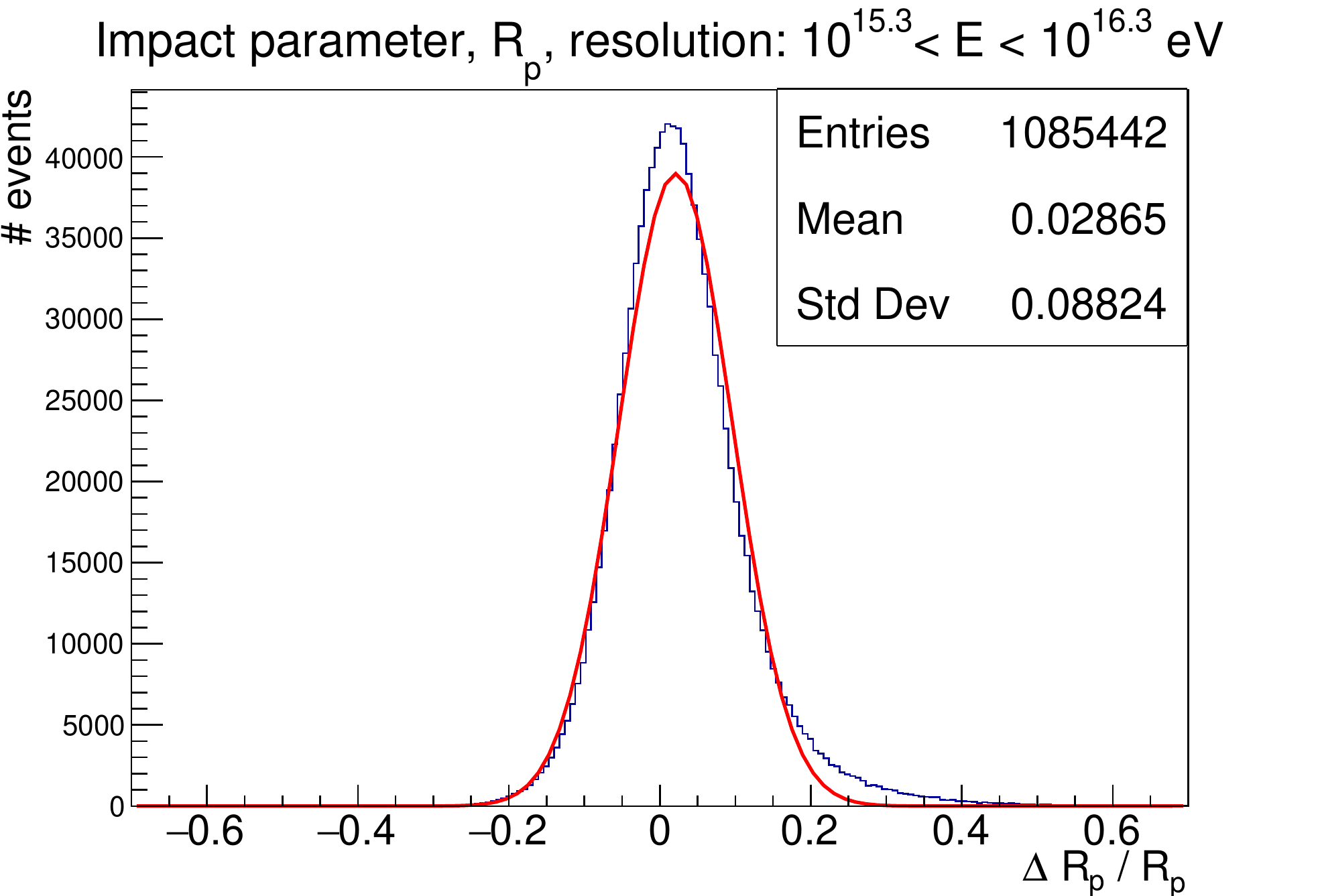}{0.33\textwidth}{(a)}
    \fig{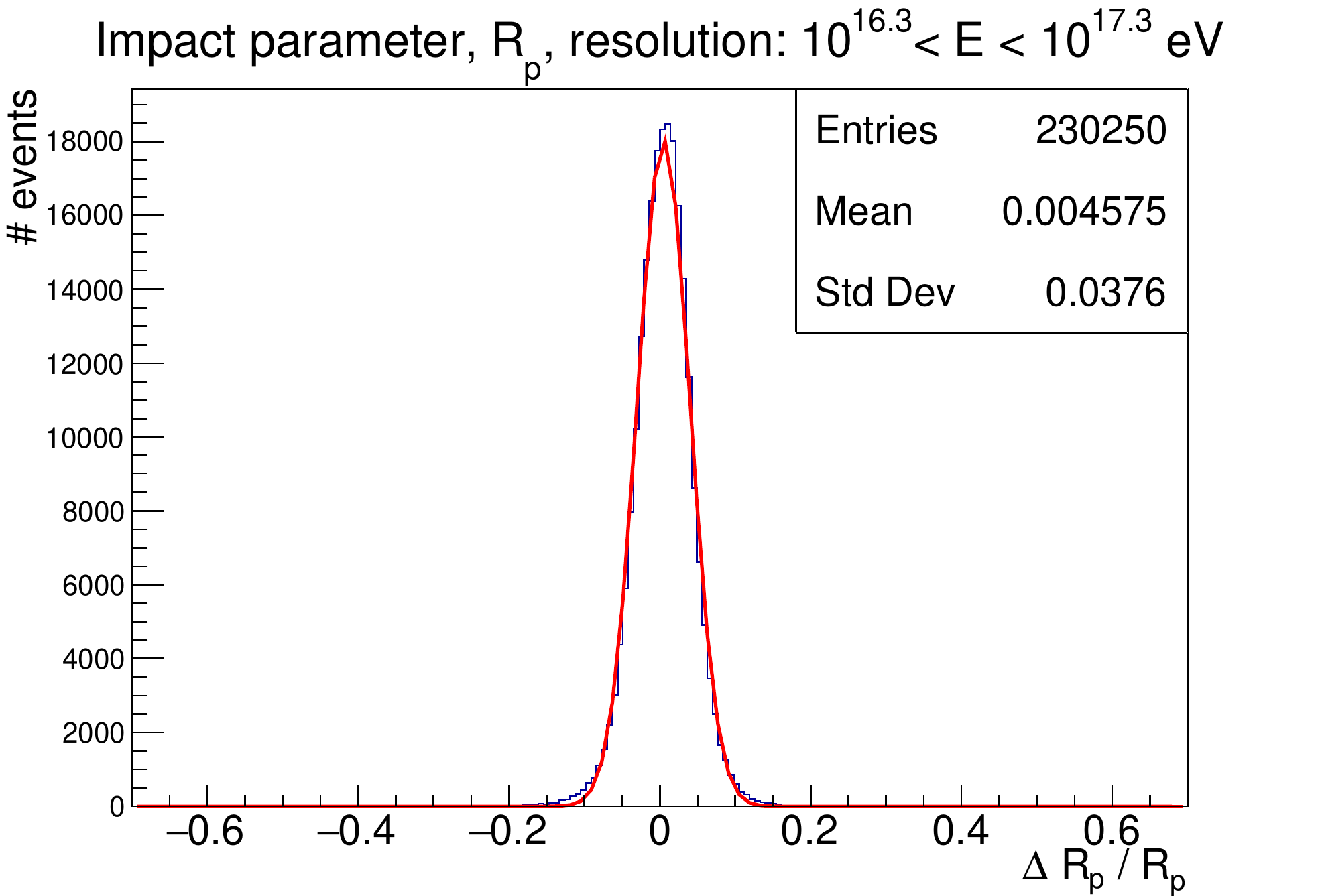}{0.33\textwidth}{(b)}
    \fig{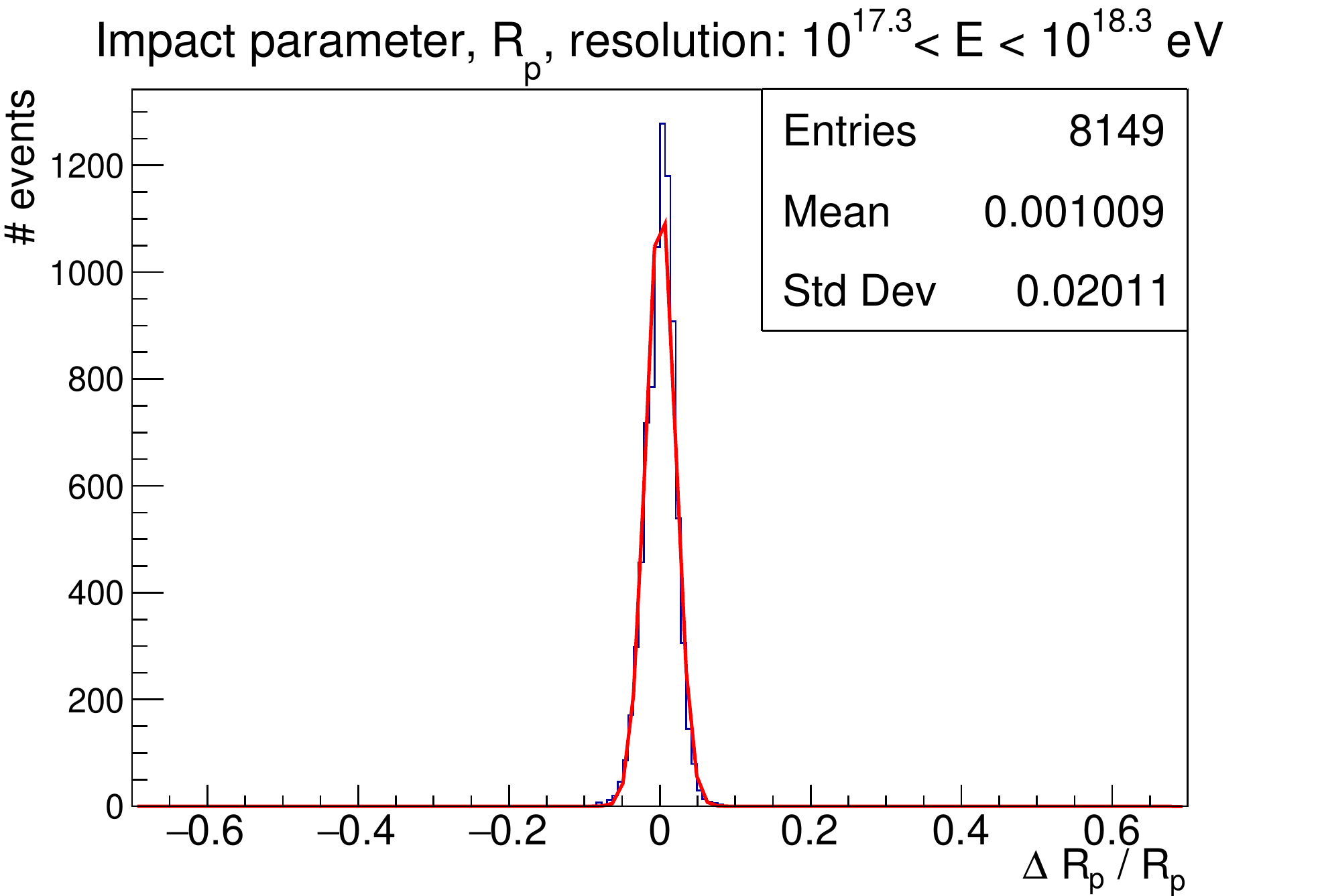}{0.33\textwidth}{(c)}
  }
  \caption{Reconstruction resolution of the shower impact parameter, $R_{p}$.
    The red curve is a Gaussian fit to the distribution of the fractional
    uncertainty,
    $\Delta R_p / R_p$.  Distribution statistics are displayed in top-right box.
    The MC uses the EPOS-LHC hadronic generator and a mixed composition,
    matching the TXF results.
    From left to right, the histograms show the distribution of the fractional
    uncertainty, $\Delta R_p / R_p$, for events reconstructed in three energy
    ranges:
    $10^{15.3}$ - $10^{16.3}$~eV, $10^{16.3}$ - $10^{17.3}$~eV,
    and $10^{17.3}$ - $10^{18.3}$~eV.
    As can be seen, the resolution improves significantly with energy.}
  \label{fig:res_plots_spectral_set_2}
\end{figure*}

\begin{figure*}[htb!]
  \centering
  \gridline{
    \fig{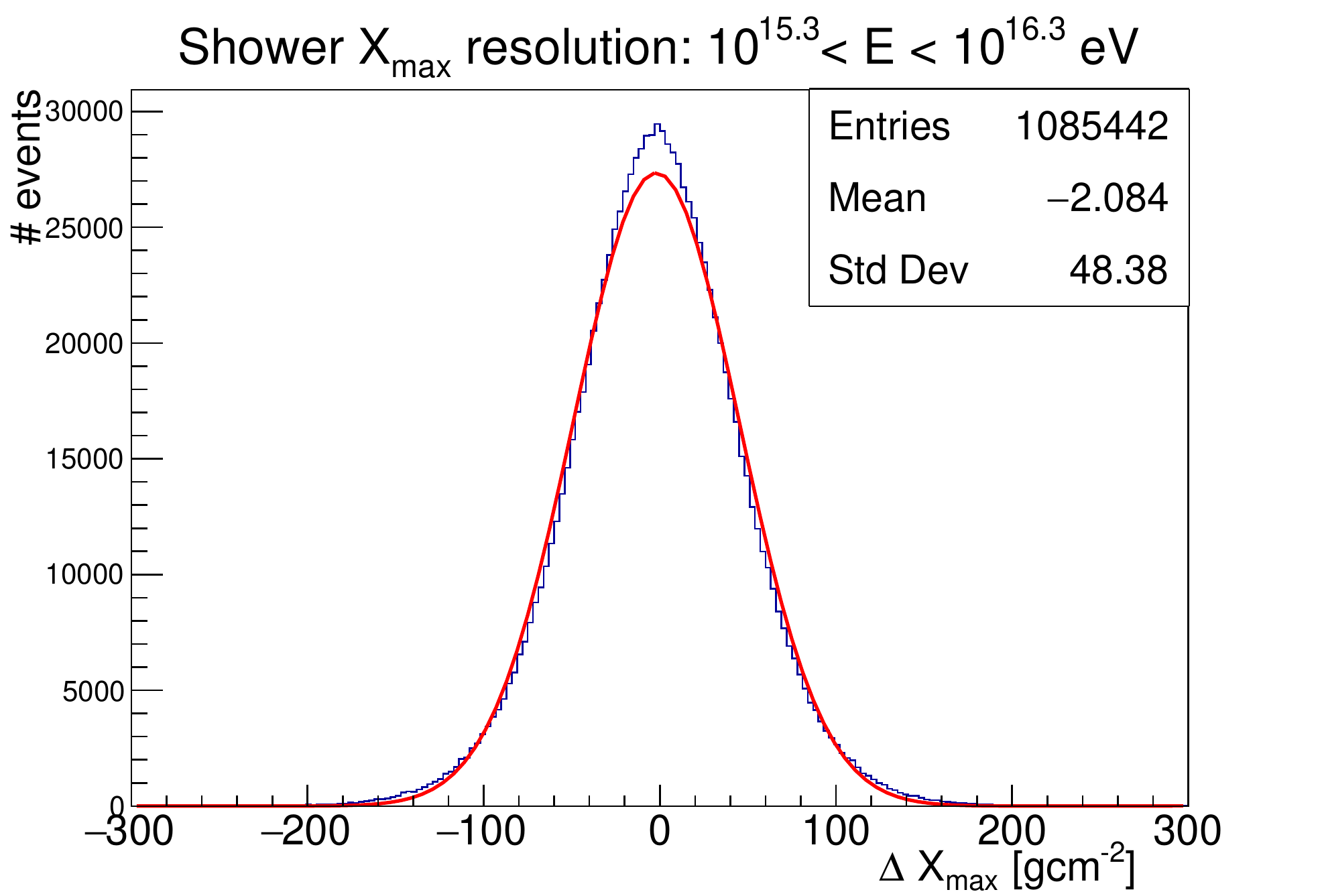}{0.33\textwidth}{(a)}
    \fig{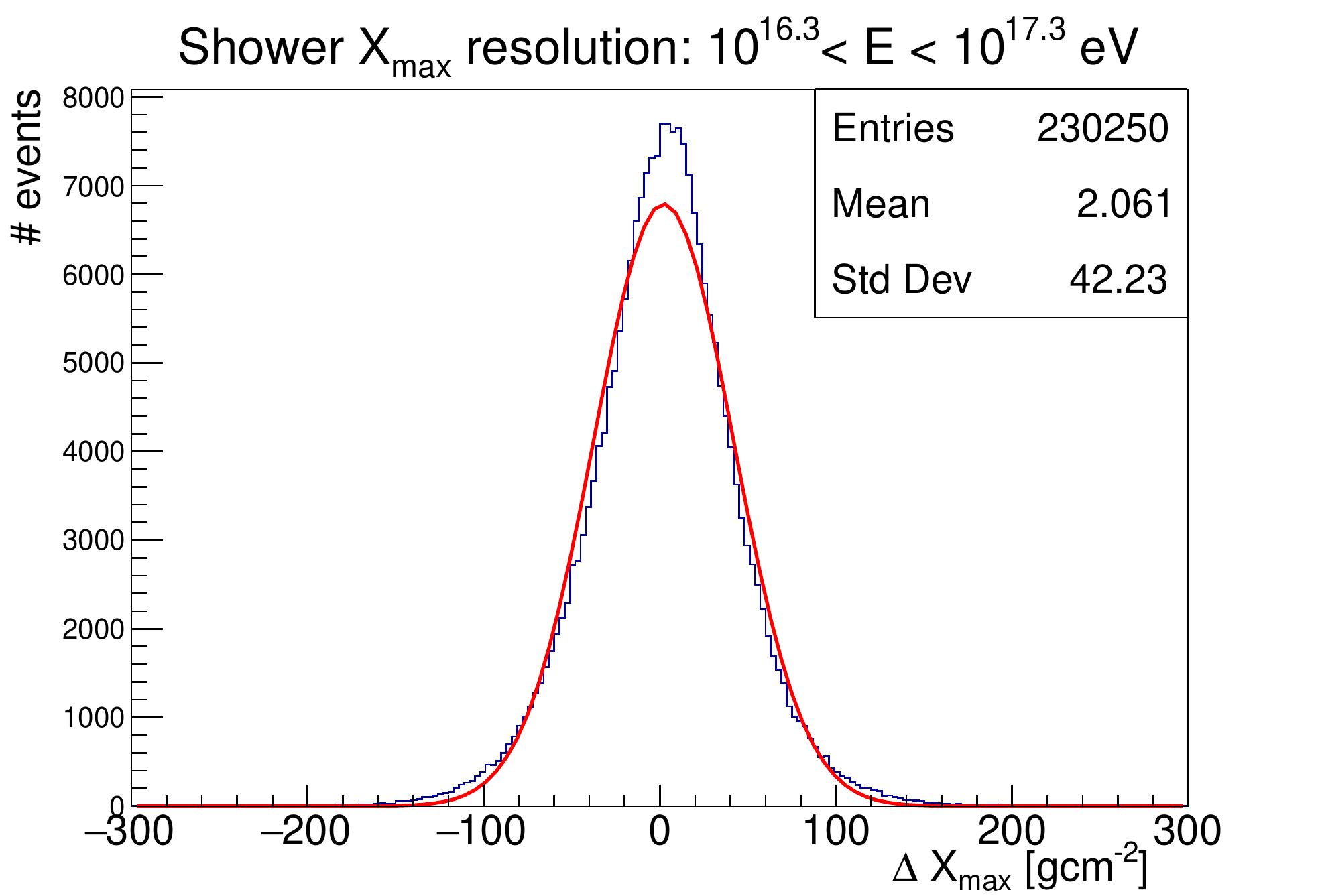}{0.33\textwidth}{(b)}
    \fig{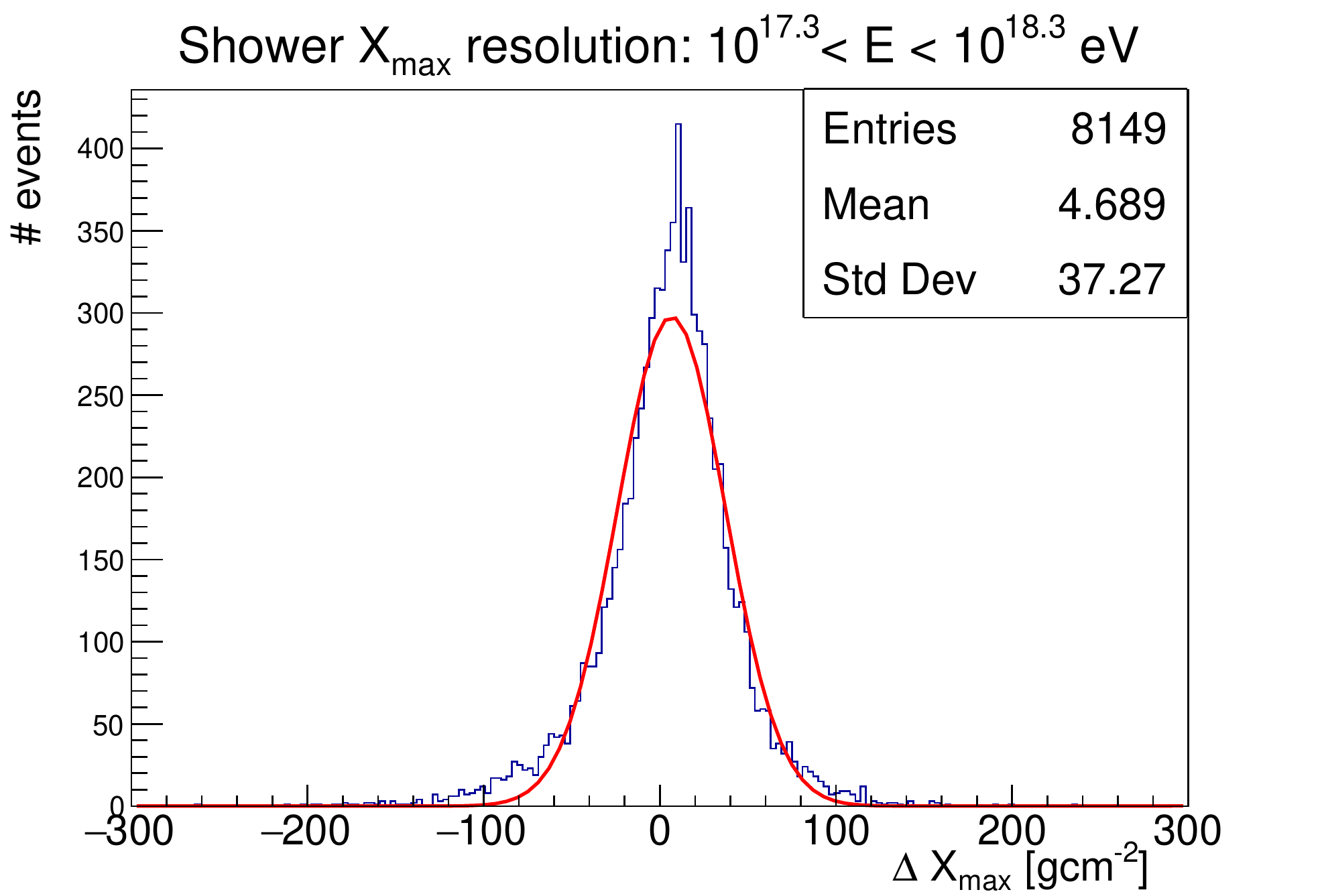}{0.33\textwidth}{(c)}
  }
  \caption{Reconstruction resolution of shower maximum, $X_{\rm max}$.
    The red curve is a Gaussian fit to the distribution of the uncertainty,
    $\Delta X_{max}~[{\rm g~cm}^{-2}]$.  Distribution statistics are displayed
    in top-right box.
    The MC uses the EPOS-LHC hadronic generator and a mixed composition,
    matching the TXF results.
    From left to right, the histograms show the distribution of the the
    uncertainty,
    $\Delta X_{max}~[{\rm g~cm}^{-2}]$, for events reconstructed in different
    energy ranges:
    $10^{15.3}$ - $10^{16.3}$~eV, $10^{16.3}$ - $10^{17.3}$~eV,
    and $10^{17.3}$ - $10^{18.3}$~eV.}
  \label{fig:res_plots_spectral_set_3}
\end{figure*}

\begin{figure*}[htb!]
  \centering
  \gridline{
    \fig{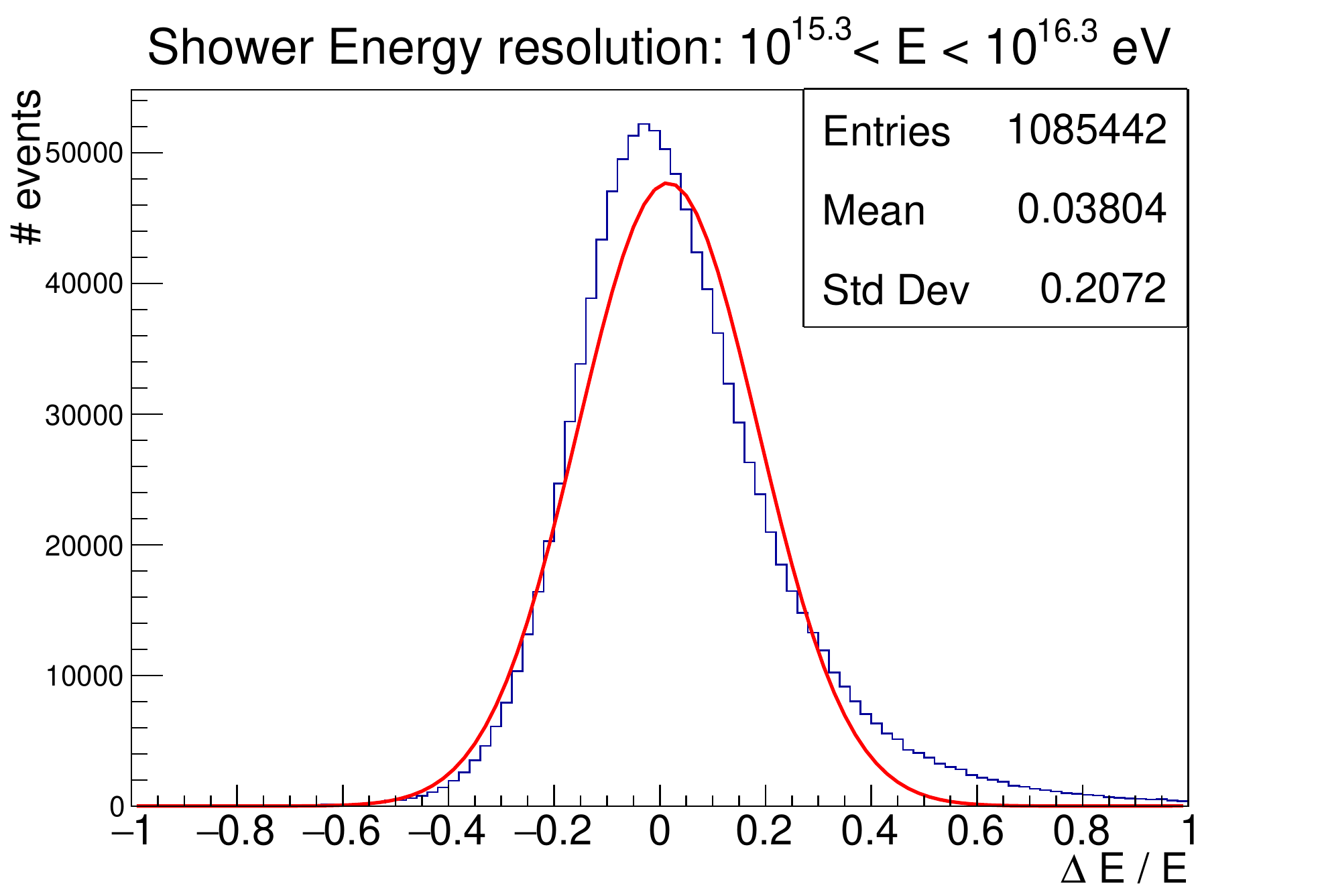}{0.33\textwidth}{(a)}
    \fig{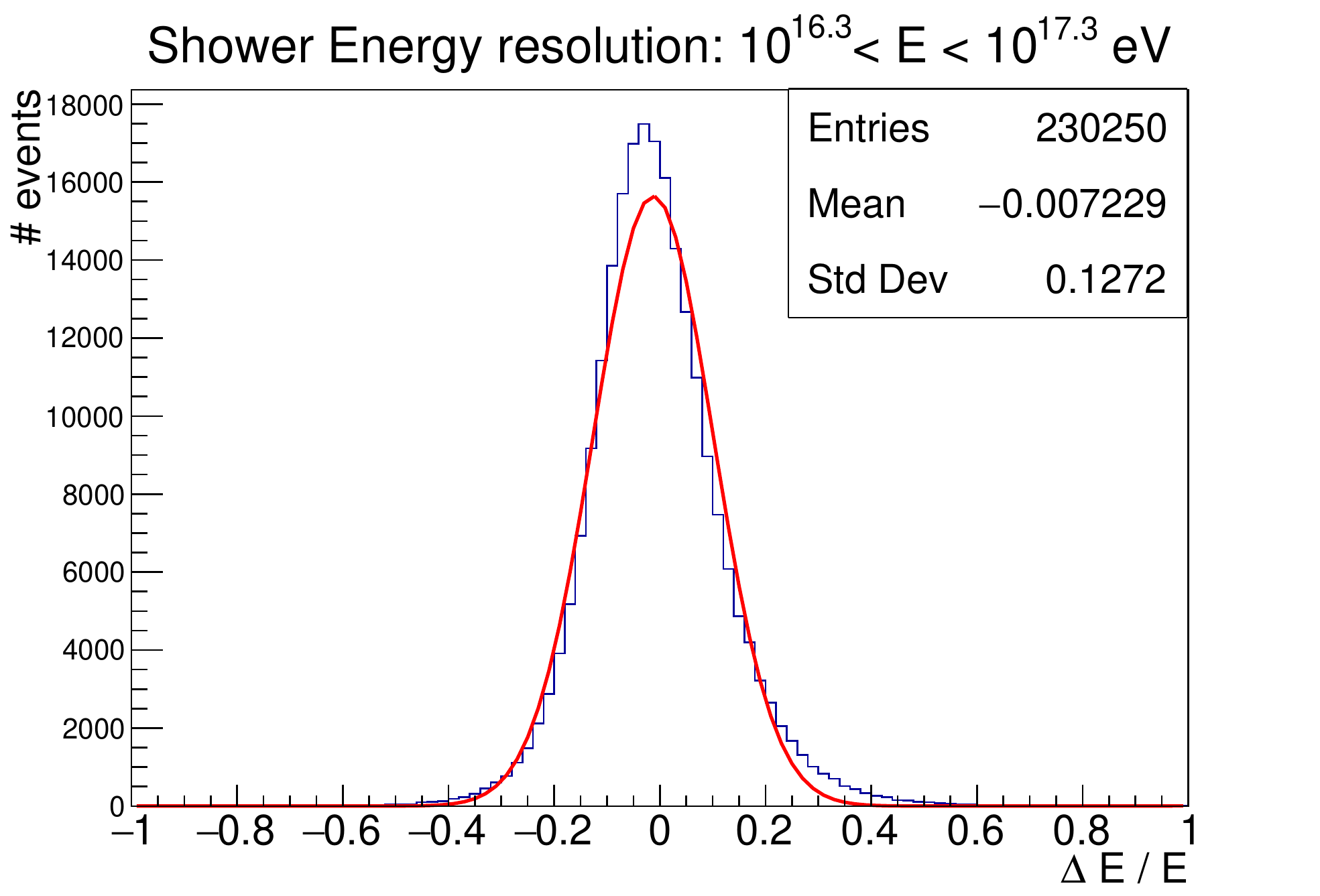}{0.33\textwidth}{(b)}
    \fig{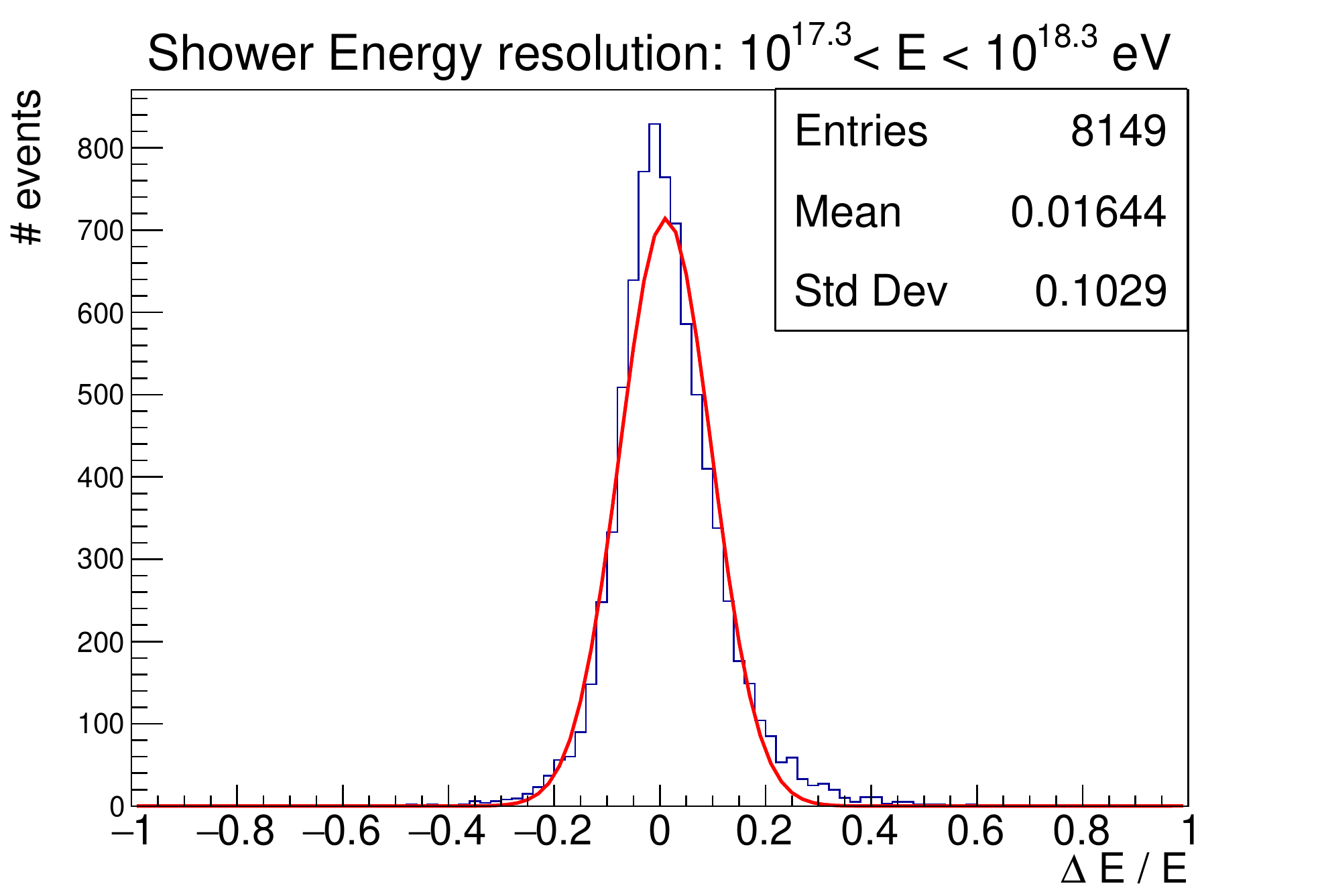}{0.33\textwidth}{(c)}
  }
  \caption{Reconstruction resolution of the shower energy, $E$.
    The red curve is a Gaussian fit to the distribution of the fractional
    uncertainty,
    $\Delta E / E$.  Distribution statistics are displayed in top-right box.
    The MC uses the EPOS-LHC hadronic generator and a mixed composition,
    matching the TXF results.
    From left to right, the histograms show the distribution of the fractional
    uncertainty, $\Delta E / E$, for events reconstructed in three energy
    ranges:
    $10^{15.3}$ - $10^{16.3}$~eV, $10^{16.3}$ - $10^{17.3}$~eV,
    and $10^{17.3}$ - $10^{18.3}$~eV.}
  \label{fig:res_plots_spectral_set_4}
\end{figure*}
 
Simulated MC showers pass through the same event selection criteria as the
real data and they are reconstructed using the same program/procedure.  
A missing energy correction is applied to both the reconstructed data and
MC showers based on the same composition assumption, with the correction for
each primary type and energy being estimated from the CONEX generated showers.

Here, the cosmic rays composition is described by the fit fractions
of the various primaries obtained in this study.  The fitting procedure
uses the measured calorimetric energy of the showers.  After the fit
fractions are calculated, this information is incorporated into the analysis
scripts to estimate the total shower energy for each event as a weighted
average over the four primaries used in the simulations, with the fit
fractions as weights.

We next present event reconstruction performance, namely resolution and bias of
reconstructed shower parameters.
The most relevant shower parameters are: 
\begin{enumerate}
\item The angle in the shower-detector plane, $\psi$
\item The shower impact parameter to the detector, $R_{p}$
\item The depth of shower maximum, $X_{\rm max}$, and  
\item The shower energy, $E$.
\end{enumerate}

The first set of results is shown for all MC showers, i.e. four primaries.
The same number of showers were generated for each primary, but the detection
and reconstruction efficiencies are different for each primary and therefore
the final number of showers is different.  Note that, the results shown
here use the EPOS-LHC simulation set.  The results using the QGSJetII-3
hadronic generator are similar.

Figures~\ref{fig:res_plots_spectral_set_1}~-~\ref{fig:res_plots_spectral_set_4}
show the difference between the reconstructed and thrown values of
simulated events, i.e. the reconstruction resolution of the shower parameters.
In light of the steeply falling number of events with
energy, each figure is shown as three separate plots, one per energy decade.

As can be seen from the figures, the reconstruction performance improves with
energy.  The $\psi$ resolution for the three energy ranges is $1.1^{\circ}$,
$0.83^{\circ}$, and $0.67^{\circ}$.  The impact parameter fractional error,
$dR_p/R_p$, expressed as a percent is 7.5\%, 3.5\%, and 2.0\%.  The
$X_{\rm max}$ resolution averaged over the four primaries is 47, 40,
and 31~g~cm$^{-2}$. In all cases, the bias in the reconstruction is small
compared to the resolution.  Figure~\ref{fig:res_plots_spectral_set_4} shows
the resolutions for the reconstructed energy, $dE/E$.
The energy resolutions are 17\%, 11\%, and 9\% for the three energy bins.  
For the full range of the data set, we see negligible bias in the
reconstructed energy values.  Note that the energy estimate here includes
the missing energy correction.

\begin{figure*}[htb!]
  \centering
  \gridline{
    \fig{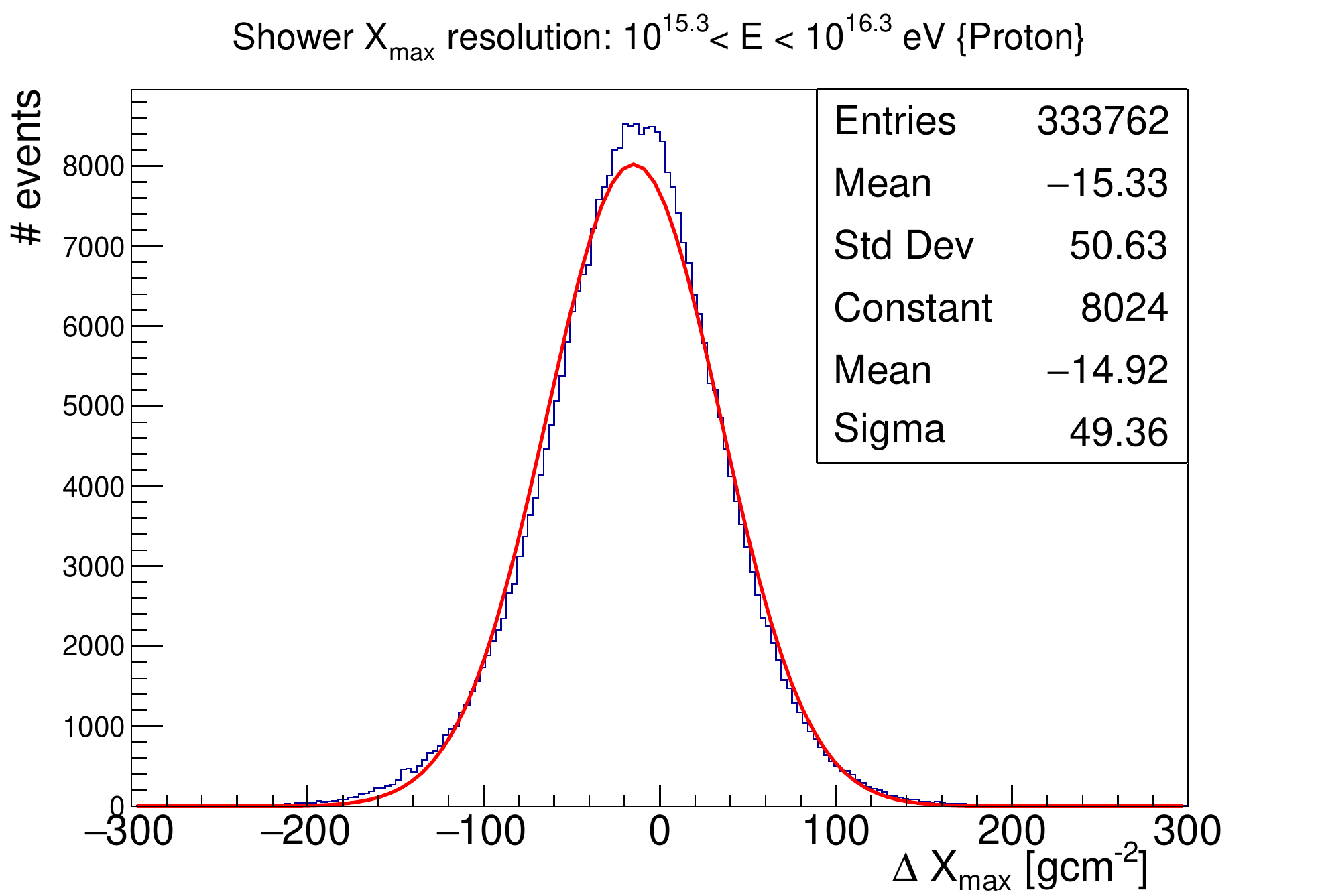}{0.33\textwidth}{(1a)}
    \fig{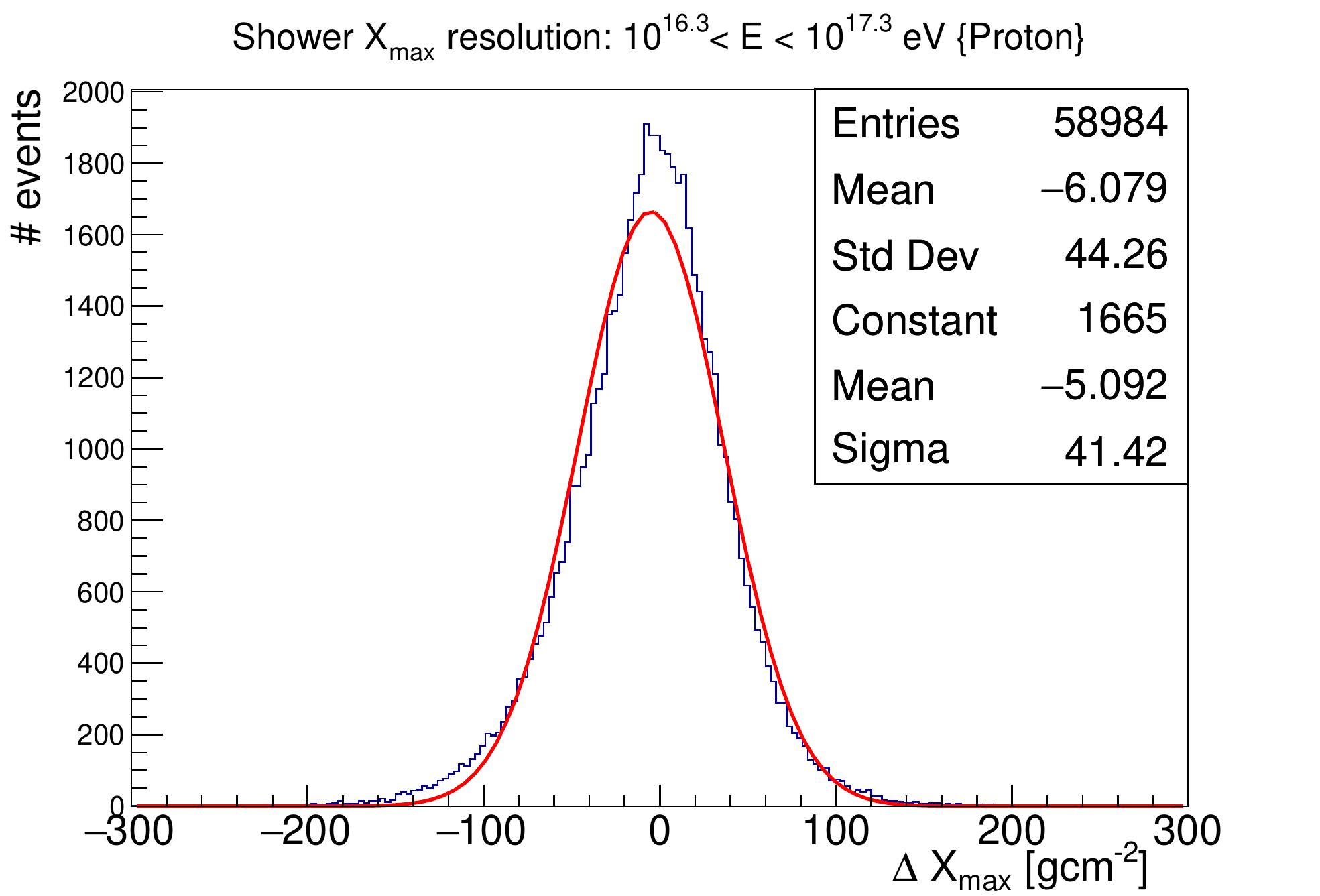}{0.33\textwidth}{(1b)}
    \fig{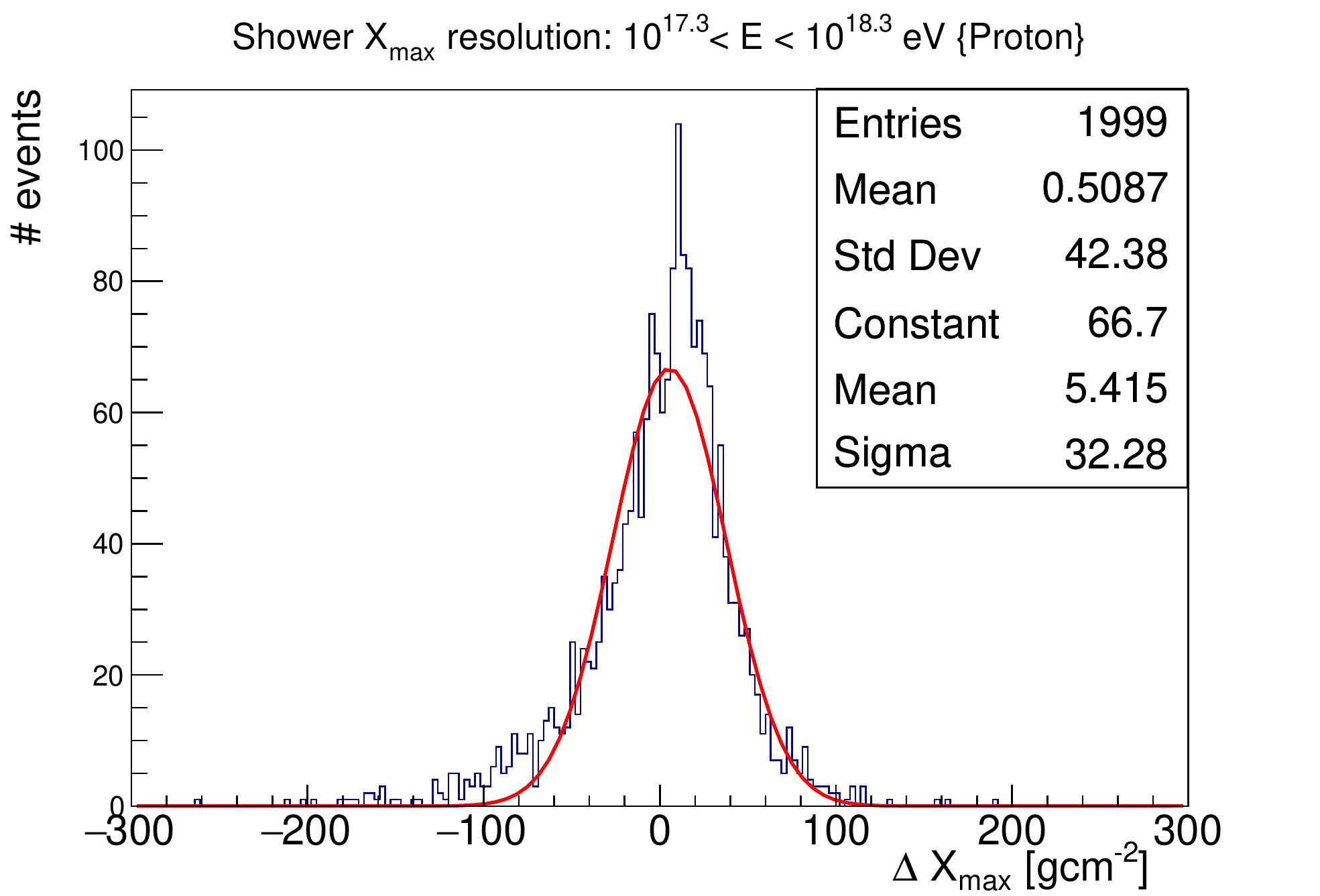}{0.33\textwidth}{(1c)}
  }
  \gridline{
    \fig{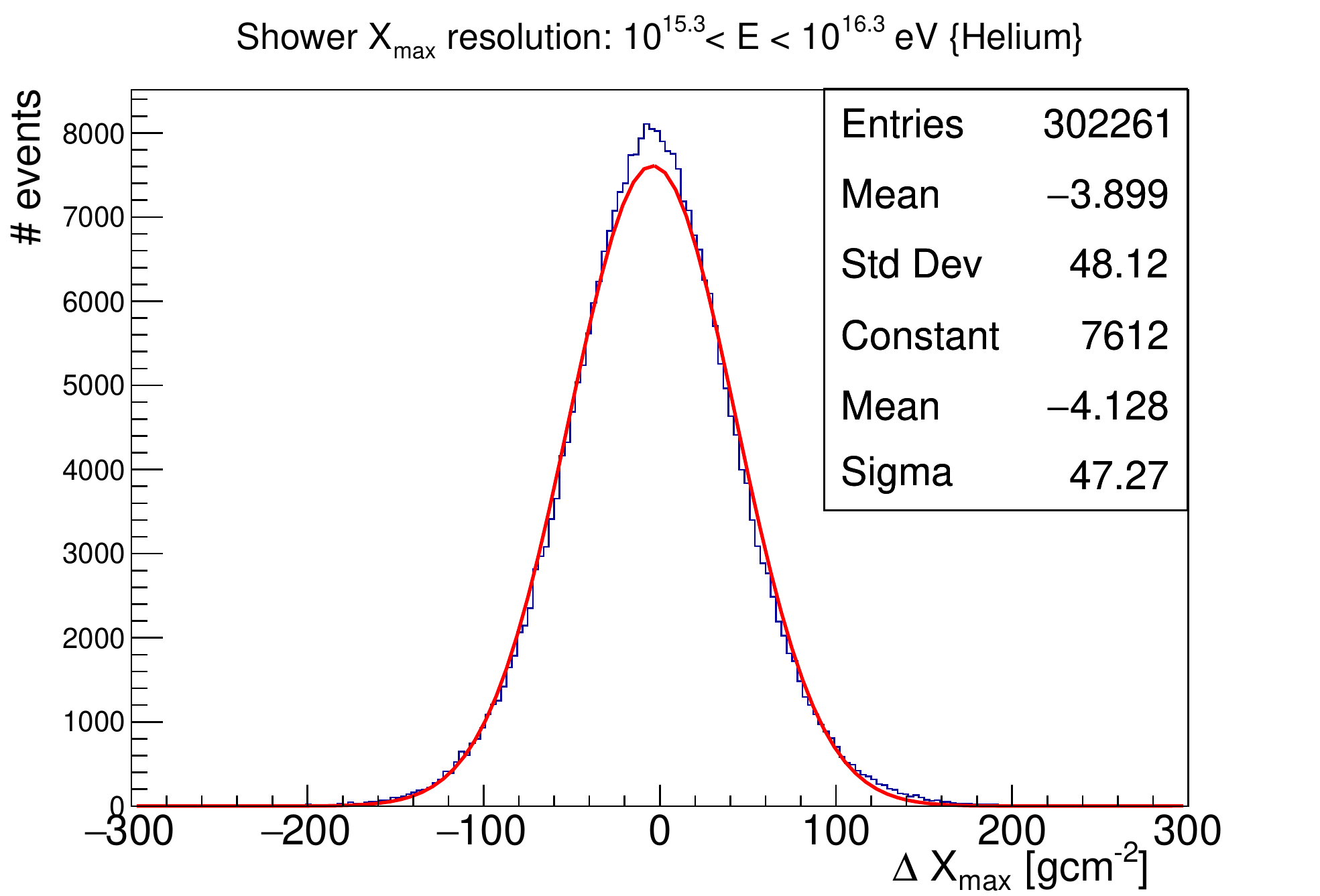}{0.33\textwidth}{(2a)}
    \fig{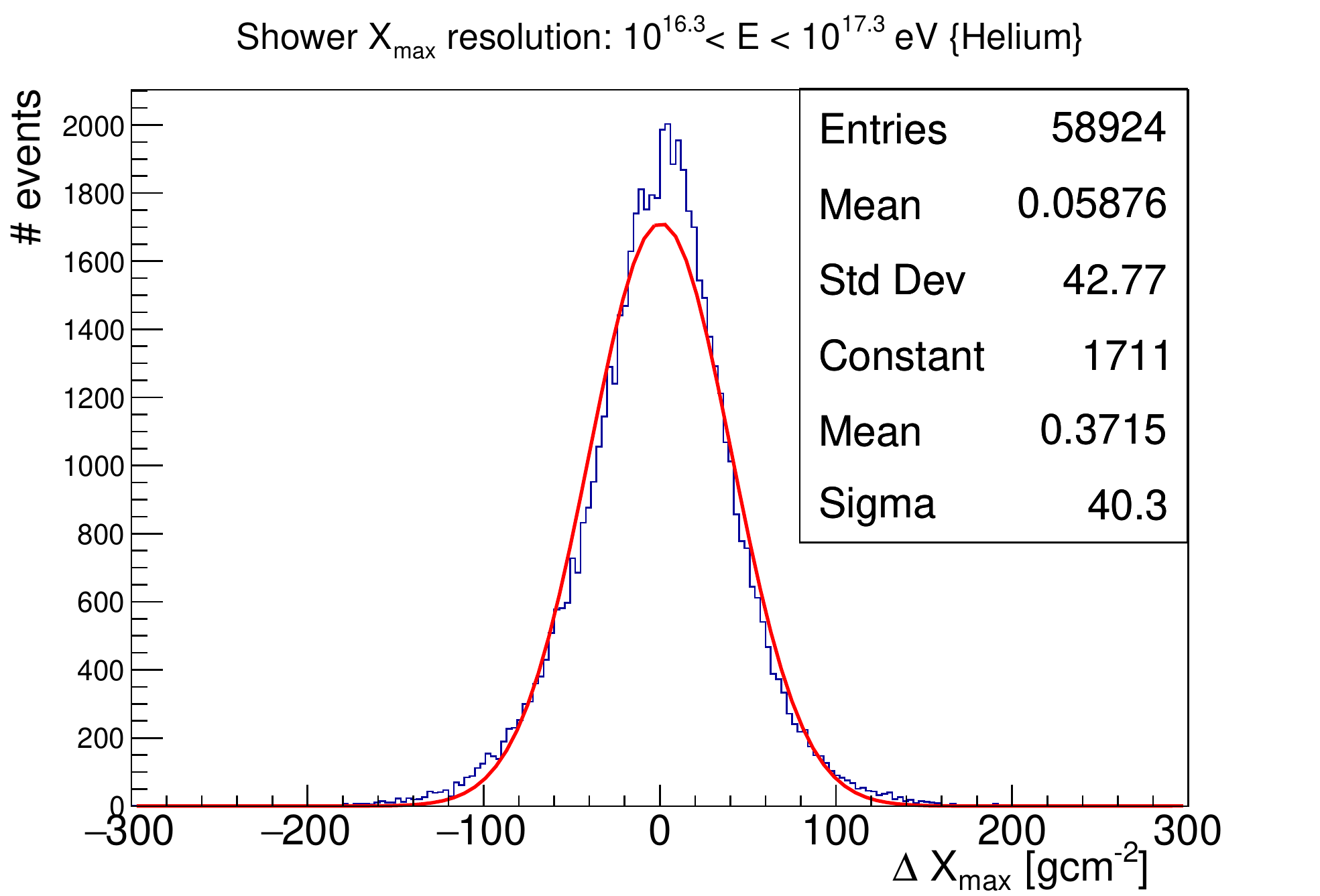}{0.33\textwidth}{(2b)}
    \fig{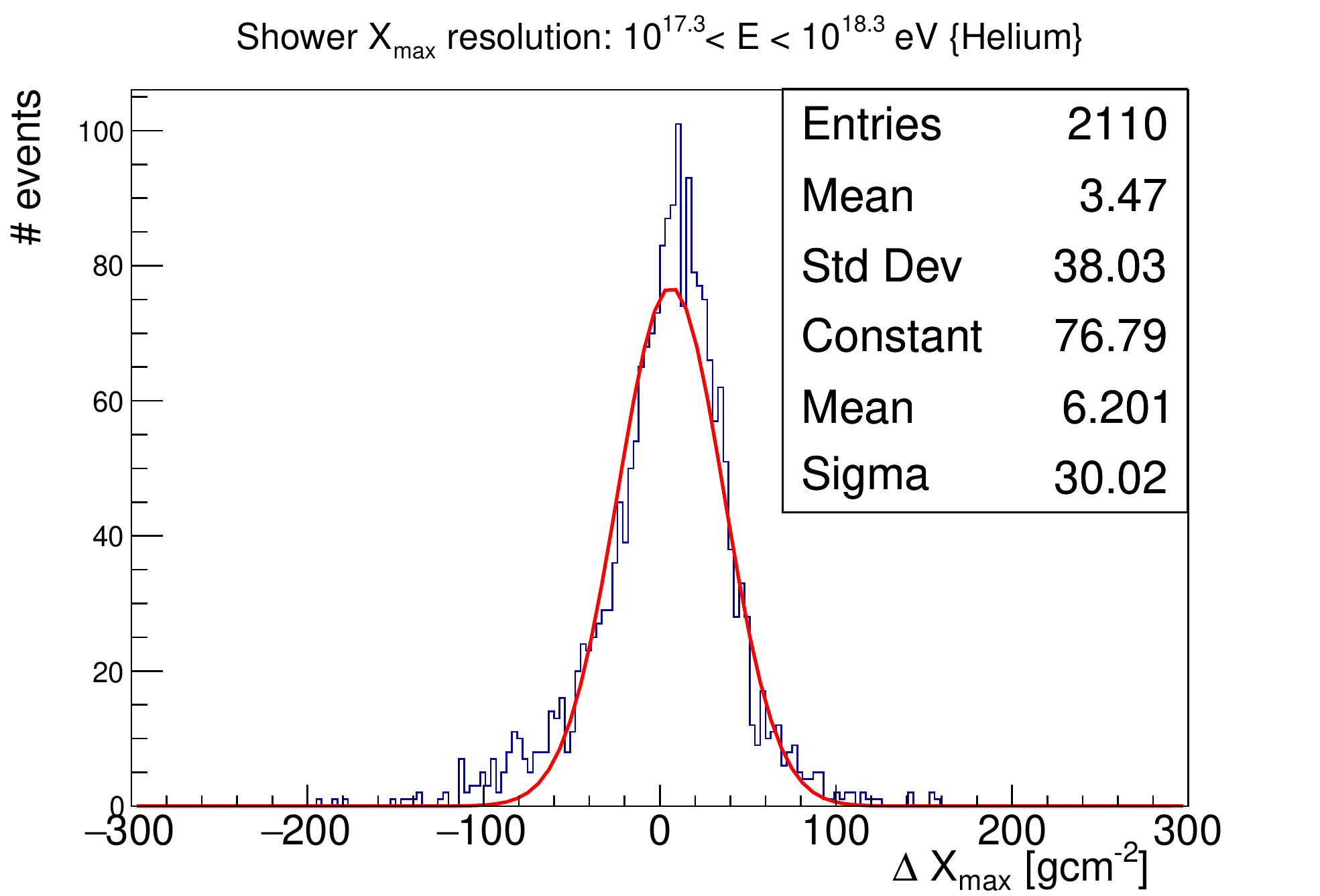}{0.33\textwidth}{(2c)}
  }
  \gridline{
    \fig{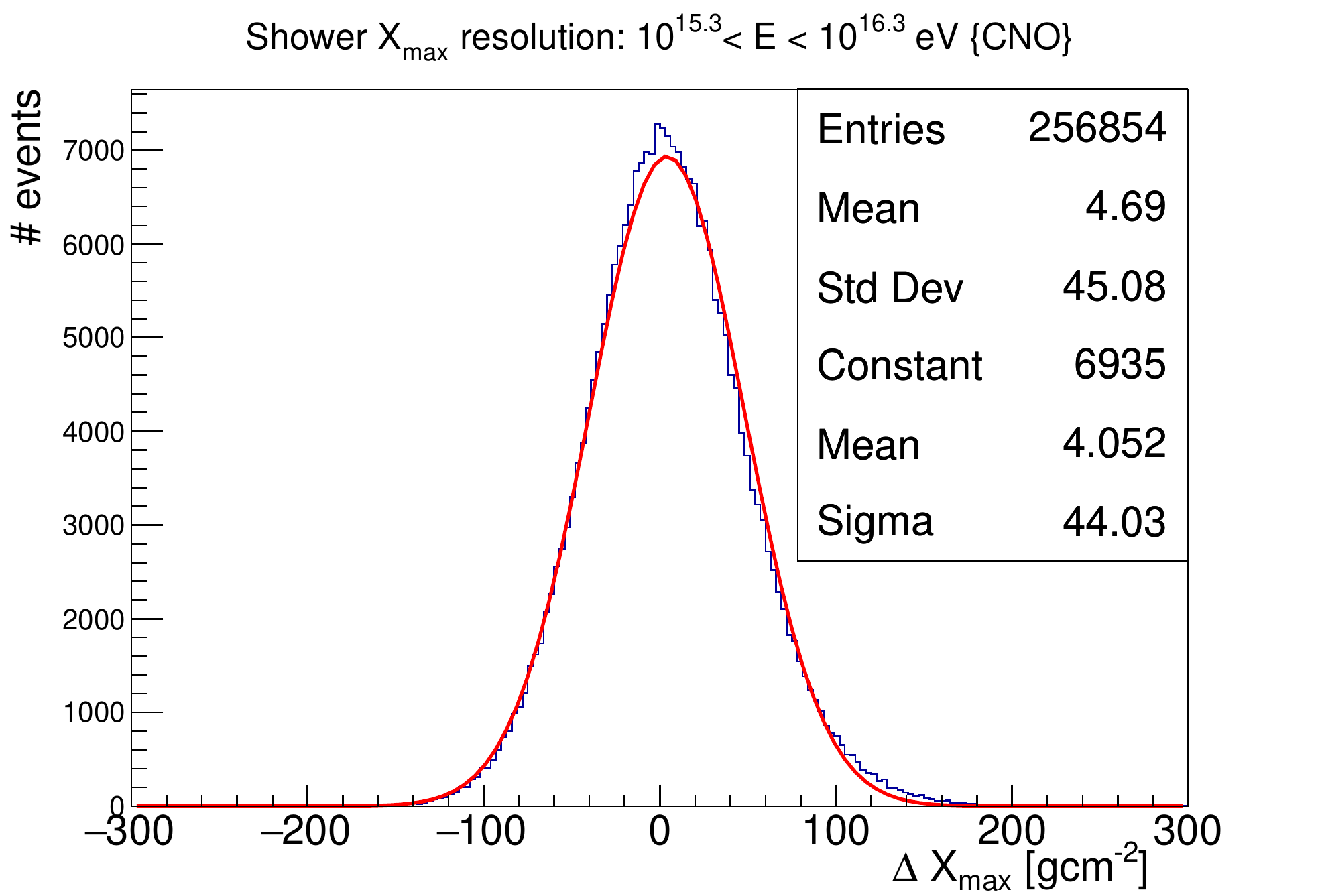}{0.33\textwidth}{(3a)}
    \fig{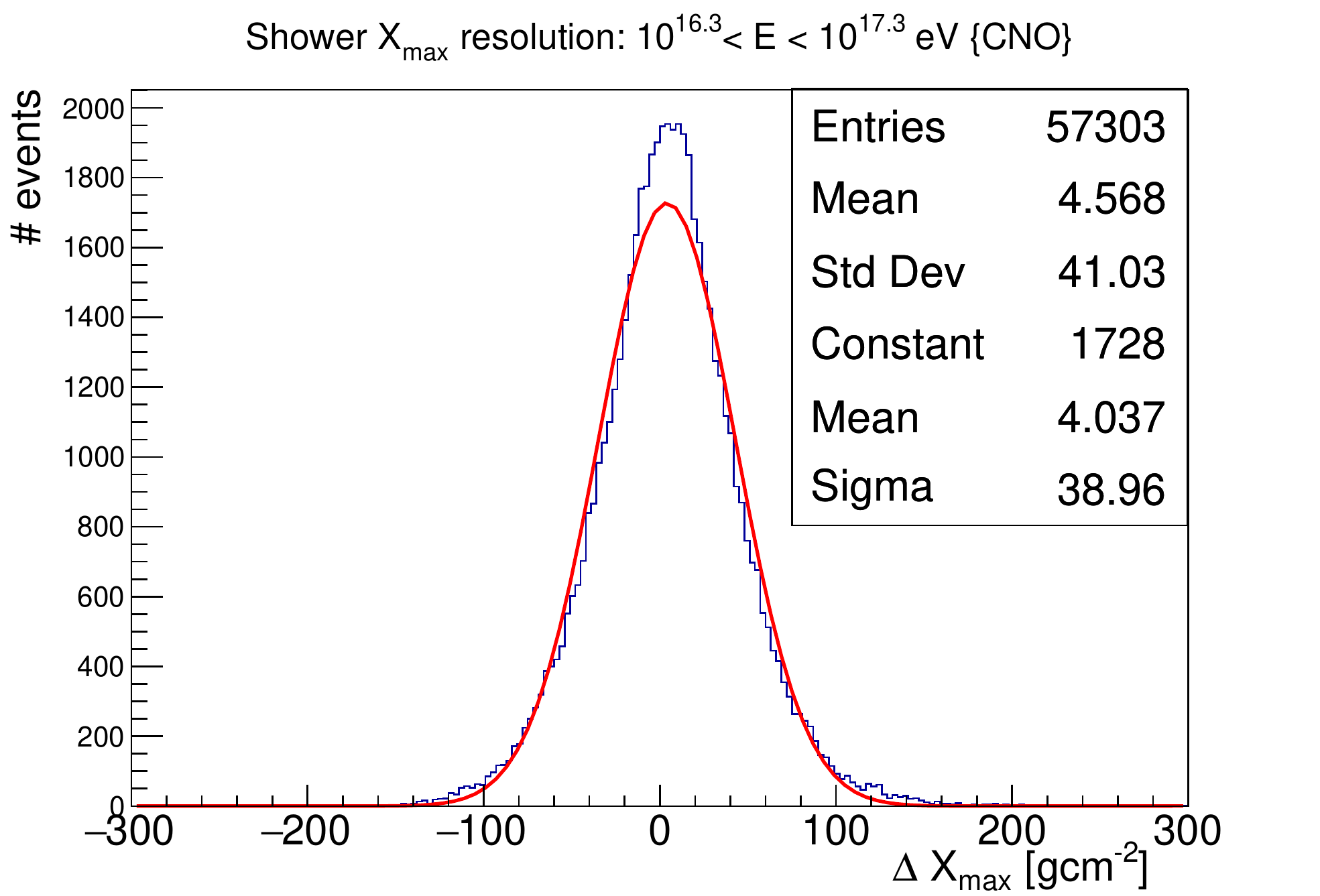}{0.33\textwidth}{(3b)}
    \fig{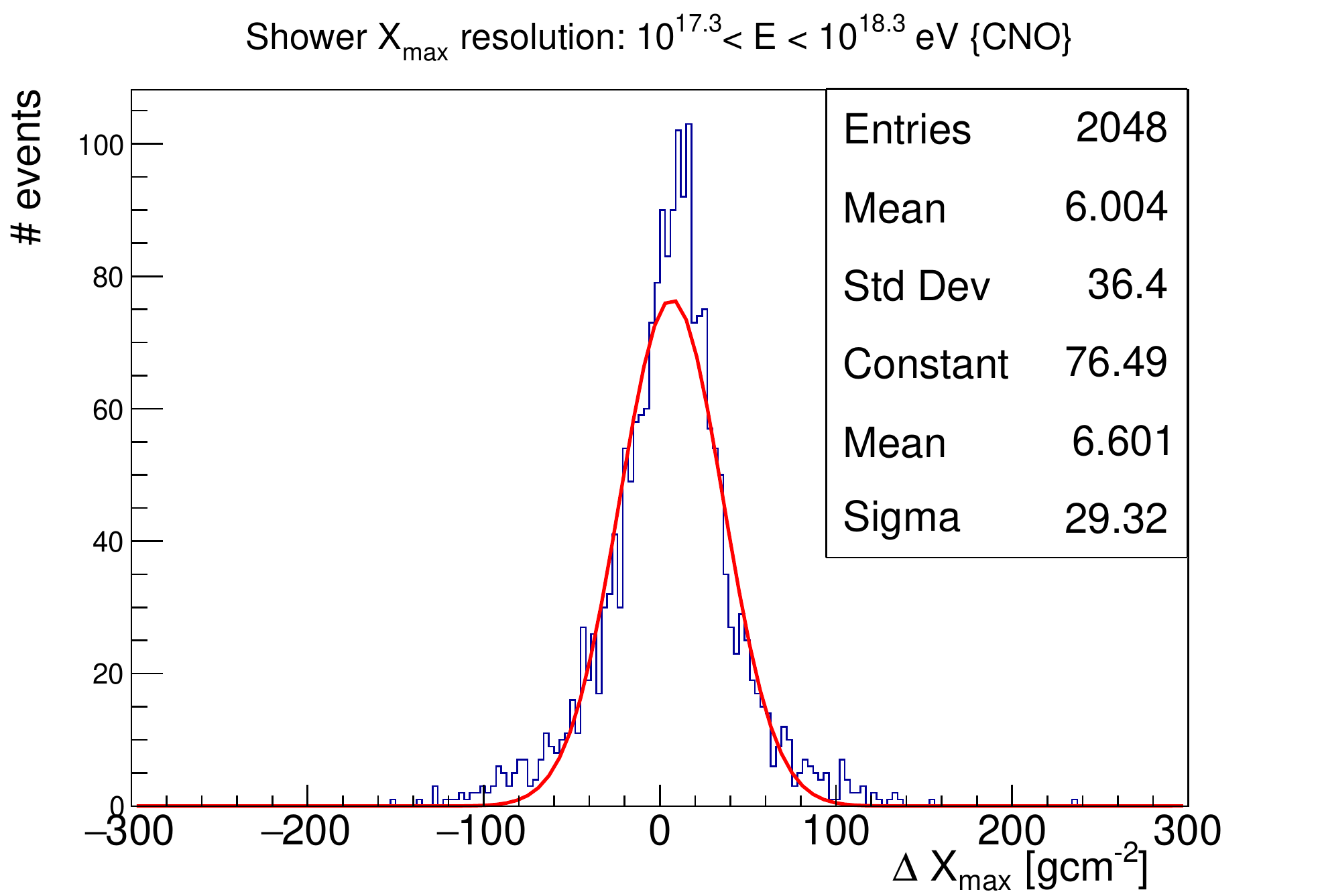}{0.33\textwidth}{(3c)}
  }
  \gridline{
    \fig{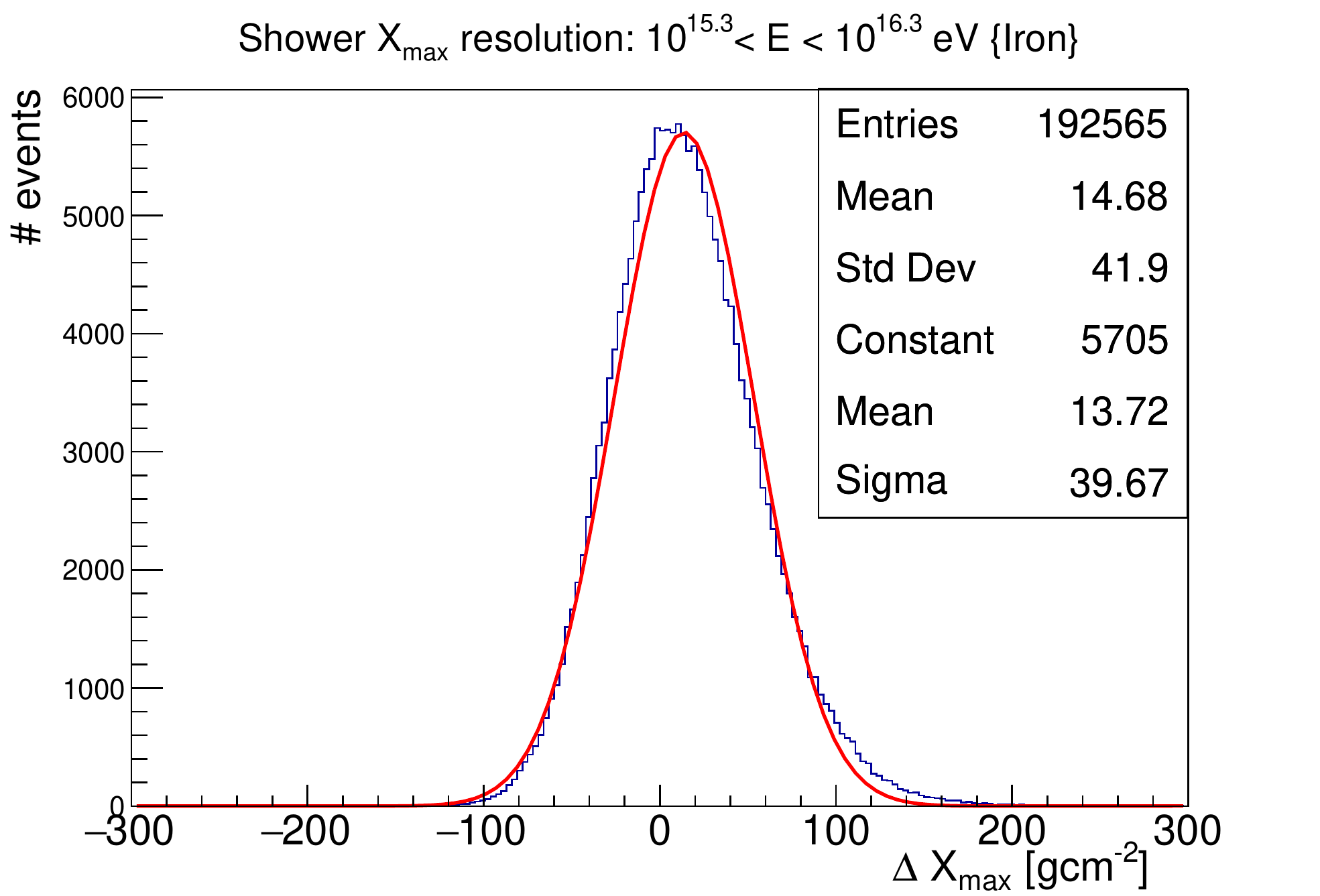}{0.33\textwidth}{(4a)}
    \fig{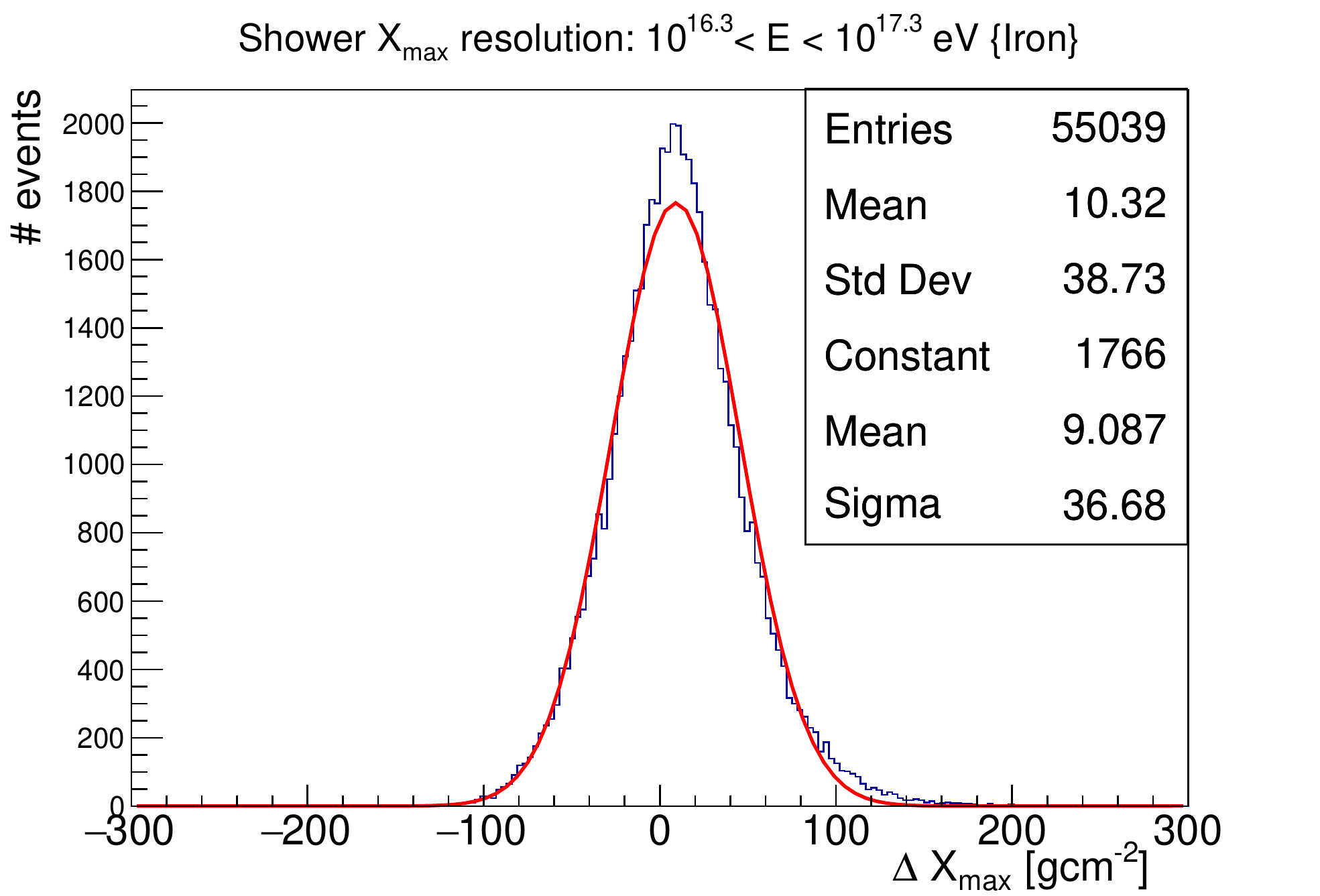}{0.33\textwidth}{(4b)}
    \fig{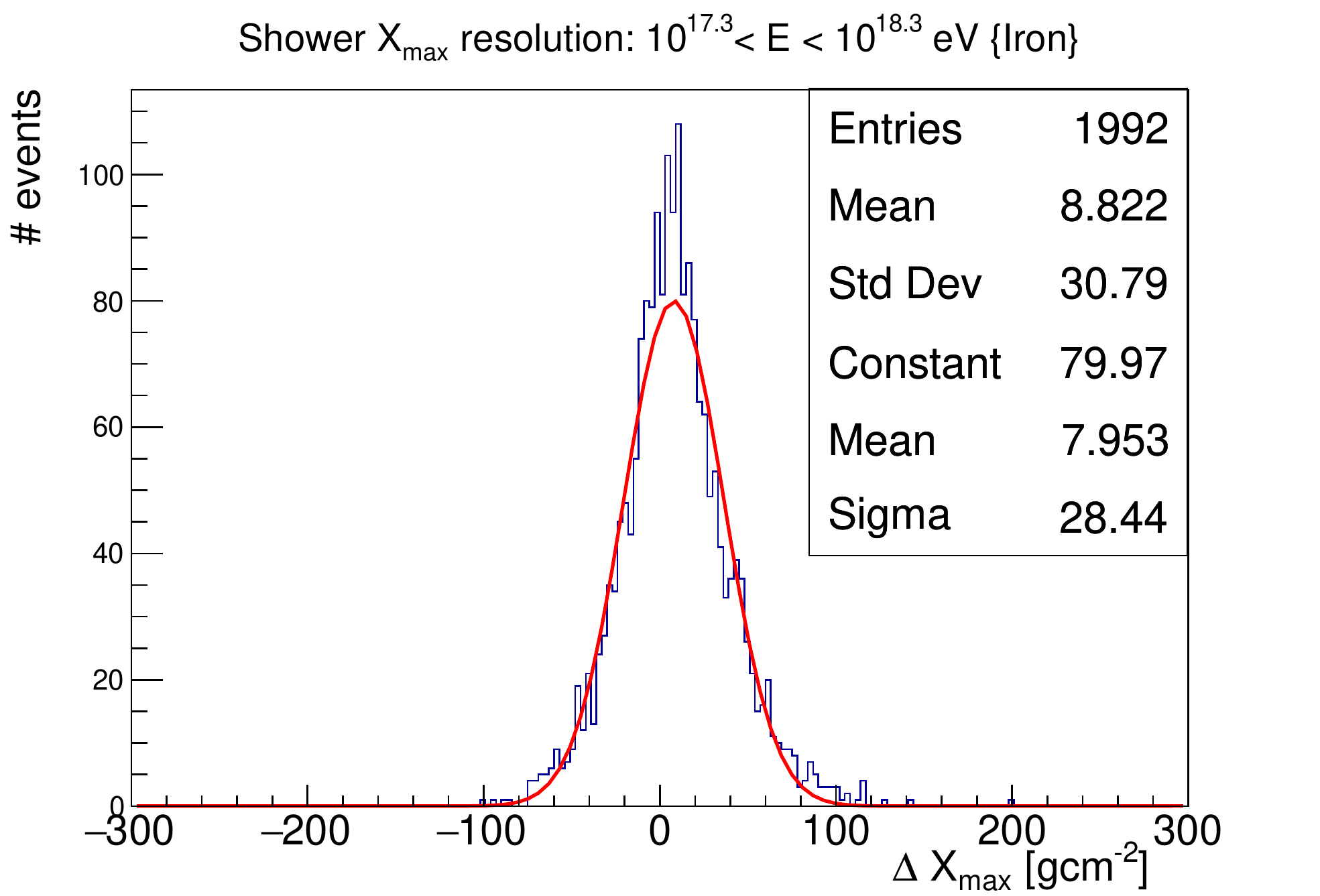}{0.33\textwidth}{(4c)}
  }
  \caption{Shower $X_{\rm max}$, reconstruction resolution.  
    The histograms (blue) show the uncertainty,
    $\Delta X_{max}~[{\rm g~cm}^{-2}]$ 
    for MC events reconstructed in three different energy ranges;  
    from the left to right: (a) $10^{15.3}$ - $10^{16.3}$~eV,
    (b) $10^{16.3}$ - $10^{17.3}$~eV, and (c) $10^{17.3}$ - $10^{18.3}$~eV.
    The red curve represents a Gaussian fit to the histogram. 
    The distributions are shown for each of the four simulated primaries, 
    from top to bottom: (1) H, (2) He, (3) N, (4) Fe.}
  \label{fig:res_plots_spectral_set_3_primaries}
\end{figure*}

The second set of results is shown for individual CR primaries.
Here we only show the $X_{\rm max}$ results,
(see Figure~\ref{fig:res_plots_spectral_set_3_primaries}), since this is the
only variable that shows any significant variance for the different primaries.
The total shower energy will, naturally, be biased since we use an average
missing energy correction.  As can be seen from 
Figure~\ref{fig:res_plots_spectral_set_3_primaries}, the $X_{\rm max}$ resolution
has similar magnitude and improves with energy for each type of primary.
The reconstruction bias, however, shows a dependence on primary type; as can
be seen by looking at the means of the distributions in
Figure~\ref{fig:res_plots_spectral_set_3_primaries}.
At lower energies, we see that $X_{\rm max}$ is underestimated for the lighter
primaries, and overestimated for heavier primaries.  The difference in bias
among the different primaries decreases with energy.

As a further check on the shower reconstruction performance, we tested an 
alternative detector response simulation procedure.  We replaced the 
calculation of shower Cherenkov photons reaching the TALE detector, 
performed by our usual MC program, with a procedure using CORSIKA with the
IACT package~\citep{Bernlohr:2008kv}.  Photons generated by CORSIKA are
``injected'' into the simulated detector at the times and into the pixels
predicted by the IACT package.  This is an independent calculation from the
one used in the reconstruction and can serve to test the validity of the
Cherenkov light modeling used in event reconstruction, as well as, to verify
our estimates for the shower reconstruction performance.  A study using
the two detector simulation procedures showed good agreement between the
estimates of the detector acceptance ($\sim 10\%$), and reconstruction
performance: energy~($\sim 5\%$), $X_{\rm max}$~($\sim 10~{\rm g~cm}^{-2}$), when  
using {\it identical sets of CORSIKA showers}.  The same reconstruction
procedure was applied to both sets.  The reader is referred
to~\citep{Abbasi:2018xsn} for more details on this study.

\section {Composition Analysis}
\label{sec:comp_analysis}

We examine both the mean depth of shower maximum, $\langle X_{\rm max} \rangle$,
as well as the full $X_{\rm max}$ distribution in order to study the
composition of cosmic rays.  The mean depth of shower maximum is known to
depend upon the cosmic ray primary type.  Therefore, the change of 
the mean $X_{\rm max}$ with energy, the \emph{elongation rate}, can be 
examined for indications of a change in composition, e.g. evolution from a
heavy to a light composition or vice versa.
Comparison of the mean $X_{\rm max}$ to that of MC showers of different primary
types allow for the inference of the dominant (if any) primary in the measured
flux.

Note that the detector acceptance and event selection and reconstruction
biases result in a primary mixture in the final data set that is different
from the true mixture, at the top of the atmosphere.
Therefore, the interpretation of the results can only be made by comparison
to MC generated showers.  The reconstructed MC showers are subject to the
same biases as the real showers.

The analysis fits the full $X_{\rm max}$ distribution histogram of the 
observed data to the weighted sum of four histograms of reconstructed MC
showers, one for each simulated primary type. The result of the fit is a
set of weights (fit fractions) that are used to produce a combined MC
histogram, as a weighted sum of the four primary MC histograms, that best
matches the data histogram.  The fractions are corrected for the detector
acceptance of each primary, using the known MC event counts, to produce
fractions which are independent of the detector acceptance.

The fit procedure starts by binning the reconstructed events in energy using 
a bin size of 0.1 in $\log_{10}(E_{cal}/{\rm eV})$, where $E_{cal}$ is the
reconstructed calorimetric shower energy.
This is the energy estimate obtained from the fit
to the PMT signals, and is independent of primary type.
In each energy bin, the data and MC $X_{\rm max}$ distributions are histogramed 
with a bin size of 10~g~cm$^{-2}$.  The fit is performed by calculating a 
weighted sum of the four MC histograms representing the reconstructed 
$X_{\rm max}$ distributions of the four 
primaries~\citep{TFractionFitter, Barlow:1993dm}.  A ``true fraction'' is 
then determined taking into account the relative detection and reconstruction 
efficiencies for each primary type.

\begin{figure*}[htb!]
  \centering
  \gridline{
    \fig{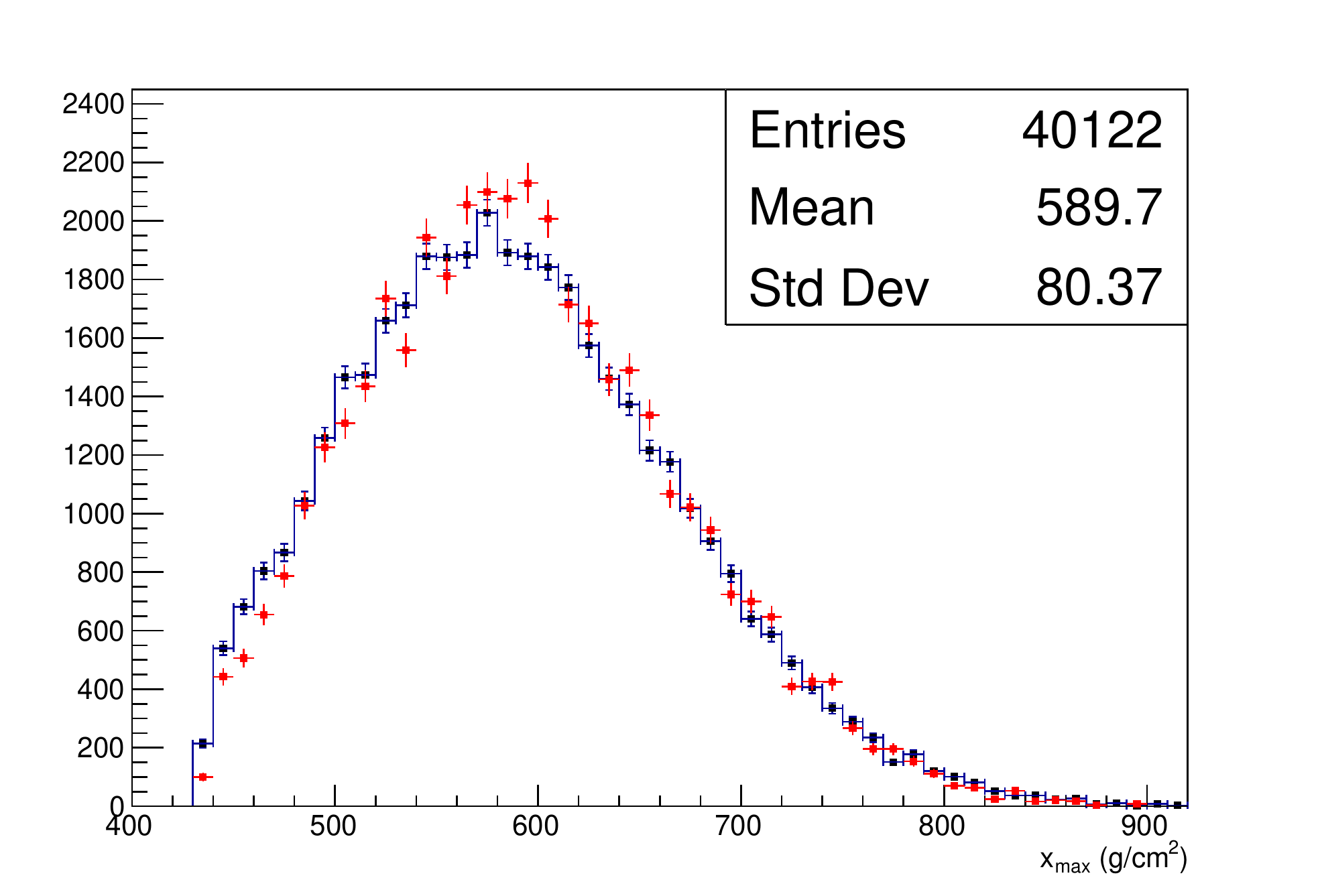}{0.43\textwidth}{}
    \fig{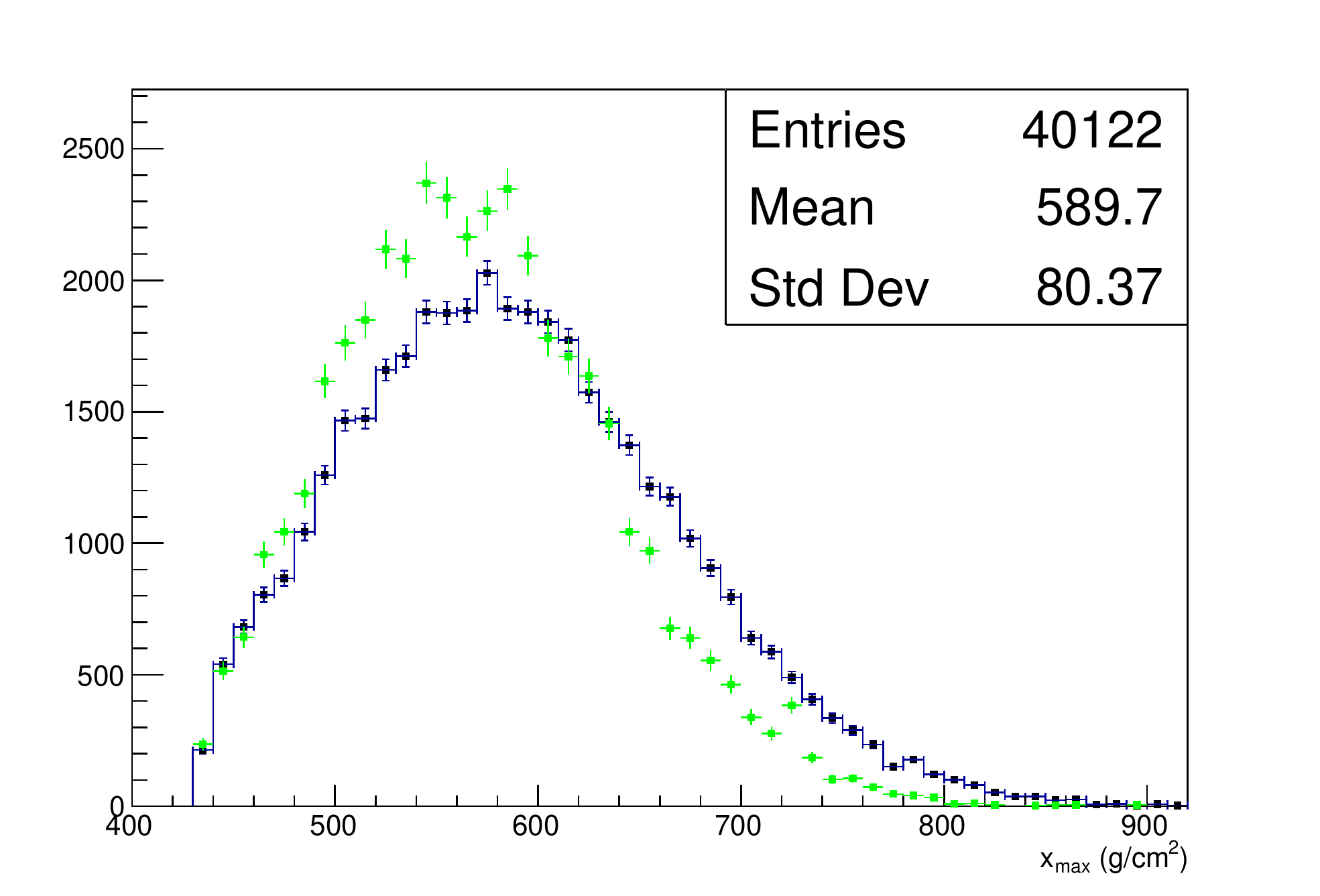}{0.43\textwidth}{}
  }
  \gridline{
    \fig{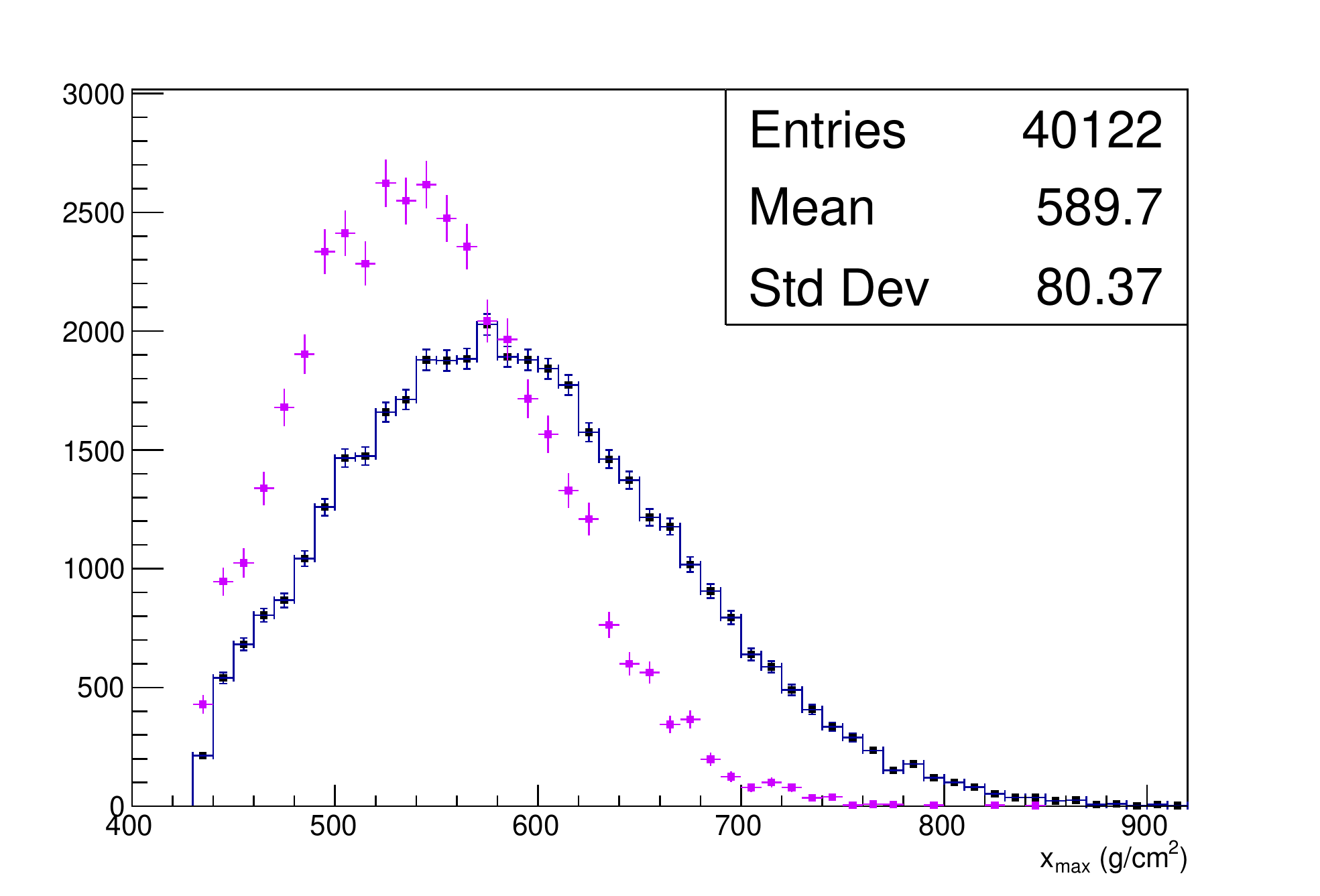}{0.43\textwidth}{}
    \fig{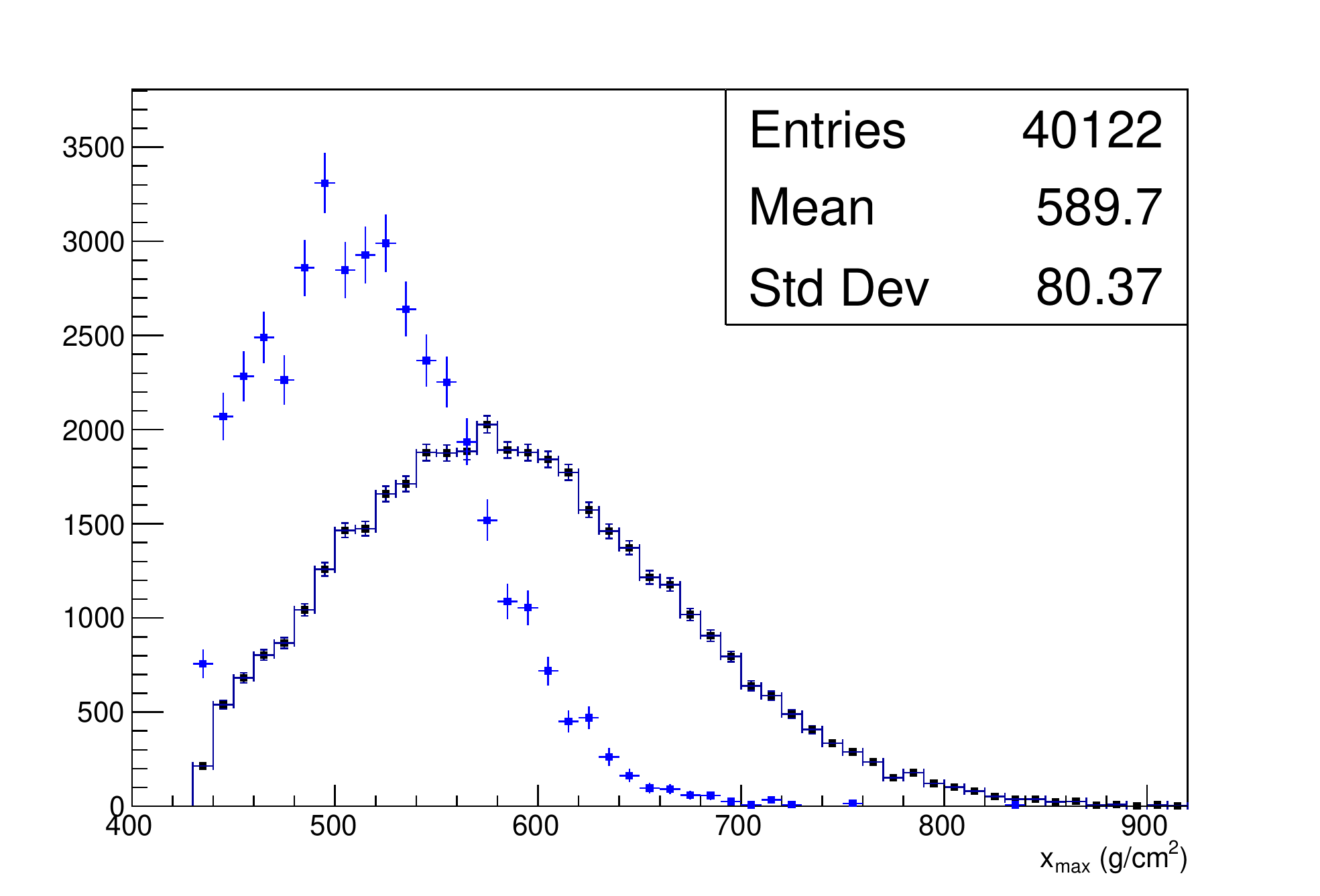}{0.43\textwidth}{}
  }
  \caption{
    Shower $X_{\rm max}$~(g/cm$^{2}$) distributions for energy bin
    $15.2 < \log_{10} (E_{cal}/{\rm eV}) < 15.3$.  Each of four plots shows
    data histogram (black points / blue line), along with MC primary
    reconstructed $X_{\rm max}$: upper left plot H (red), upper right He (green),
    lower left CNO (violet), and lower right Fe (blue)
  }
  \label{fig:ebin153_xmax_primaries}
\end{figure*}

\begin{figure}[htb!]
  \centering
  \includegraphics[height=2.4in]{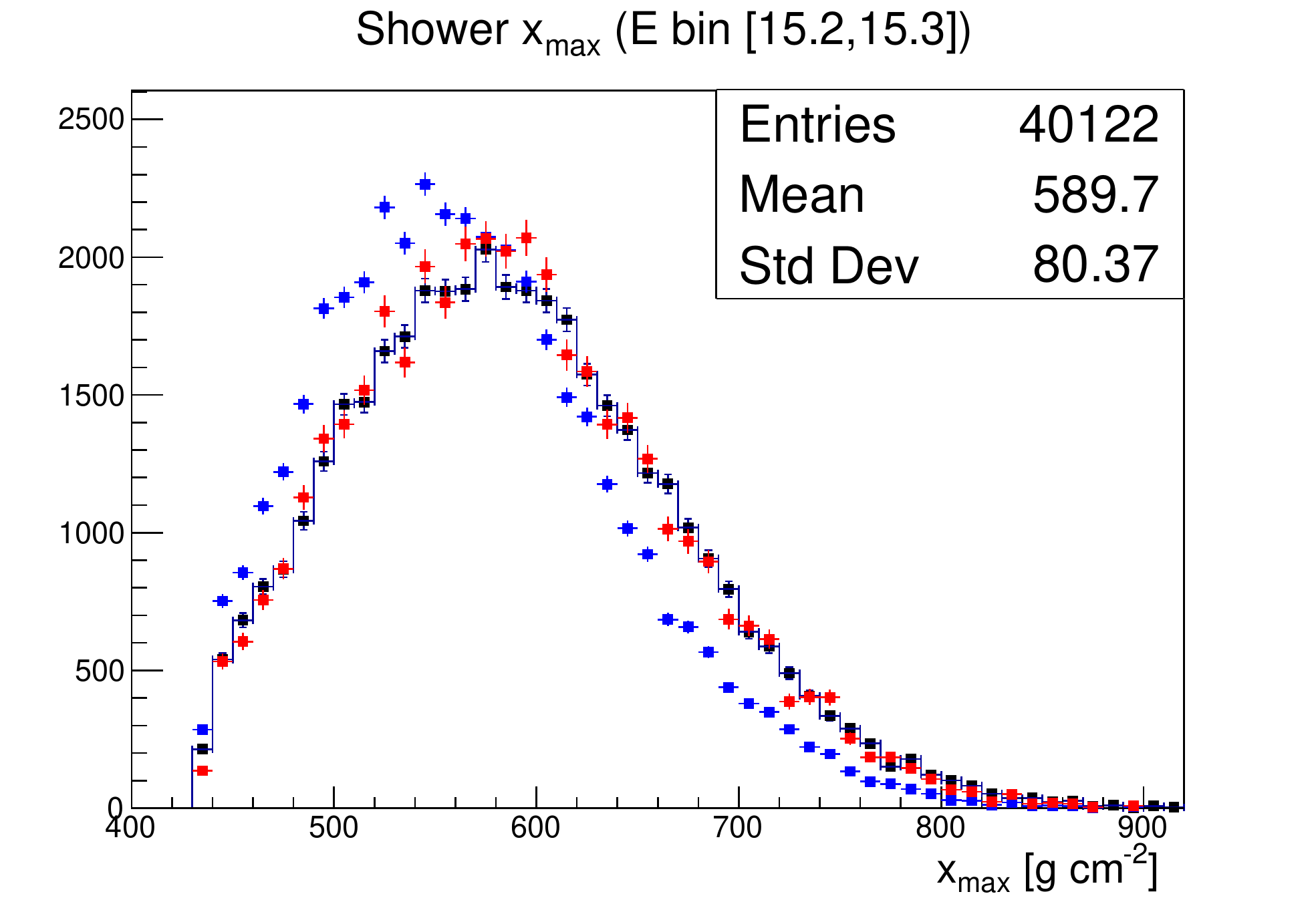}
  \caption{
    Shower $X_{\rm max}$~(g/cm$^{2}$) distributions for energy bin
    $15.2 < \log_{10} (E_{cal}/{\rm eV}) < 15.3$. Black points represent the data,
    the $X_{\rm max}$ distribution fit result (TXF) is shown in red.
    Using the H4a~\citep{Gaisser:2011cc} composition model as input
    to the TALE detector MC results in the distribution shown in blue.
    \explain{replaced figure with ``identical figure'' except for displayed value
      of Std Dev, in order to be consistent with Figure 9.  Old Std Dev = 80.33,
      new value = 80.37.  Difference comes from rounding errors resulting from
      saving ROOT figures in different file formats, namely as a PDF or as a C++
      macro.}
  }
  \label{fig:ebin153_data_mc_xmax}
\end{figure}

As is well known from shower simulations, there is significant overlap in the 
$X_{\rm max}$ distributions obtained from different cosmic ray primary particles 
with the same energy.  A practical consequence of this fact is that attempts 
to fit a measured distribution using MC generated distributions do not benefit 
from including many primaries in the fit.  On the contrary, the fit becomes 
unstable, and gives results with highly correlated fit parameters, i.e. primary 
fractions, making the physical interpretation of the results difficult.  For 
this reason, we chose to use only four primaries for this analysis, namely H, 
He, N (CNO), and Fe that cover the range of interest. An example of the
measured $X_{\rm max}$ distributions, and MC primary reconstructed
$X_{\rm max}$ distributions in the same energy bin,
$15.2 < \log_{10} (E_{cal}/{\rm eV}) < 15.3$, is shown in
Figure~\ref{fig:ebin153_xmax_primaries}.  We find that that the choice of 
four primaries is sufficient to provide a good fit to the data at all energies. 
This is shown in Figure~\ref{fig:ebin153_data_mc_xmax} for the same energy bin.
The figure also shows the reconstructed $X_{\rm max}$ distribution that results
from using the H4a~\citep{Gaisser:2011cc} composition model as input to the
TALE detector simulation.

\section{Systematic Uncertainties}
\label{sec:systematics}

The main sources of systematic uncertainties on the $\langle X_{\rm max} \rangle$
measurement are the energy scale uncertainty and possible
uncertainty in the detector acceptance calculations in the MC.
The main source of uncertainty in the cosmic ray composition measurement,
i.e. primary fractions estimates, comes from the selection of hadronic model
used in the shower simulations.

The TALE detector total energy scale uncertainty was estimated to be $\pm$15\%,
including a $\pm$10\% contribution from the shower missing energy 
correction~\citep{Abbasi:2018xsn}.  This implies that the uncertainty on the
reconstructed shower calorimetric energy is $\sim10\%$.  To estimate the
systematic uncertainty on the fit to the primary fractions and quantities
derived from them, we propagate the uncertainty in the calorimetric energy,
as explained below.

Systematic uncertainty due to detector acceptance effects are investigated
below, however, they are not folded into the final systematic uncertainty.
This is due to the fact that some of contributions are small enough to be
ignored, while others are contained within the calorimetric energy uncertainty
and are therefore already accounted for.

The choice of hadronic model determines the predicted $X_{\rm max}$ distribution
for each of the primary particles which are used to fit the data $X_{\rm max}$
distributions.  To a first approximation, we can consider the differences in
the predictions of each hadronic model for the mean value of $X_{\rm max}$ of
each primary as a measure of the systematic uncertainty introduced by the
choice of a particular model.  An examination of the shower simulations 
using various post-LHC models~\citep{Pierog:2017awp} has shown that the
predictions
of the different models for the mean $X_{\rm max}$ lie in an interval of about
20~g~cm$^{-2}$, with EPOS-LHC producing results in the middle of those predicted
by QGSJetII-04 and Sybil2.3-c~\citep{Engel:2017wpa}.  We therefore, estimate 
the uncertainty on the $\langle X_{\rm max} \rangle$ of simulated CR primaries to
be $\pm~10~{\rm g~cm}^{-2}$ around those used in our EPOS-LHC MC set.

The data analysis was also performed using the QGSJetII-03 model, producing an
equivalent set of results which can be compared to the results using EPOS-LHC.
A comparison of the results using the two models includes the effects not only 
of differences in the $\langle X_{\rm max} \rangle$, but the full $X_{\rm max}$
distributions.  In addition, the comparison introduces a shift in the energy
scale due to the different missing energy correction.

The systematic uncertainty on the $\langle X_{\rm max} \rangle$ measurements 
was calculated by shifting 
the reconstructed event (total) energy by $\pm~15\%$, while also shifting the
reconstructed event $X_{\rm max}$ by $\pm~10~{\rm g~cm}^{-2}$.  The $\pm$ sign
in both
shifts is chosen to move the $\langle X_{\rm max} \rangle$ versus energy in the
same direction.  The $\pm~10~{\rm g~cm}^{-2}$ in this case is attributed to
detector
acceptance bias, and to reconstruction bias introduced by Cherenkov light
modeling~\citep{Abbasi:2018xsn}.

\begin{figure}[htb!]
  \centering
  \epsscale{1.0}
  \plotone{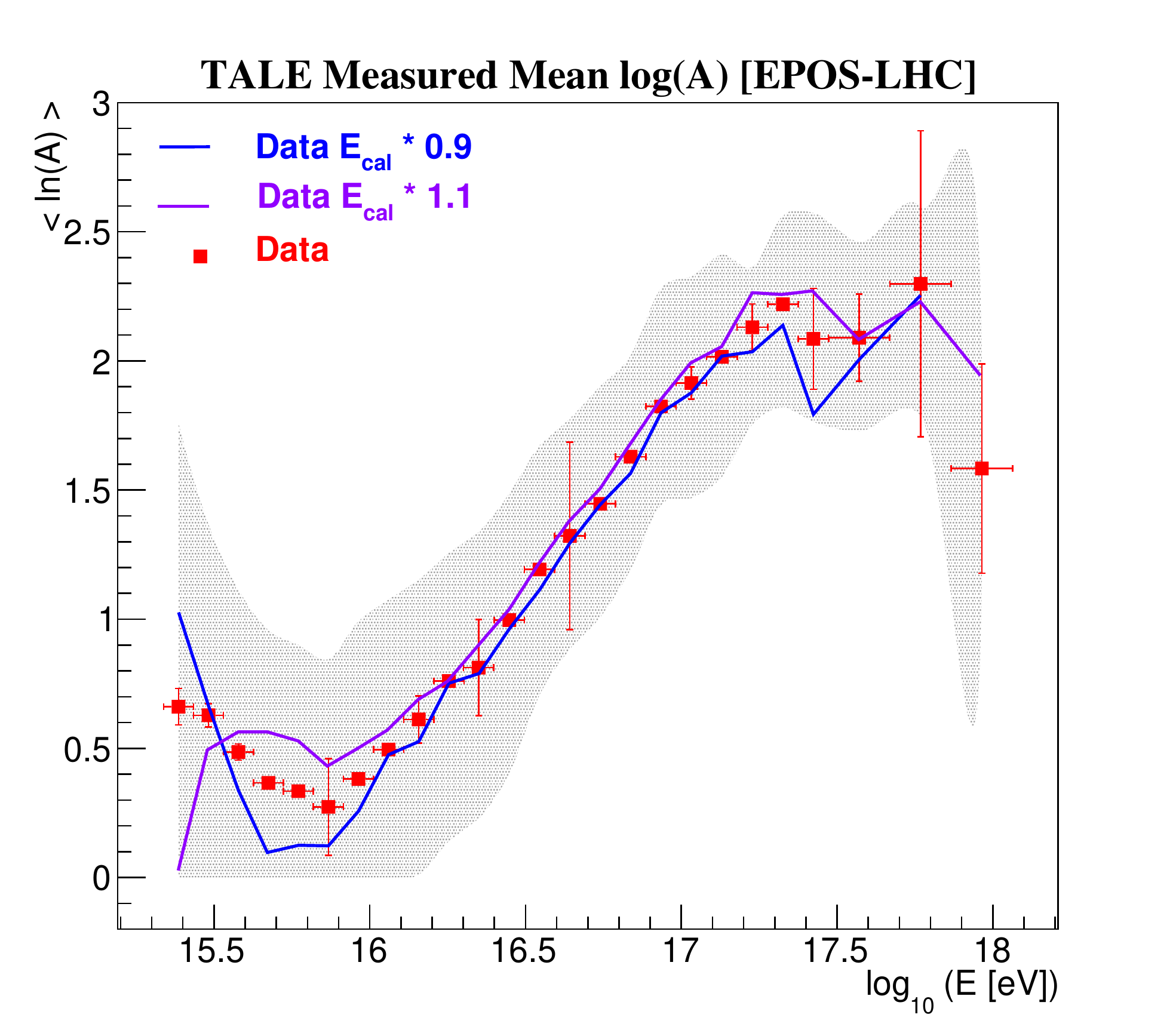}
  \plotone{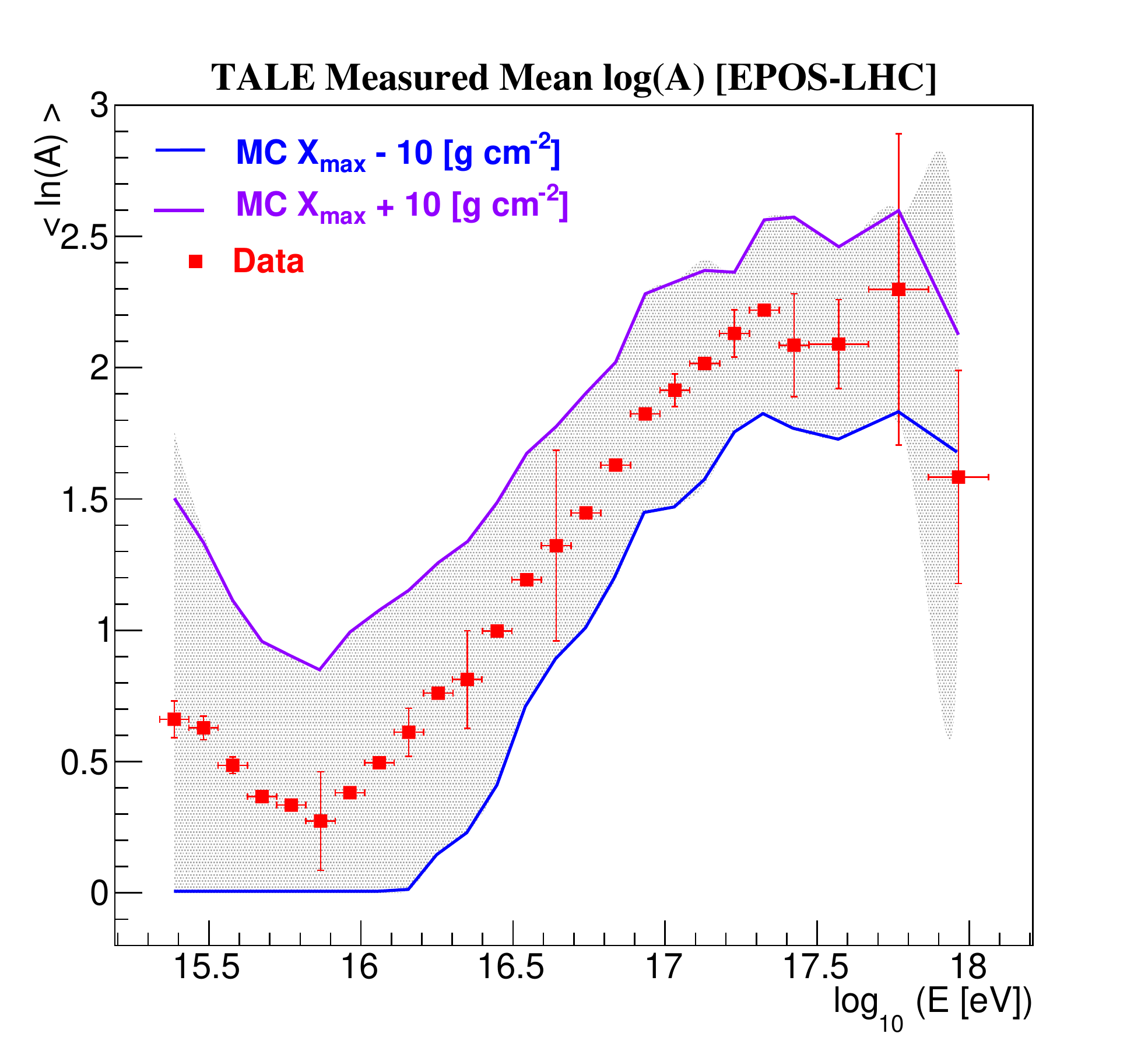}
  \plotone{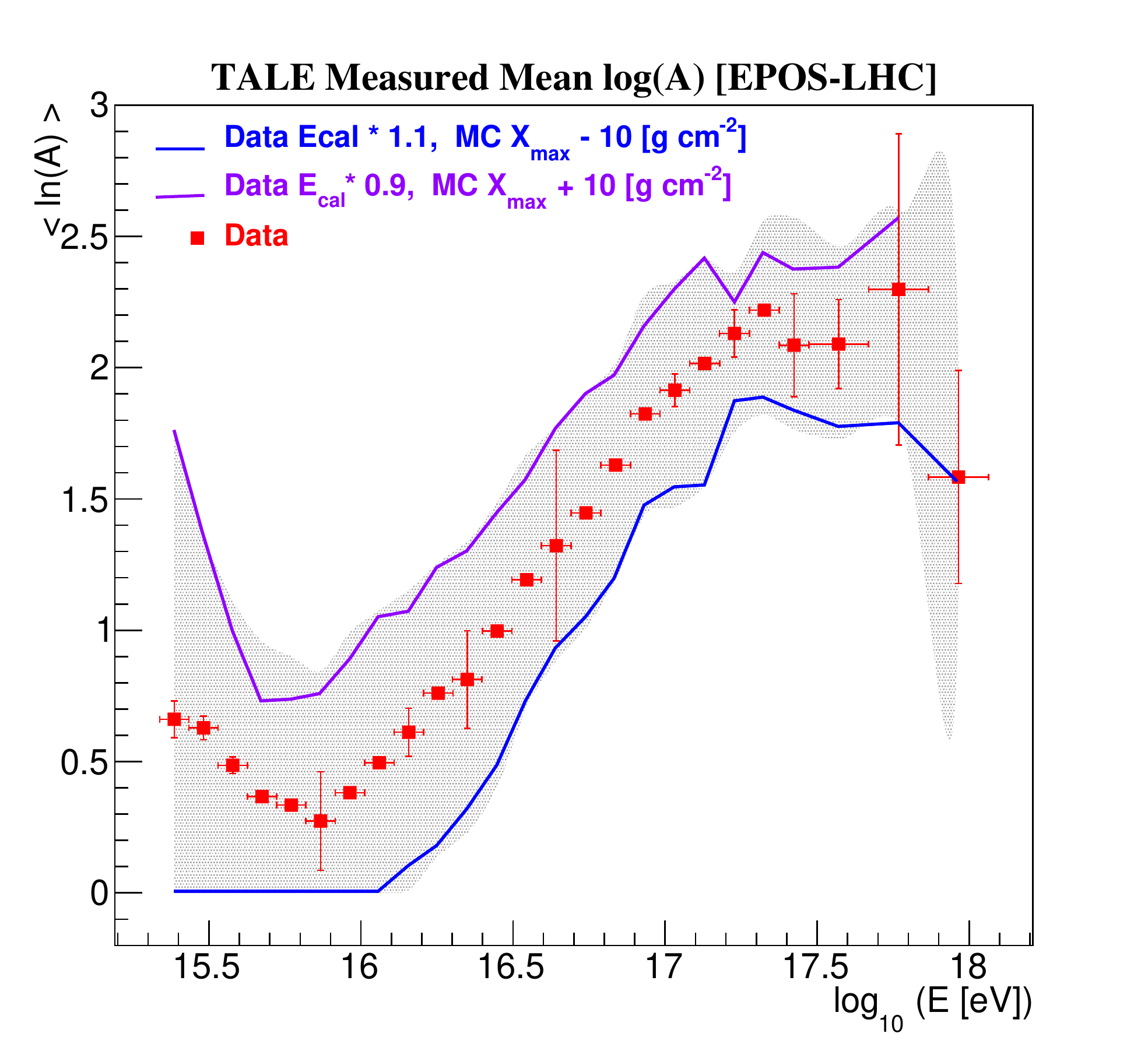}
  \caption{The $\langle \ln(A) \rangle$ systematics band (shown in gray)
    compared
    to the effect on the estimated $\langle \ln(A) \rangle$ of shifting the
    reconstructed data events energy by $\pm 10\% $ (top), or shifting
    reconstructed $X_{\rm max}$ values of MC showers by $\pm 10~{\rm g~cm}^{-2}$
    (middle),
    or the combination of the two (bottom).  In all three panels, points
    with error bars are TALE data with no systematic shifts to
    event energy or MC showers $X_{\rm max}$.
  }
  \label{fig:sys_check_energy_hmodel}
\end{figure}

The systematics bands displayed in the primary fractions obtained by fitting
the data $X_{\rm max}$ distributions were calculated by repeating the fitting
procedure with some variations: (1) Shift the calorimetric energy of the
data by $\pm~10\%$. (2) Shift the MC $X_{\rm max}$ distributions by
$\pm~10~{\rm g~cm}^{-2}$, a common shift is applied for the four components.
(3) Combine these shifts when they have an additive effect on the resulting 
shift in the fit fractions for the different primaries.
We examined six different sets of fits and set the bounds on each primary
fraction at the minimum and maximum values obtained by any of the six shifted
sets.

We can summarize a set of four fit-fractions as a single number using the
definition: $\langle \ln(A) \rangle  = \sum_{ip}{f_{ip}*\ln(A_{ip})}$ where
$ip$ stands for one of \{H, He, N, Fe\}.  In the following discussion, we use
this quantity to examine the overall systematic uncertainty of the composition
measurement.  We start with Figure~\ref{fig:sys_check_energy_hmodel}, showing
the six different combinations of energy and $X_{\rm max}$ shifts, discussed
above, along with the overall systematics band.

To estimate the size of the uncertainty due to acceptance, we divide the data
into
multiple subsets and redo the analysis on these subsets. We also vary some of
the event quality cuts values and examine how the results change with these
modifications.  

\begin{figure}[htb!]
  \centering
  \epsscale{1.0}
  \plotone{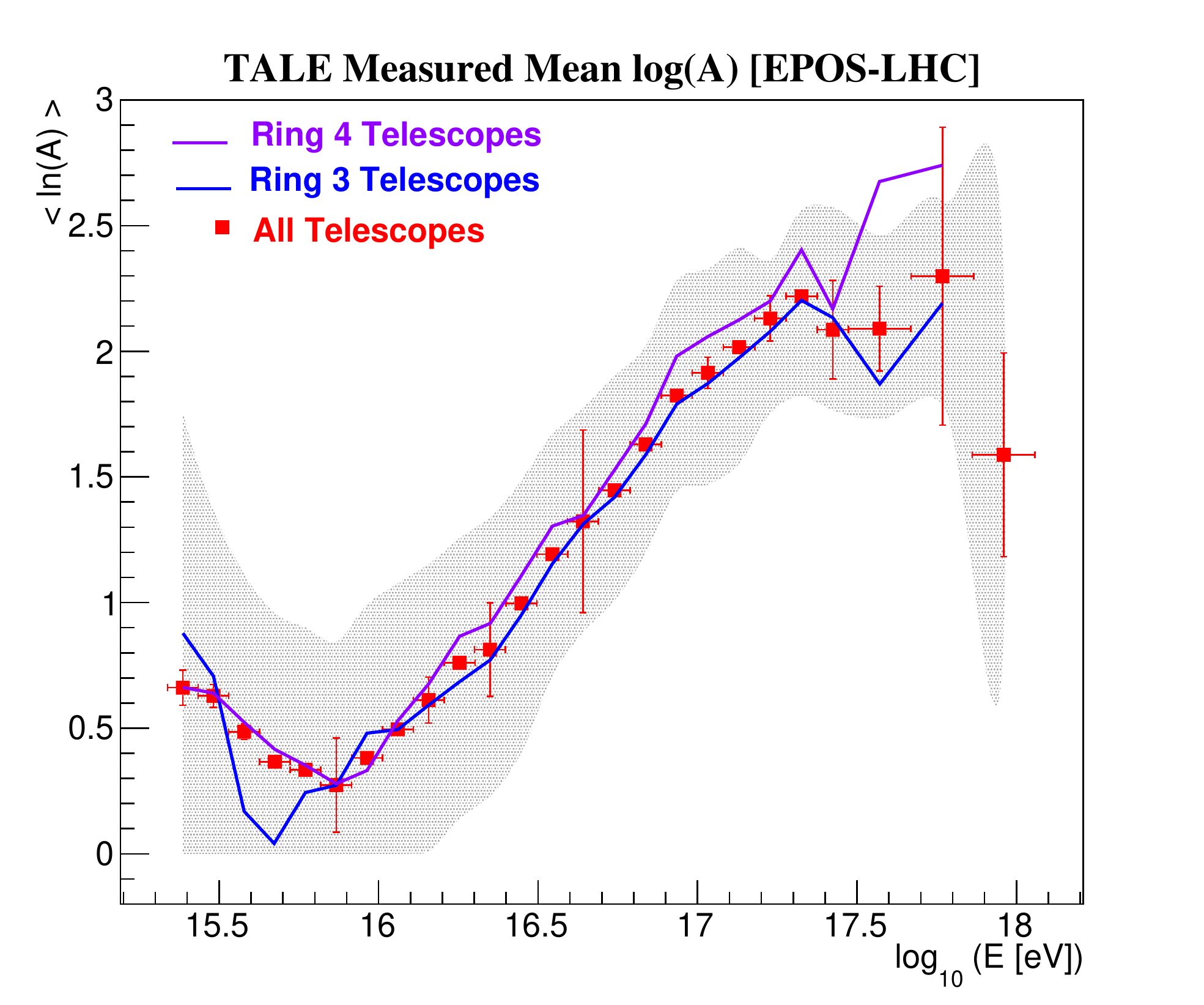}
  \caption{The $\langle \ln(A) \rangle$ for the full TALE dataset compared to
    subsets
    using Ring 3 (31-45$^{\circ}$ elevation) and Ring 4 (45-59$^{\circ}$ elevation)
    events separately.  For most of the energy range, the difference between the
    two subsets is small relative to other systematics.  Near the bottom and 
    top of the energy range the difference is comparable to other systematics.
  }
  \label{fig:sys_check_ring_elon4}
\end{figure}

The vast majority of the events in the final data set are single telescope
events.  Thus, a possible way to create subsets of the data is to 
divide the set by telescope.  ``Ring 4'' telescopes view higher elevation 
angles (45-59$^{\circ}$ elevation), therefore, they are more likely to trigger
on heavy primaries, due to shorter path through the atmosphere, than
``Ring 3'' telescopes (viewing 31-45$^{\circ}$ elevation).  Division by
telescope 
``ring'' is related to division by shower zenith angle, since Cherenkov
dominated 
events must have a direction that is close to the pointing angle of the 
observing telescope.  The comparison of the two is shown in 
Figure~\ref{fig:sys_check_ring_elon4}.  As can be seen in the figure, the
difference between the two subsets is small relative to other systematics for
most of the energy range.  Near the ends of the energy range the difference is
comparable to other systematics.

\begin{figure}[htb!]
  \centering
  \epsscale{1.0}
  \plotone{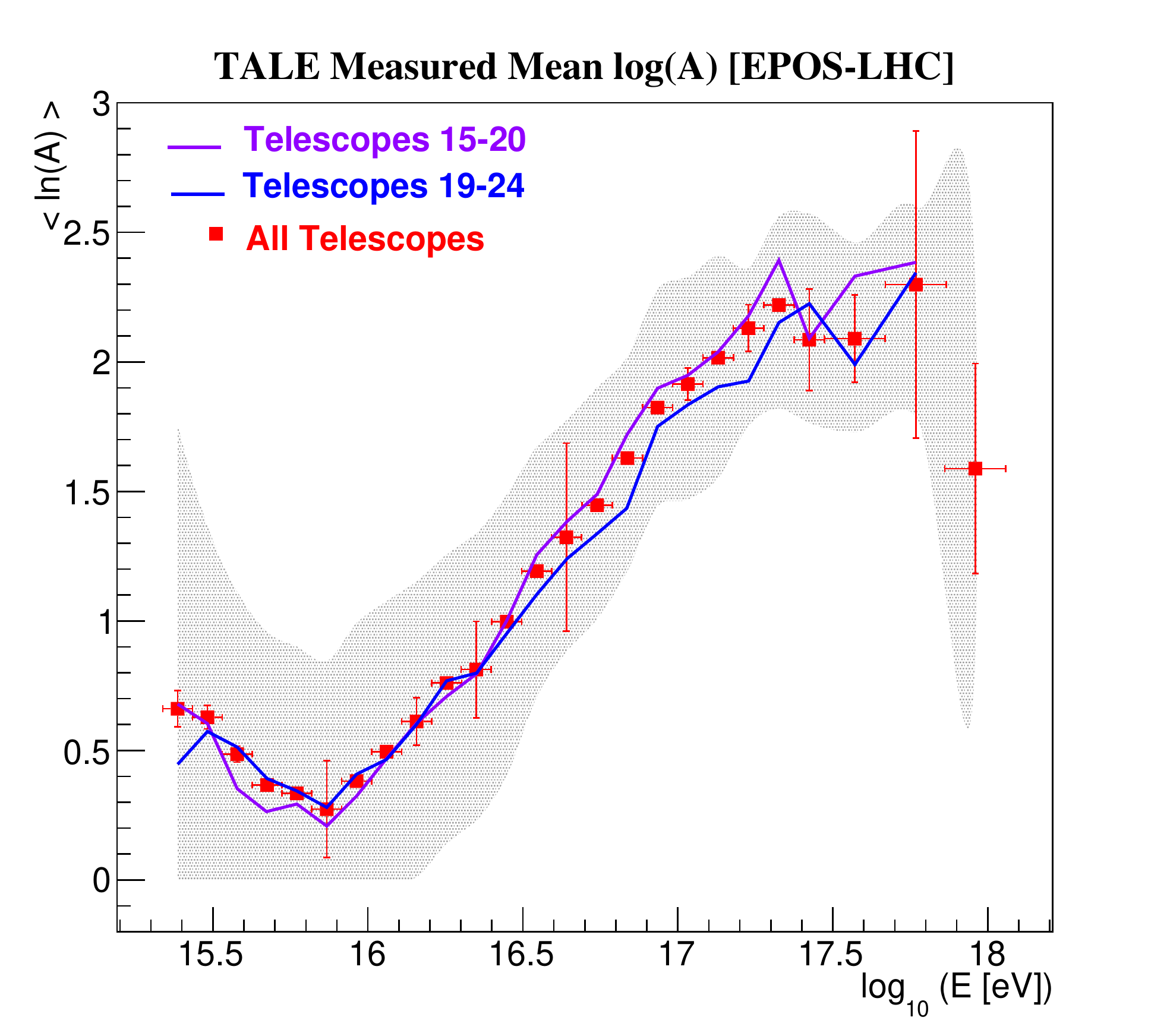}
  \caption{The $\langle \ln(A) \rangle$ for the full TALE dataset compared to
    subsets
    using telescopes with different azimuthal viewing directions.  Telescopes
    15-20 point farther east, while 19-24 point farther west; two
    telescopes 19, 20
    are shared by the two subsets. As can be seen in the plot, there is a
    relatively small azimuthal effect at higher energies, but it is not a major
    contributor to the overall systematic uncertainty.
  }
  \label{fig:sys_check_east_west}
\end{figure}

An east, west division of telescopes with the central two telescopes included
in both sets, checks for any geomagnetic effect and different sky noise
background.  Results of a comparison are shown in
Figure~\ref{fig:sys_check_east_west}.  As can be seen in the figure, there is
a relatively small azimuthal effect at higher energies, but it is not a major
contributor to the overall systematic uncertainty.

\begin{figure}[htb!]
  \centering
  \epsscale{1.0}
  \plotone{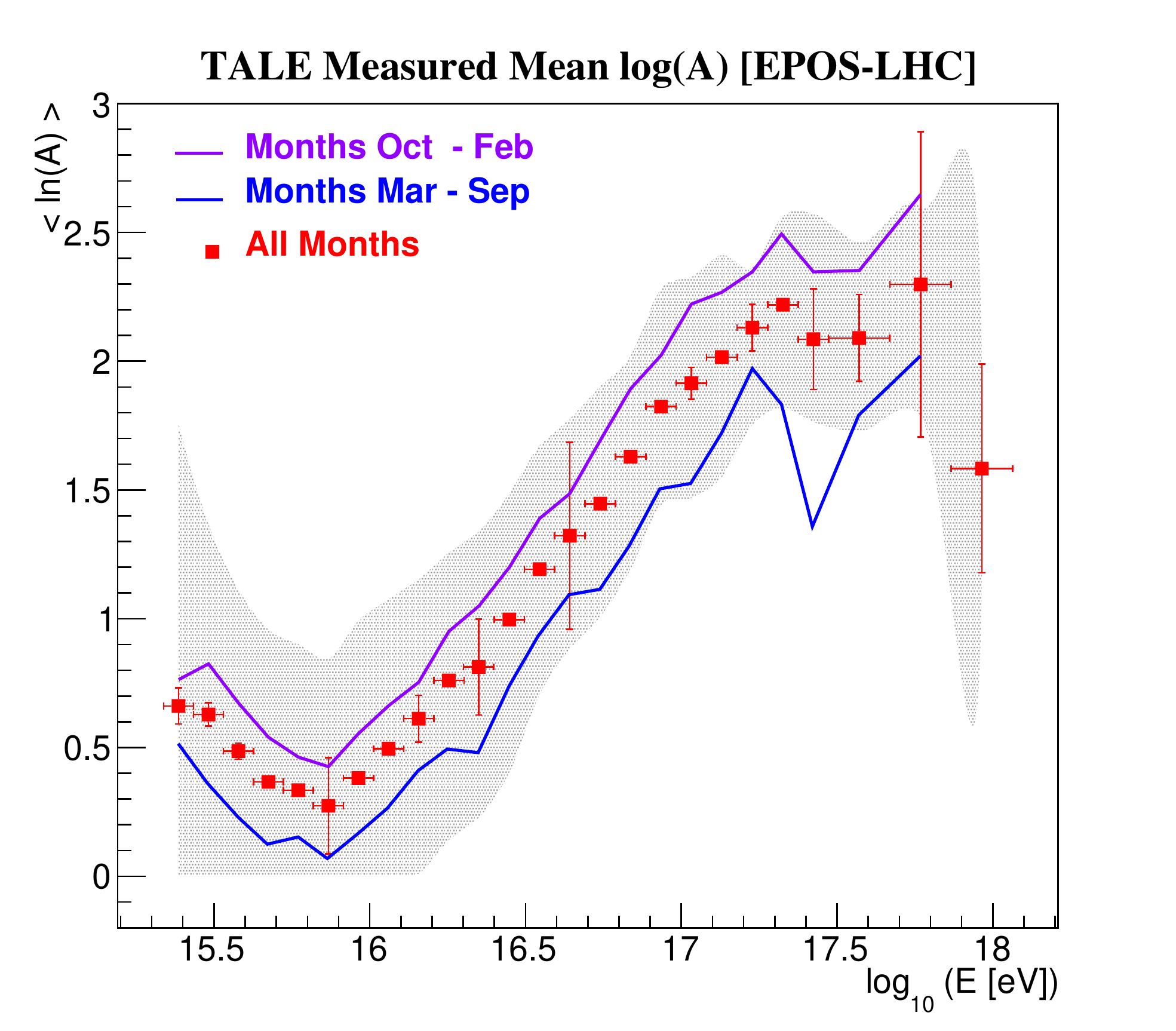}
  \caption{The $\langle \ln(A) \rangle$ for the full TALE dataset compared to
    subsets
    using ``winter'' months October through February and the
    months of March through September.  This break down gives similar number
    of events in each subset.  Despite the
    use of hourly GDAS atmospheric pressure profiles in the simulation and
    event reconstruction, we still find a significant difference in the
    predictions based on season.
  }
  \label{fig:sys_check_season}
\end{figure}

Another source of uncertainty on detector acceptance are time dependent 
effects such as atmospheric clarity or sky noise background that may not be 
accurately reflected in the detector simulation.  To examine these effects, we 
divide the data in time, namely in time of year.  Winter months usually allow
for longer run periods, and so we divided the data into a set collected during
the months of October through February, and another from March through
September.  This division showed the largest difference in the absolute
value of $\langle \ln(A) \rangle$ of all the various checks we performed.
The comparison is shown in Figure~\ref{fig:sys_check_season}.  Despite the
use of hourly GDAS atmospheric pressure profiles in the simulation and
event reconstruction, we still find a significant difference in the predictions
based on season.

A possible cause for the difference is, the seasonal variation in the average
concentration of atmospheric aerosols.  The nightly, or even hourly,
density of aerosols is variable, and difficult to measure continuously.  It
can be
treated on average however.  For the TALE analysis, an average concentration
characterized by the {\em vertical aerosols optical depth}, VAOD, is used.
Aerosols attenuate the light signal reaching the detector from the shower.
Therefore, a variation in the aerosols concentration results in a variation
in the amount of light reaching the detector from the shower.  An increase in
the
light attenuation can cause some showers to either fail to trigger the detector,
or otherwise, to pass some reconstruction or quality cut.  Summer months in Utah
tend to have poorer air quality, i.e. more aerosols, and therefore more light
attenuation.  Showers created by heavier primaries, larger $\ln(A)$, develop
higher
up in the atmosphere, and light produced by these showers will travel further in
the atmosphere, through more aerosols, to reach the detector.  We speculate that
the effect of increased average aerosols concentration will be stronger for
heavier primaries than light primaries, resulting in a decrease of the
fraction of
heavy primaries in the data and therefore a smaller $\langle \ln(A) \rangle$.
This
is the observed effect for TALE, as can be seen in
Figure~\ref{fig:sys_check_season}.

\begin{figure}[htb!]
  \centering
  \epsscale{1.0}
  \plotone{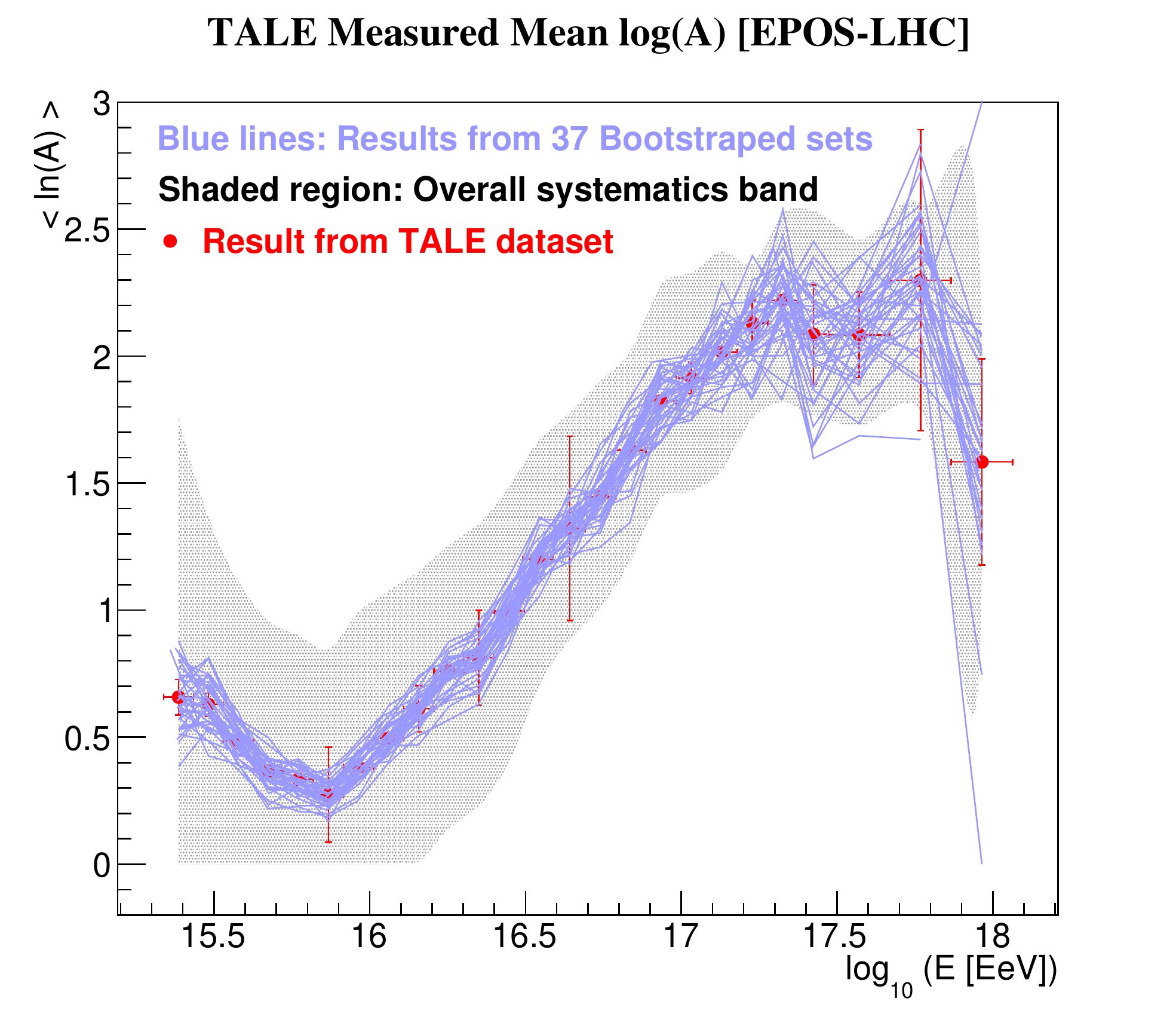}
  \caption{The $\langle \ln(A) \rangle$ for the full TALE dataset compared to 37
    artificial sets generated using the bootstrap method, with sampling over 
    run months (58 months in total).  By randomly sampling the run months we
    get a measure of the overall effect of atmospheric variation on the
    composition result.  We see that, when averaged over the entire
    data set, the effect of atmospheric variations on the composition result
    is likely smaller than the expectation based on the seasons check, shown in
    Figure~\ref{fig:sys_check_season}. 
  }
  \label{fig:sys_check_bootstrap}
\end{figure}

\begin{figure}[htb!]
  \centering
  \epsscale{1.0}
  \plotone{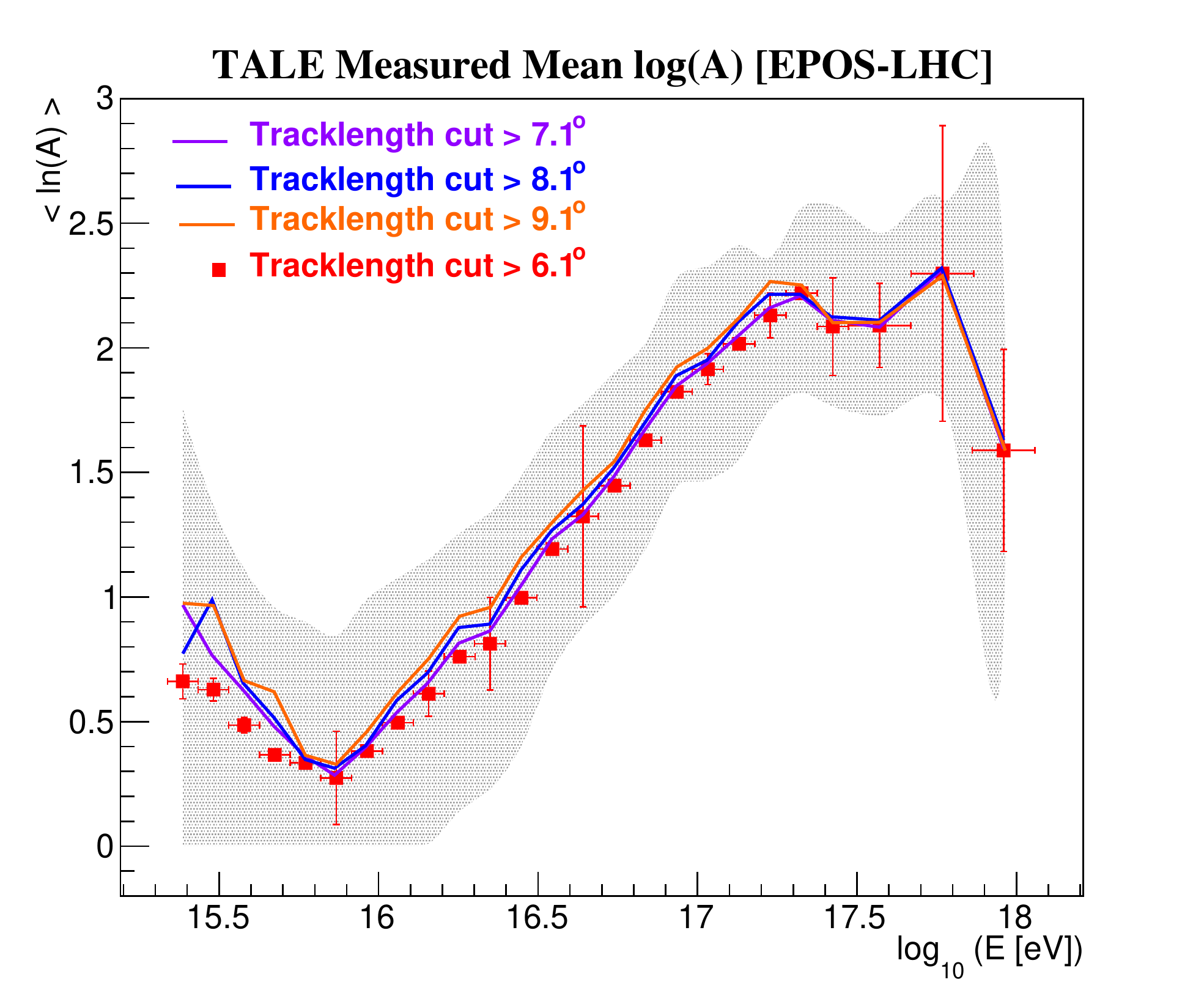}
  \caption{The $\langle \ln(A) \rangle$ for TALE data using standard quality 
    cuts compared to changes of cut value on the event angular-tracklength.
    As can be seen, changing the track-length cut has a minimal effect on the
    composition results.  Similar results were observed for other cut parameters
    examined as part of the data analysis.
  }
  \label{fig:sys_check_trk_elon4}
\end{figure}

Another approach to look at time dependence of the result is to use the
Bootstrap method~\citep{efron1979}, to sample different run months and form
a data set comparable to the actual set in terms of the number of events.
The complete observation period is comprised of 58 run months.  By sampling
run months instead of individual events, we maintain the correspondence
between the simulation and real data included in the sampled set.  We
performed 37 iterations to obtain a measure of the stability of the result
and to get a sense of the expected spread of the result due to inclusion or
exclusion of certain run periods.  Results are shown in
Figure~\ref{fig:sys_check_bootstrap}.  By randomly sampling the run months we
get a measure of the overall effect of atmospheric variation on the
composition result.  We see that, when averaged over the entire
data set, the effect of atmospheric variations on the composition result
is likely smaller than the expectation based on the seasons check, shown in
Figure~\ref{fig:sys_check_season}. 

Finally, we examined varying some of the quality cuts applied to the data and
simulation sets.  As an example, the effect of changing the cut value of the
angular track-length is shown in Figure~\ref{fig:sys_check_trk_elon4}.  As can
be seen in the figure, changing the track-length cut has a minimal effect on the
composition results.  Similar results were observed for other cut parameters
examined as part of the data analysis.

\deleted{ and event selection and reconstruction
  biases }
\explain{deleted spurious sentence fragment accidentally copy/pasted from
  line 450}

\section{Results and Discussion}
\label{sec:results_and_discussion}

We present the results of the analysis in the following forms:
\begin{enumerate}
\item Measured $\langle X_{\rm max} \rangle$ evolution with shower energy.
  These values are for the final event sample and do not include corrections
  for detector acceptance bias or other biases related to event selection or
  event reconstruction.  We also show the results for reconstructed MC showers
  for comparison with the data.
\item Estimated cosmic ray primary fractions based on the full measured 
  $X_{\rm max}$ distribution, using a four component fit.  The primary 
  fractions in this case are corrected for biases in detector acceptance, 
  event selection, and event reconstruction.  
\item Resulting $\langle \ln(A) \rangle$ from the bias corrected four
  component fit.  This result can be thought of as a condensed form of the
  four component fit result.
\item Bias corrected $X_{\rm max}$ using EPOS-LHC fit fractions and the
  unbiased EPOS-LHC MC prediction for the mean $X_{\rm max}$ of the four
  primary particles used in the analysis.
\end{enumerate}

Where applicable, the above results are shown separately for the two hadronic
models, QGSJetII-03 and EPOS-LHC.

\begin{figure}[htb]
  \centering
  \includegraphics[height=2.4in]{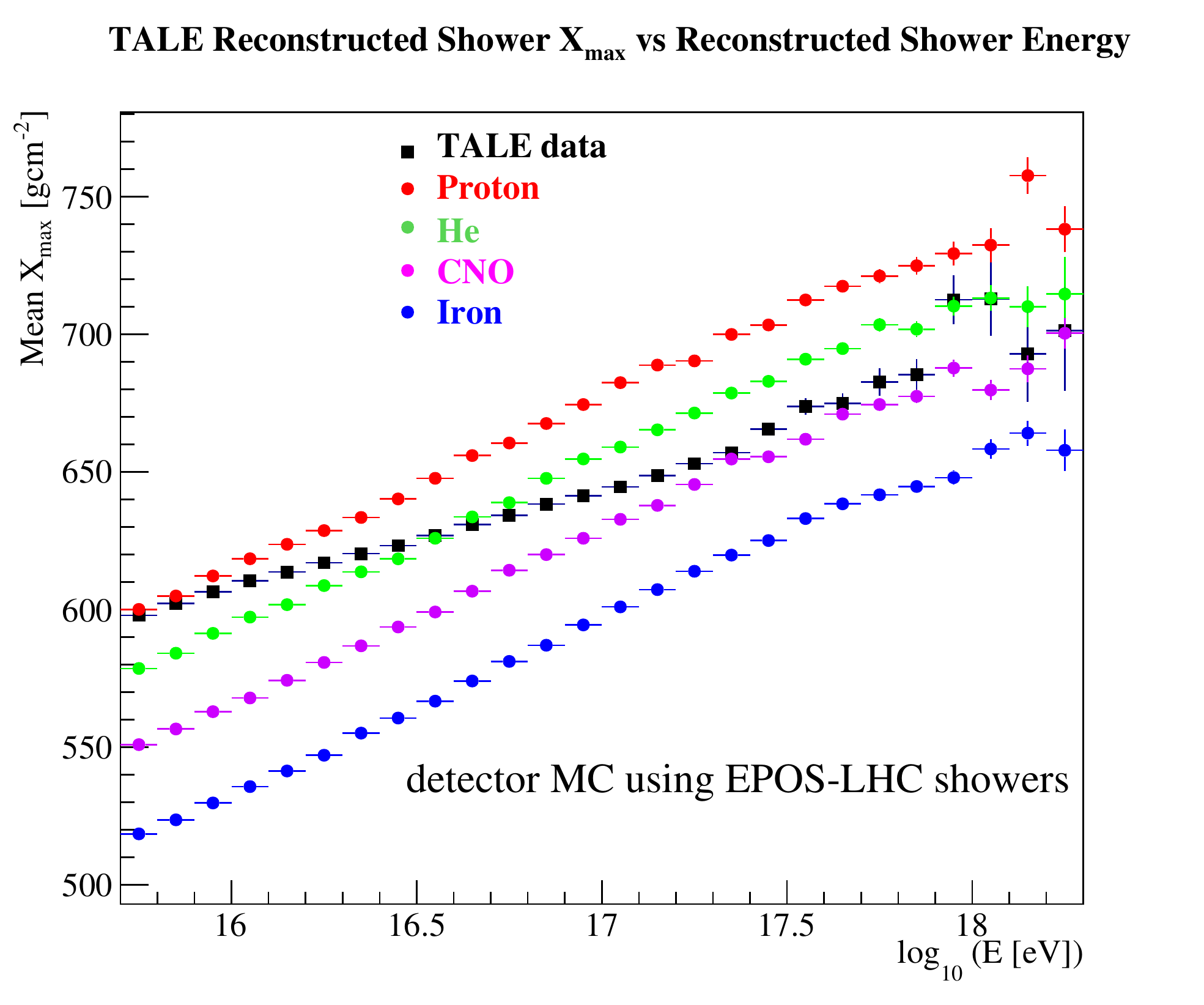}
  \includegraphics[height=2.4in]{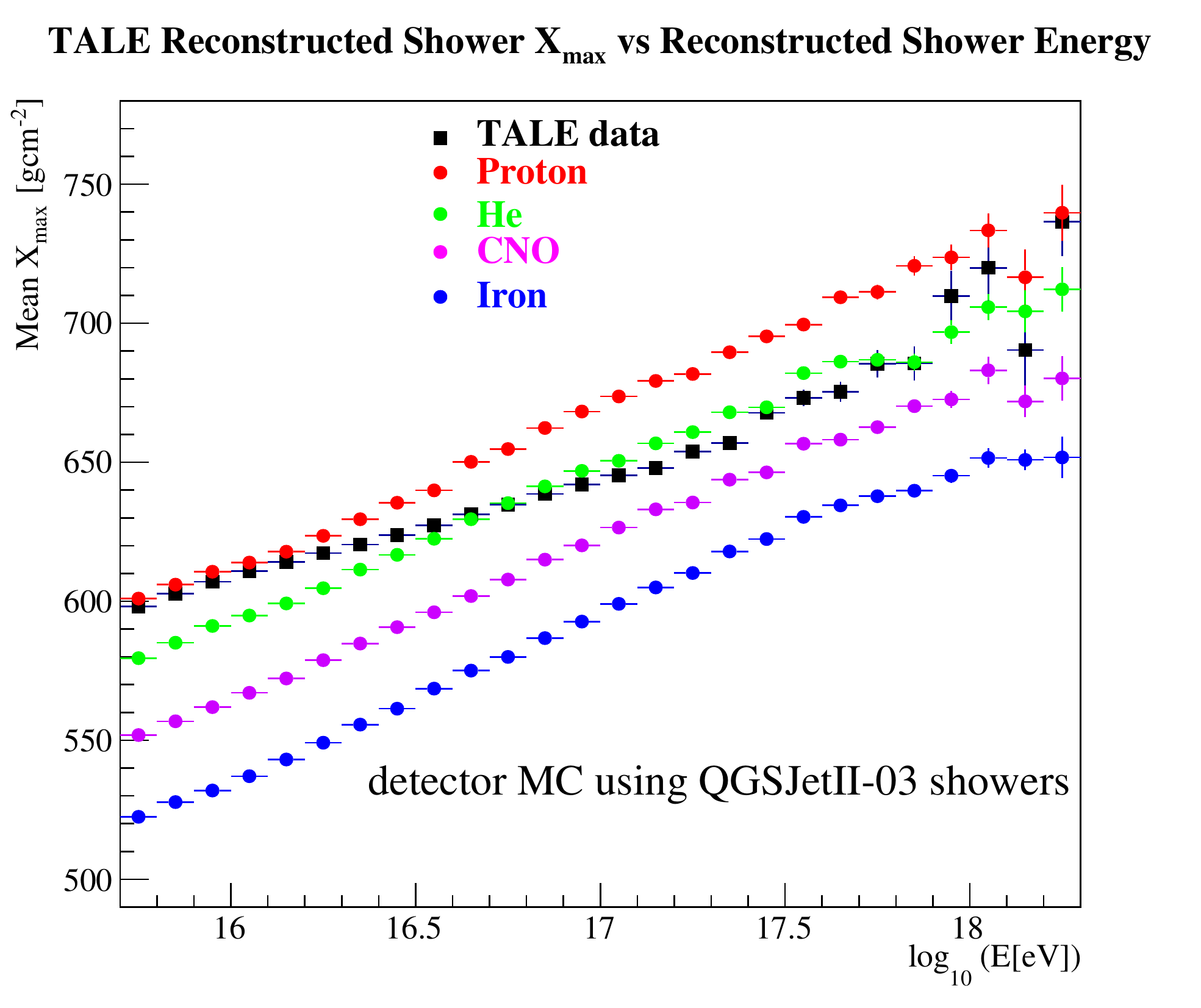}
  \caption{Mean $X_{\rm max}$ as a function of energy for the reconstructed TALE
    events.  
    Shower missing energy correction estimate using EPOS-LHC is shown
    in the upper plot, and using QGSJetII-03 in the lower plot.
    The reconstructed $X_{\rm max}$ values for the four MC primaries, generated
    and reconstructed using the corresponding hadronic model, are
    shown alongside the data for comparison.  The data appears to be getting
    heavier
    with energy in this range for both models.
  }
  \label{fig:xmax_elon1}
\end{figure}

\begin{figure}[htb]
  \centering
  \includegraphics[height=2.4in]{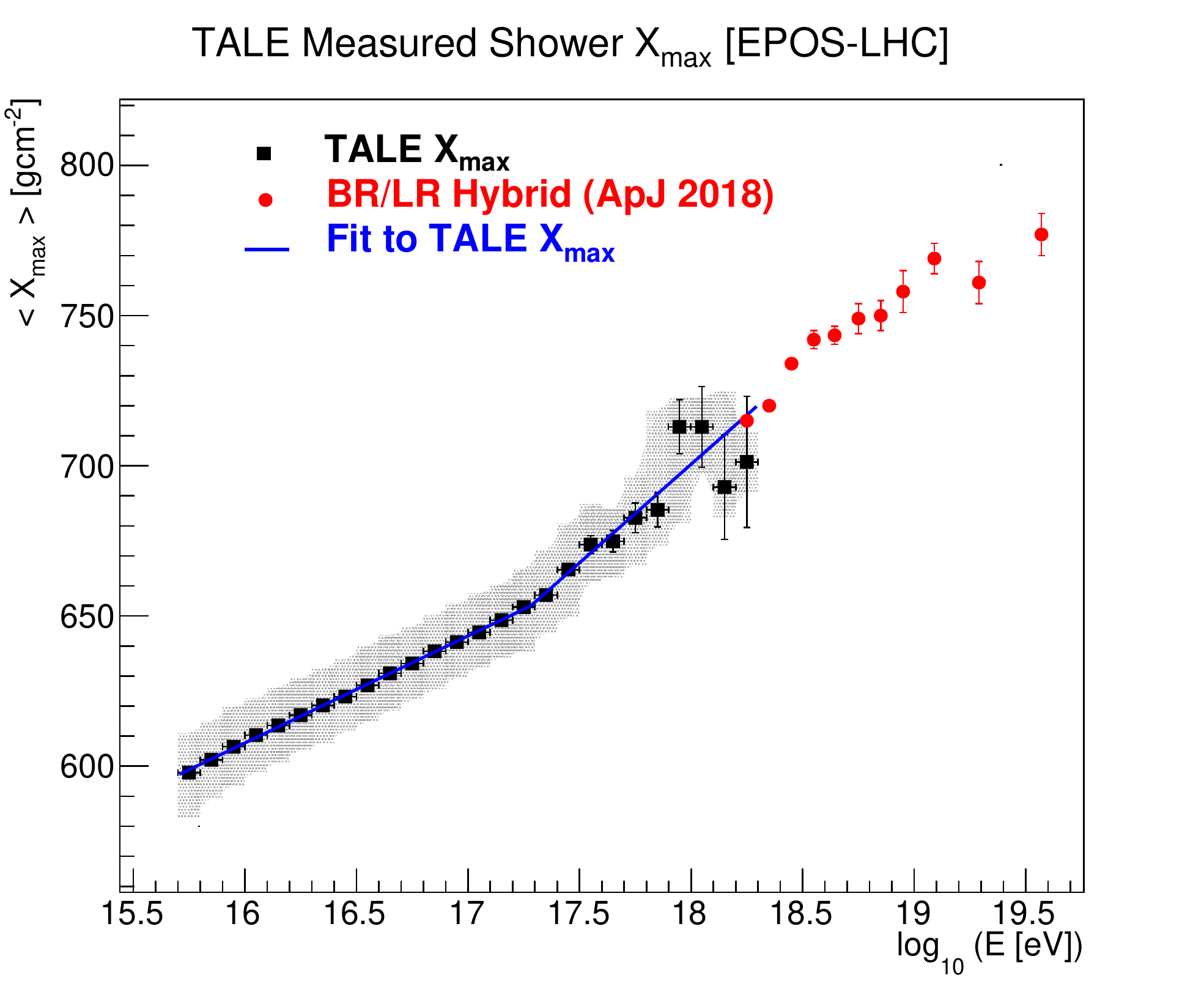}
  \includegraphics[height=2.4in]{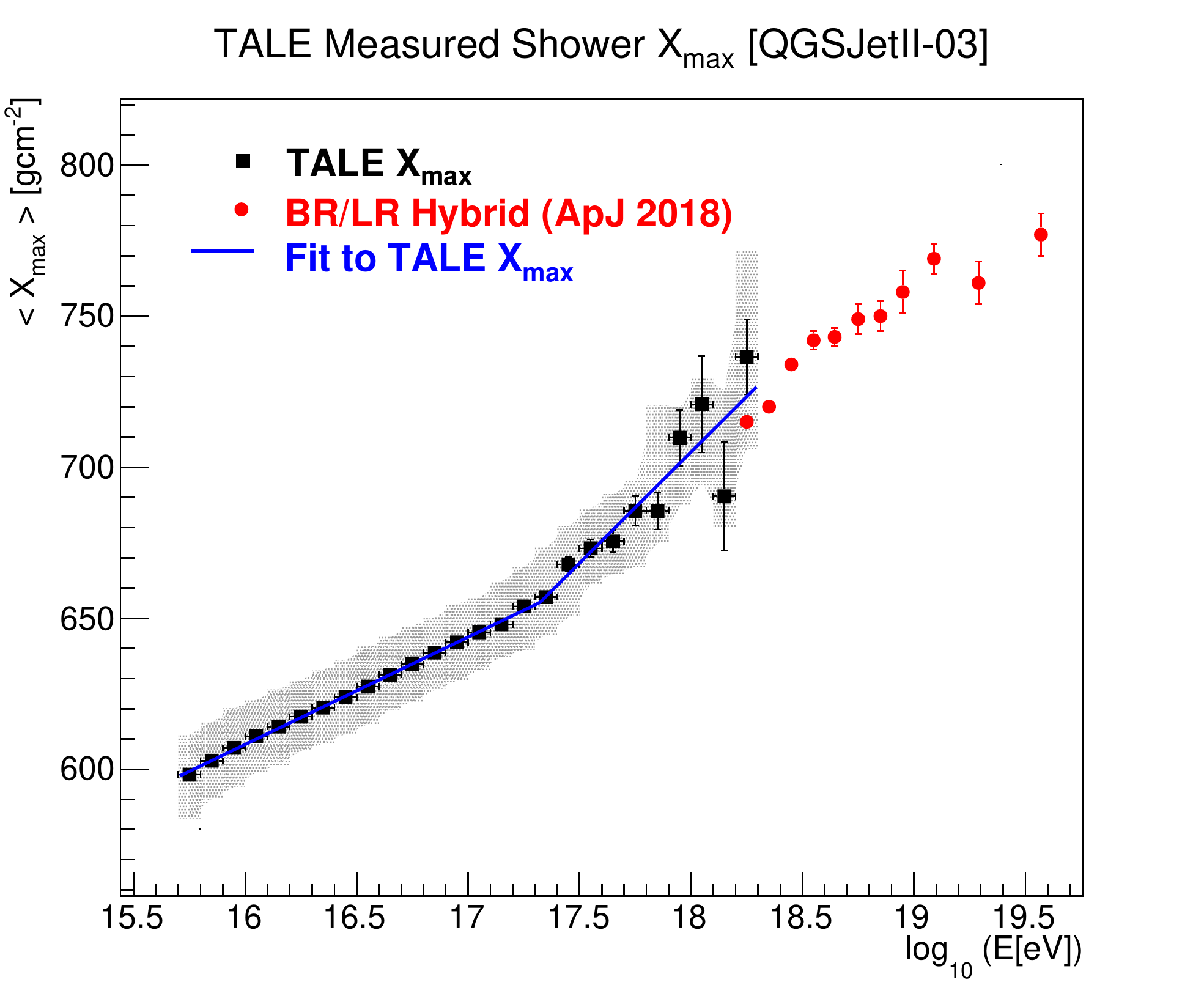}
  \caption{Mean $X_{\rm max}$ as a function of energy for the reconstructed TALE
    events.  Shower missing energy correction estimate using EPOS-LHC is shown
    in the upper plot, and using QGSJetII-03 lower plot.  A broken line fit
    is shown for each plot.  Fit parameters are listed in
    Table~\ref{table:elongation_fit_results}.  Red points at higher energies
    come from a hybrid measurement by the main Telescope Array
    telescopes~\citep{Abbasi:2018nun}.
  }
  \label{fig:xmax_elon2}
\end{figure}

Figure~\ref{fig:xmax_elon1} shows the mean $X_{\rm max}$ values of TALE data 
along with those of simulated showers.
At $\sim 10^{18}$, it can be 
seen that the event statistics becomes low. The data is presented in
Tables~\ref{table_xmax_epos},~\ref{table_xmax_qgs} in Appendix~A.


\begin{table}[htb!]
  \centering
  \caption{Fit parameters to a broken line fit to TALE $X_{\rm max}$ elongation
    rate\replaced{.}{:} \added{Break point energies are expressed as
      $\log_{10}(E/eV)$, and the slopes have units of $(g~cm^{-2}) / decade$.
      Uncertainties are reported as
      $value \pm \sigma_{stat.} + \sigma_{sys.} - \sigma_{sys.} $.}
    The upper set of measurements are for the EPOS-LHC, the lower set
    is for QGSJetII-03.\deleted{Uncertainty reported as
      $value \pm \sigma_{stat.} + \sigma_{sys.} - \sigma_{sys.} $.}
  }
  \label{table:elongation_fit_results}
  \begin{tabular}{|l|l|r|}
    \hline

    EPOS- &  break point  &  17.291  $\pm  0.060 + 0.077 - 0.084$ \\
    LHC   &  slope before &  35.863  $\pm  0.294 + 1.481 - 0.536$ \\
          &  slope after  &  65.413  $\pm  6.655 + 0.000 - 3.269$ \\
    \hline
  QGSJet- &  break point  &  17.310  $\pm  0.049 + 0.052 - 0.179$ \\
  II-03   &  slope before &  35.784  $\pm  0.298 + 1.337 - 0.667$ \\
          &  slope after  &  70.860  $\pm  6.508 + 0.000 - 11.387$ \\ \hline
  \end{tabular}
\end{table}


The elongation rate at energies below $10^{17.3}$~eV indicates that
the composition is getting heavier in this energy range.
A change in the elongation rate (slope of the line fit to $X_{\rm max}$ vs 
energy) is clearly seen for energies greater than $10^{17}$~eV that is not
present in the MC showers for any one primary type.  This change in slope
can be interpreted as a change in composition.  A broken line fit (one floating
break point) to the slope is used to determine the energy at which the slope
changes.  The fit is shown in Figure~\ref{fig:xmax_elon2}; Fit results are 
presented in Table~\ref{table:elongation_fit_results}.  The fit line is in
agreement with the mean $X_{\rm max}$ values measured by the Telescope Array
detectors at EeV energies~\citep{Abbasi:2018nun}, as can be seen in 
Figure~\ref{fig:xmax_elon2}.

\begin{figure}[htb]
  \centering
  \includegraphics[height=2.4in]{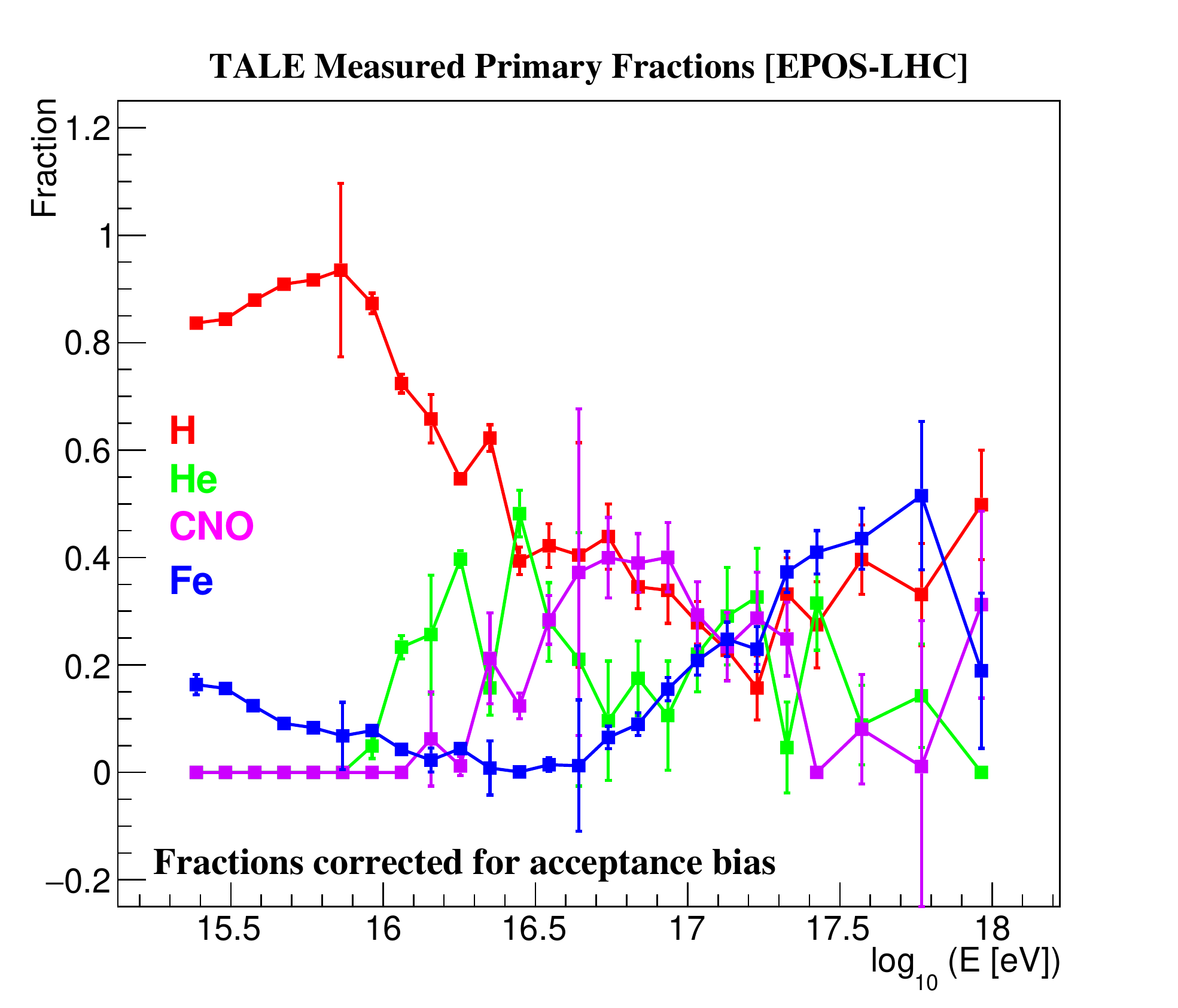} 
  \includegraphics[height=2.4in]{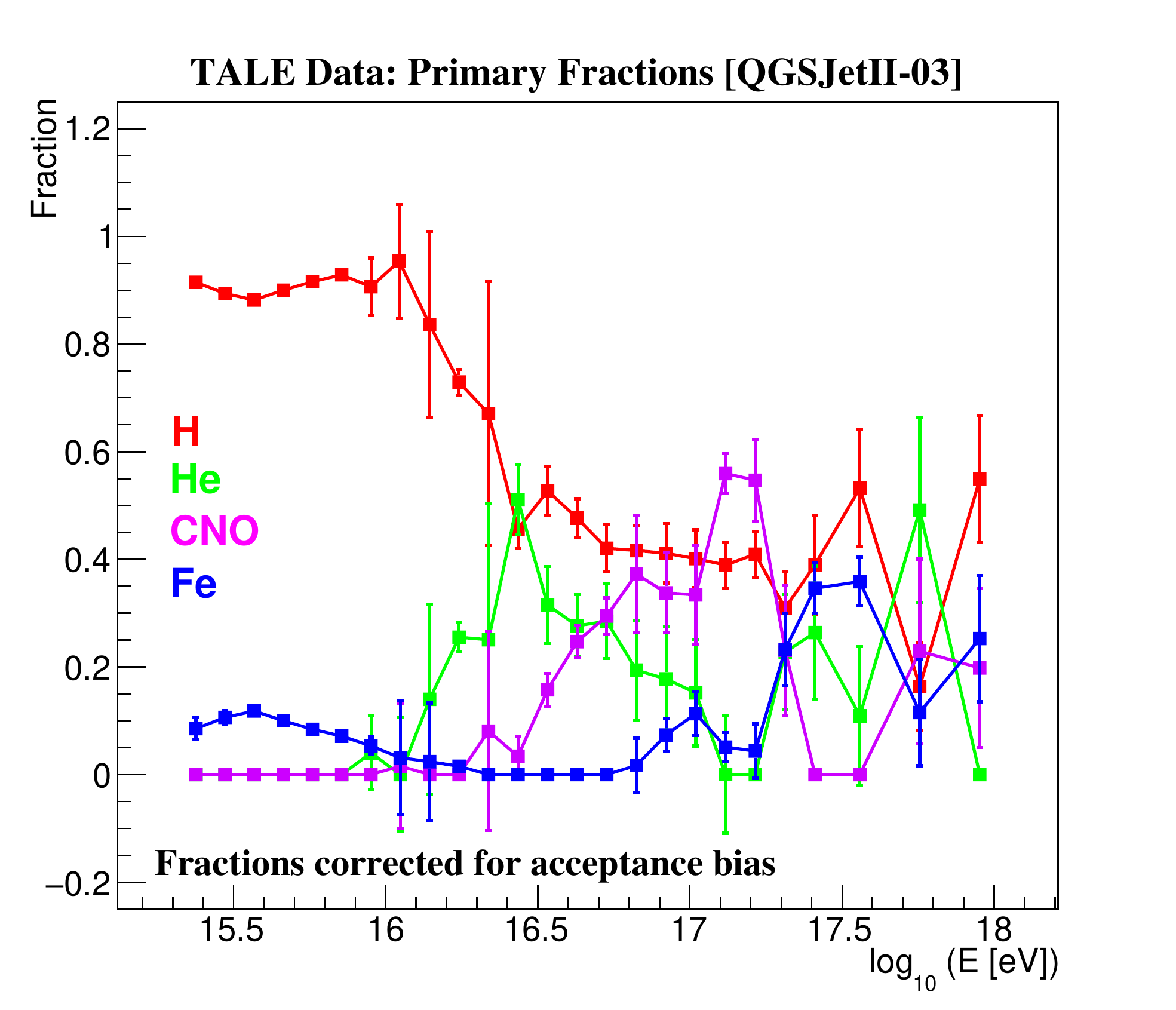} 
  \caption{Fit results to the data $X_{\rm max}$ distributions (per energy bin) to 
    a four component MC distributions. Primary fractions using the EPOS-LHC
    based simulations are shown in the top plot.  Those using QGSJetII-03, 
    in the bottom plot.
  }
  \label{fig:xmax_fractions}
\end{figure}


\begin{figure*}[htb!]
  \centering
  \gridline{
    \fig{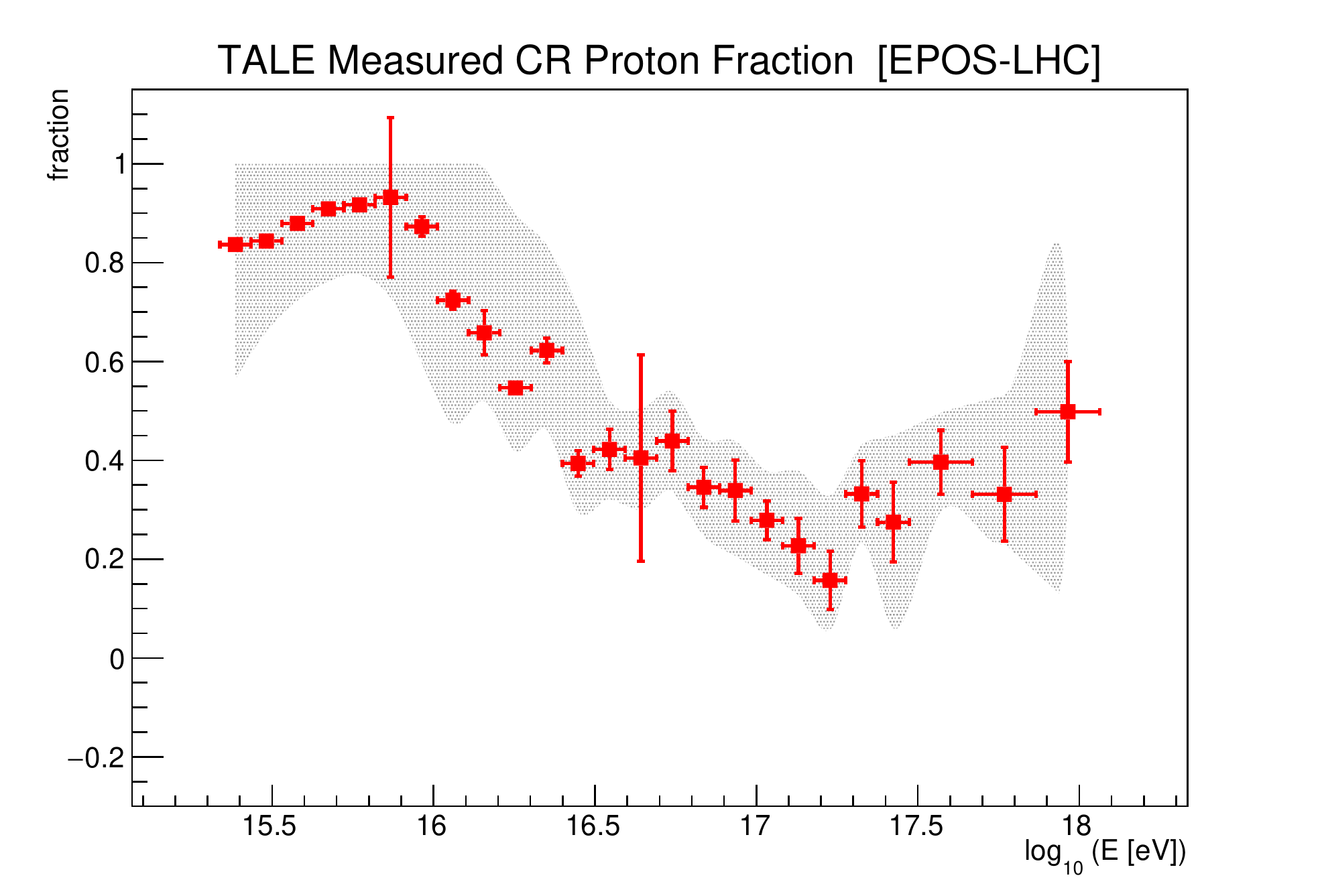}{0.45\textwidth}{}
    \fig{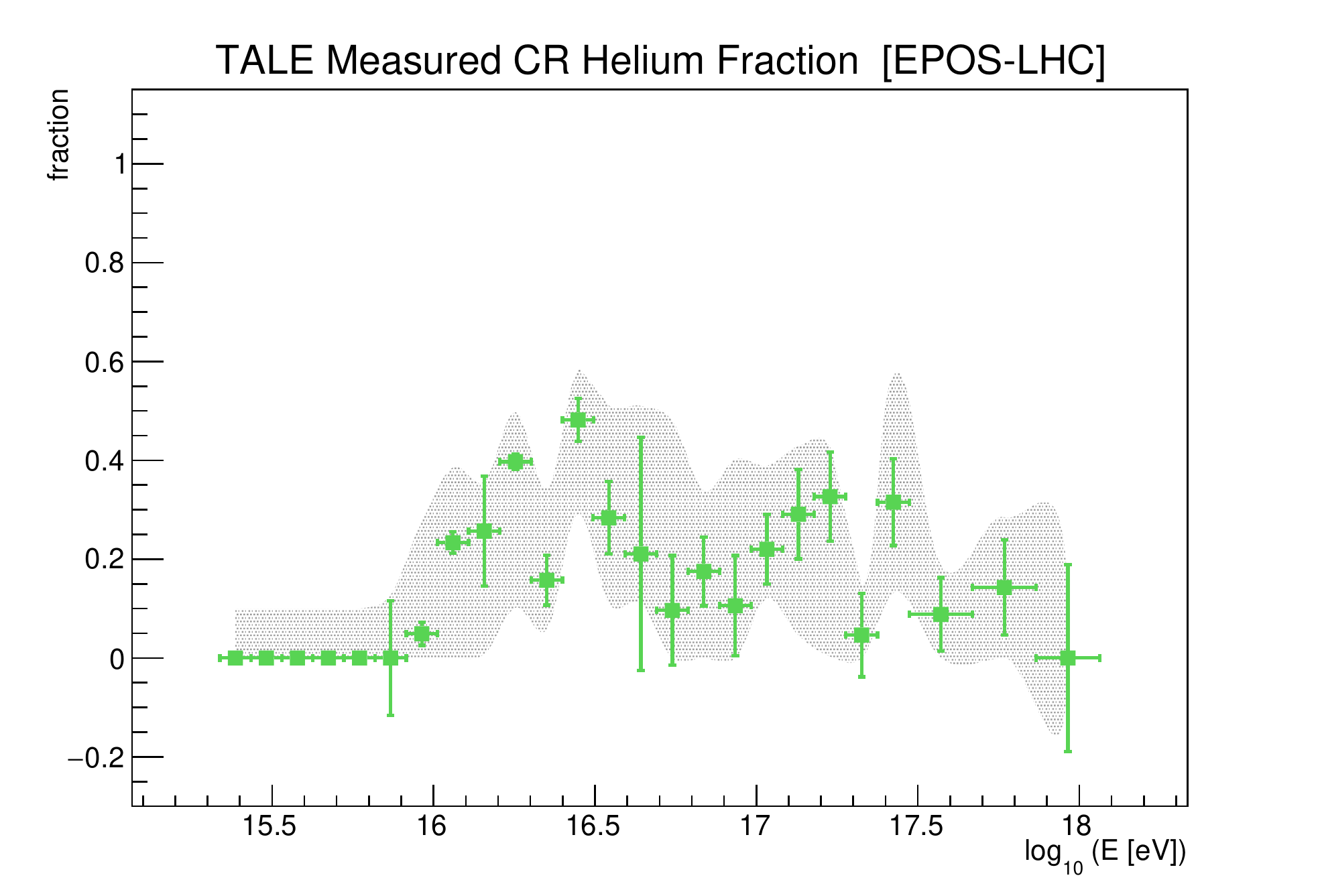}{0.45\textwidth}{}
  }
  \gridline{
    \fig{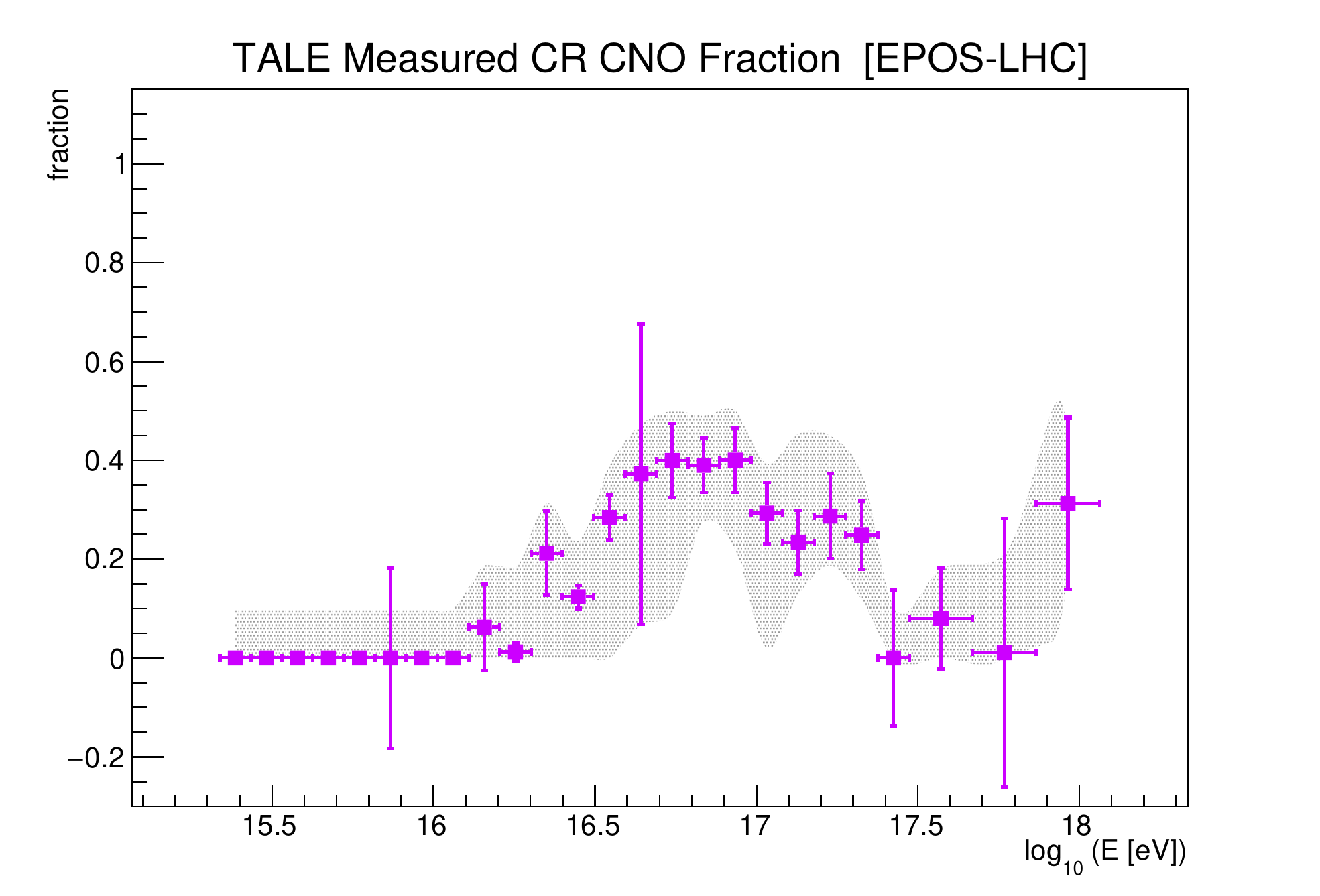}{0.45\textwidth}{}
    \fig{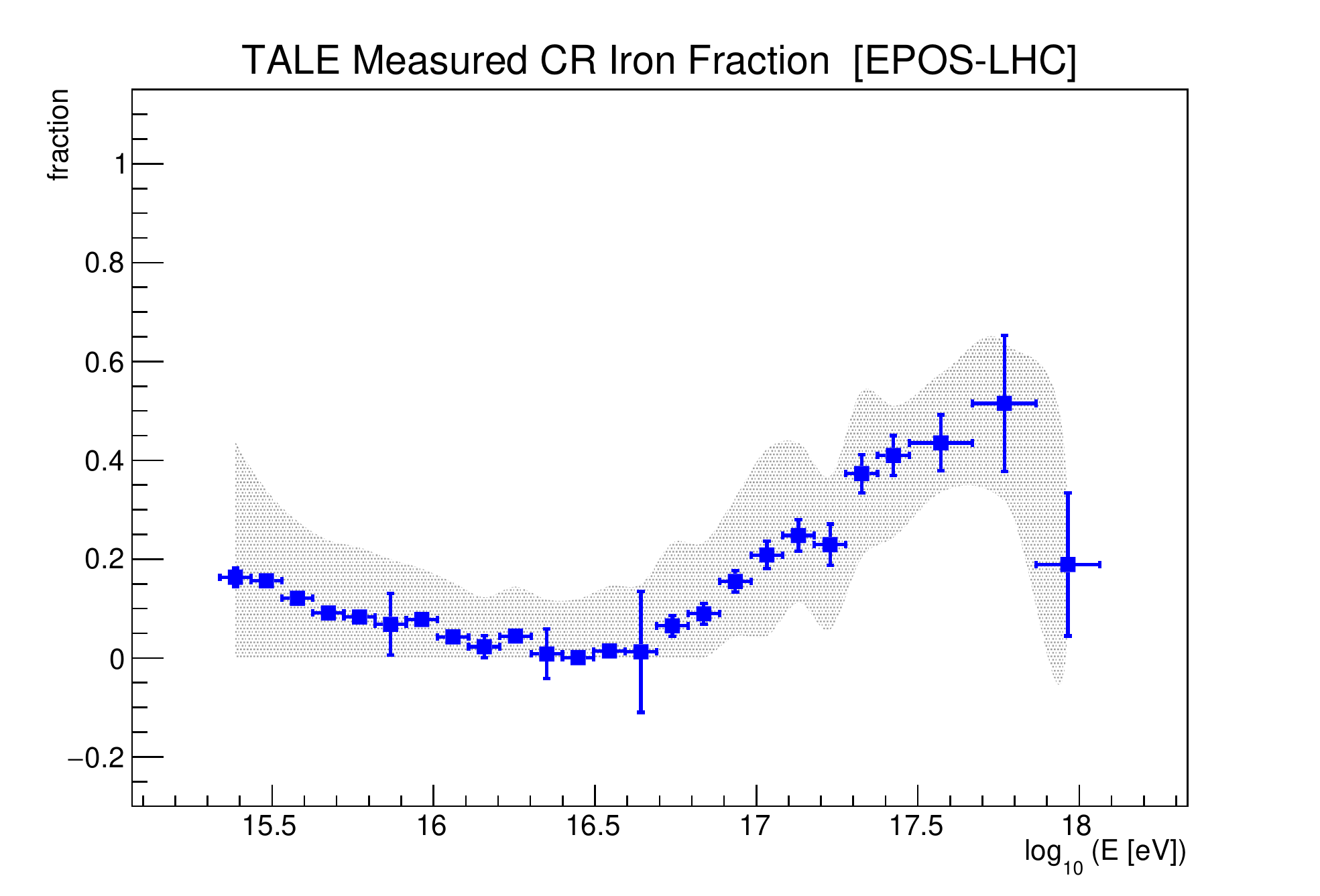}{0.45\textwidth}{}
  }
  \caption{Primary fractions using the EPOS-LHC based simulations
    are shown for a four primary composition.
  }
  \label{fig:xmax_fractions_sys_band}
\end{figure*}

\begin{figure}[htb]
  \centering
  \includegraphics[height=2.4in]{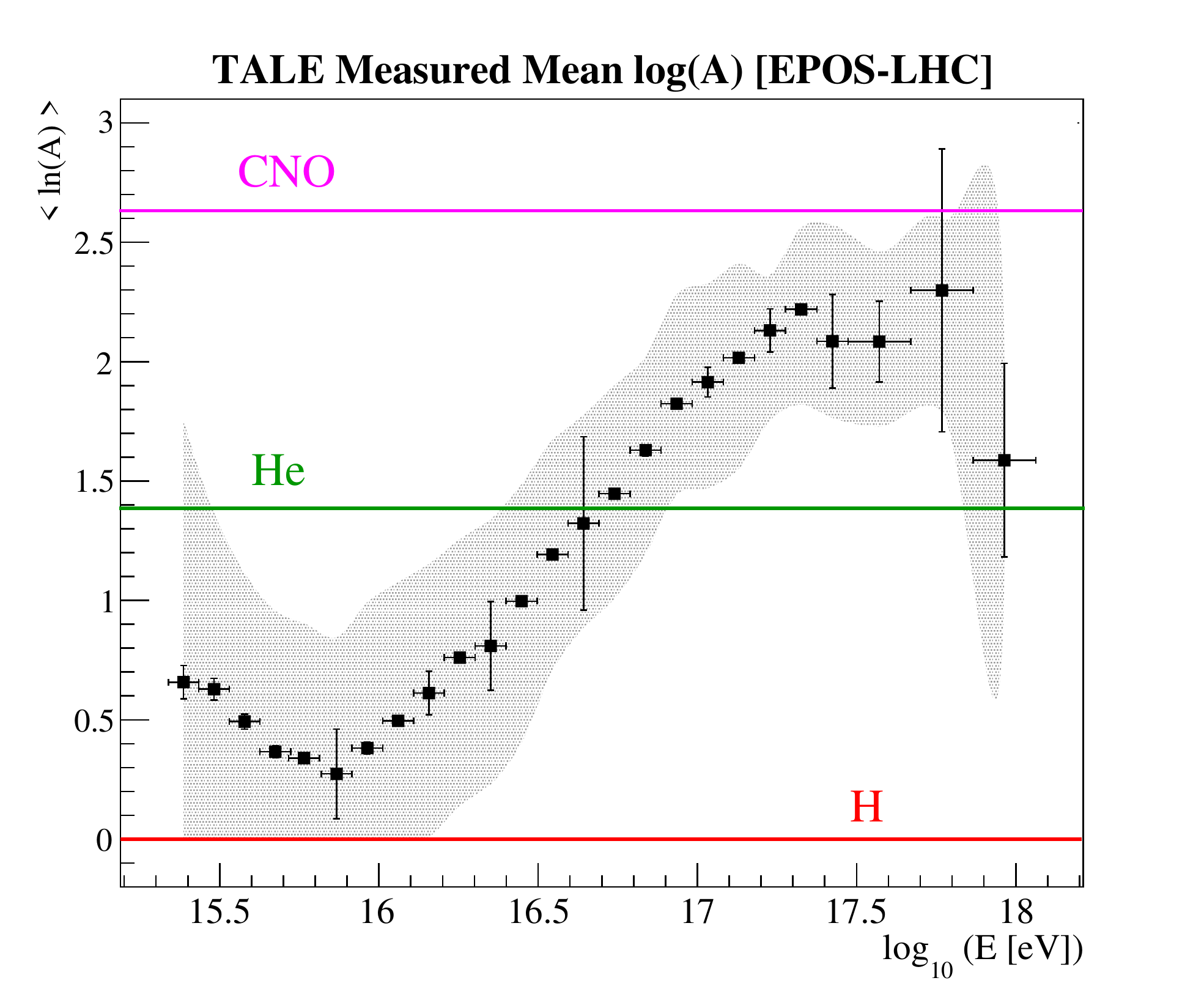} 
  \caption{
    Estimated $\langle \ln(A) \rangle$ from four component fits to TALE data.
    Horizontal lines show calculated $\ln(A)$ values for H, He, and N, for
    reference.
  }
  \label{fig:mean_logA_sys_band}
\end{figure}

Data $X_{\rm max}$ distributions were created for each energy bin, using a bin 
width of 0.1 in $\log_{10}(E)$ up to an energy of $10^{17.4}$~eV, then 0.2 in 
$\log{E}$ up to $10^{18}$~eV.  These distributions were fit using MC 
distributions created using either of two hadronic interaction models, and 
containing four different primaries.  Fit results are shown in 
Figure~\ref{fig:xmax_fractions}.  The small number of observed
events with energies greater than $10^{18}$~eV was not sufficient to
extend the fits beyond $10^{18}$~eV.

\added{
  Figure~\ref{fig:xmax_fractions} shows that the proton fraction is dominant
  at energies below~$\sim 10^{16}$~eV. Only a small contribution from Iron is
  seen, and Helium and CNO fractions are too small to measure.  The Helium
  fraction only becomes appreciable at an energy~$\sim 10^{16}$~eV, and is
  followed by contributions from CNO and Iron at higher energies as
  expected~\citep{Peters:1961}.  The observed dominance of protons at energies
  below $\sim 10^{16}$~eV is in contrast with recent observations by
  IceTop~\citep{IceCube:2012nn} and IceCube~\citep{Aartsen:2016nxy},
  which report a more even mixture of the four components,
  (H, He, O, Fe), as shown in FIG.~15 of~\citep{IceCube:2019hmk}.
  KASCADE~\citep{Antoni:2003gd}, 
  on the other hand seems to favor a dominant role for Helium over protons in
  this energy region~\citep{Antoni:2005wq}.  Extrapolation of lower energy
  measurements by CREAM~\citep{Yoon:2017qjx}, also lead to the expectation
  that the Helium contribution should be comparable to, if not larger than,
  that of protons at TALE observed energies below~$\sim 10^{16}$~eV.

  Due to the significant overlap of the proton and Helium $X_{\rm max}$
  distributions, and the hadronic model dependence of the predicted
  distributions, it is possible that some of the Helium contribution to
  the CR flux is getting misidentified as proton contribution.  We studied
  the event reconstruction performance and the measurement's systematic
  uncertainties, to estimate the size of this effect.  In addition to the
  reconstruction performance and systematics studies discussed above,
  additional checks were made to test the proton fraction for the lower
  energy bins.  The fit to the full $X_{\rm max}$ distributions was modified
  in two ways:
  First, the range of accepted shower $X_{\rm max}$ was restricted to values
  less than 760~g~cm$^{-2}$ (compare to Figure~\ref{fig:ebin153_data_mc_xmax}).
  This should cut any bias on the fit result due to the distribution tail;
  expected to be all protons.  This change resulted in an insignificant
  change to the fit result, for all energy bins below~$\sim 10^{16}$~eV.
  Second, in a separate check, the {\em recorded} proton fit fraction was
  constrained to be smaller than $80\%$.  In this case, as expected, the Fitter
  set the proton fit fraction ``at the limit'', i.e. $80\%$, and increased
  the Helium fit fraction to compensate for the missing protons.  For most low
  energy bins, the resulting fit $\chi^{2}$ was larger by roughly three sigma's
  than the $\chi_{min}^{2}$ obtained for the unconstrained fit.  The fit quality
  got worse for smaller proton fraction limits.

  Resolving the proton and Helium components of the CR flux may
  have involved some subtle effects that escaped our scrutiny; and, 
  one can take a more conservative viewpoint, and treat the combined
  proton and Helium fractions as representing the ``light'' component
  of the cosmic ray flux.  We do, however, present the fit results for
  the individual primaries, along with estimated systematic uncertainties,
  as our best estimate of the primary fractions in the CR flux.
}

To incorporate the systematic uncertainty into the presentation of the results
we focus our attention on the EPOS-LHC based analysis.  The primary fractions
are shown separately in Figure~\ref{fig:xmax_fractions_sys_band},
followed by the estimated $\langle \ln(A) \rangle$ displayed in
Figure~\ref{fig:mean_logA_sys_band}.  Similar to the trend found for
$\langle X_{\rm max} \rangle$, figures~\ref{fig:xmax_fractions_sys_band}
and~\ref{fig:mean_logA_sys_band} also indicate that the composition is
getting heavier in the $10^{16}$~eV decade, and that there is a further
change just above $10^{17}$~eV.

\begin{figure}[htb]
  \centering
  \includegraphics[height=2.4in]{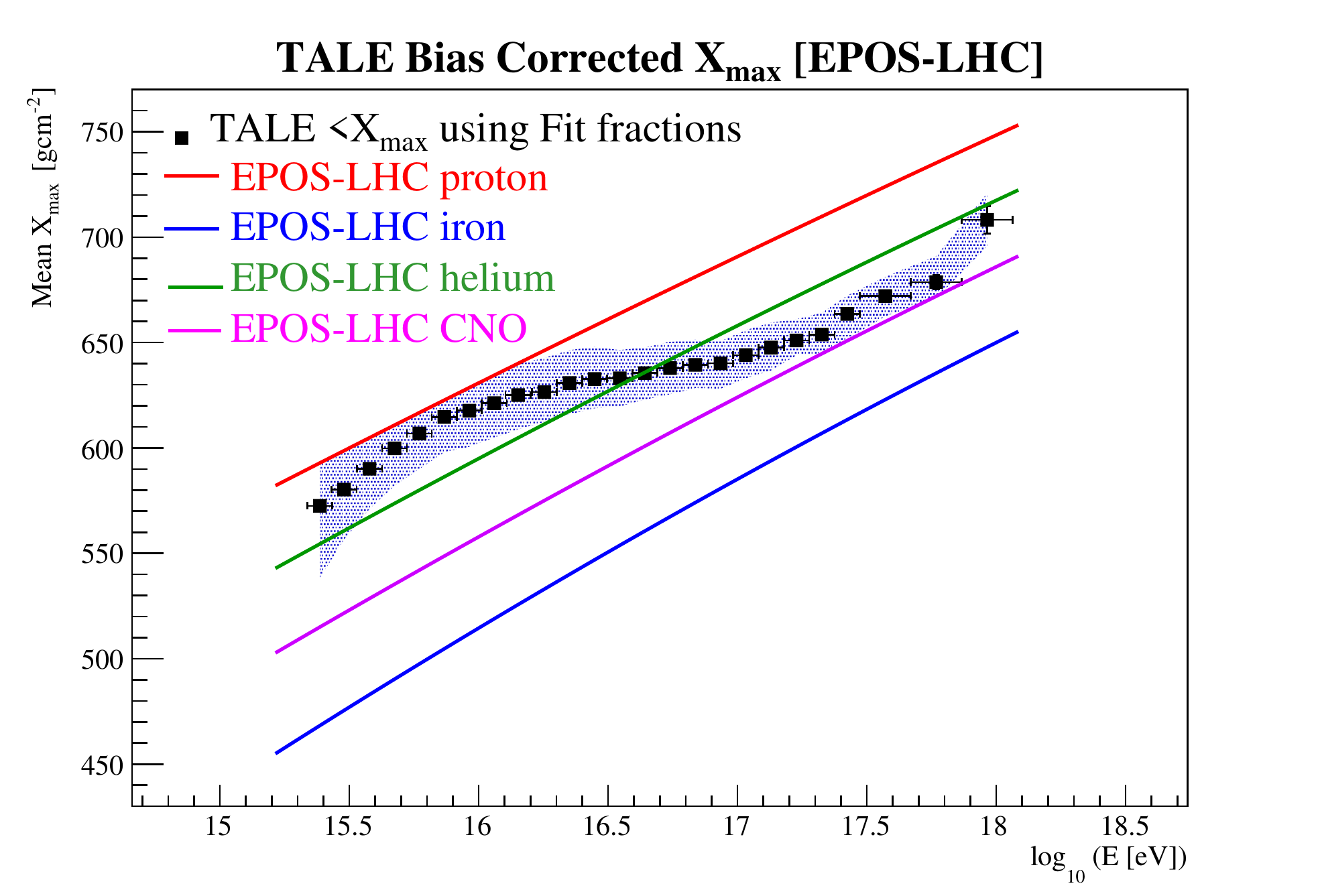} 
  \caption{
    Bias corrected $X_{\rm max}$ using EPOS-LHC fit fractions and the unbiased EPOS-LHC
    MC prediction for the mean $X_{\rm max}$ of the four primary particles used in the
    analysis.  These results include a first order correction to the detector
    acceptance bias.
  }
  \label{fig:xmax_no_bias}
\end{figure}

Finally, similar to the calculation of $\langle \ln(A) \rangle$, the bias
corrected fractions were used to calculate a no-bias
$\langle X_{\rm max} \rangle = \sum_{ip}{f_{ip}*\langle X^{(ip)}_{\rm max} \rangle}$,
where $ip$ stands for one of \{H, He, N, Fe\}, and
$\langle X^{(ip)}_{\rm max} \rangle$ is the EPOS-LHC predicted (MC thrown)
$\langle X_{\rm max} \rangle$ of primary $ip$.  These results are displayed in
Figure~\ref{fig:xmax_no_bias}, with data in Table~\ref{table_xmax_no_bias}.

\section{Summary}
\label{sec:summary}
We presented the results of a measurement of the cosmic rays composition in
the energy range of $10^{15.3}$ - $10^{18.3}$~eV using data collected by the
TALE detector over a period of roughly four years.  An examination of the
mean $X_{\rm max}$ versus energy,
\replaced{Figures~\ref{fig:xmax_elon2}~\ref{fig:xmax_elon1}}
         {(Figures~\ref{fig:xmax_elon1},~\ref{fig:xmax_elon2})},
shows a composition that
is getting heavier, followed by a change in the $X_{\rm max}$ elongation rate
at an energy of $\sim 10^{17.3}$~eV.  This ``break'' in the elongation rate
is likely correlated with the observed break in the cosmic rays energy
spectrum \citep{Abbasi:2018xsn}.

We also fit the data $X_{\rm max}$ distributions, per energy bin, to 
reconstructed MC showers generated for four primary particle types.  These fits 
show a light mostly proton and Helium composition at the lower energies 
becoming more mixed near $10^{17}$~eV.
In this analysis, we do not have sufficient statistics 
to comment on the composition for cosmic rays with energies greater
than $10^{18}$~eV.  These results are shown as the fractions themselves,
\added{(}Figures~\ref{fig:xmax_fractions},~\ref{fig:xmax_fractions_sys_band}\added{)},
as a derived mean $\ln(A)$, Figure~\ref{fig:mean_logA_sys_band}, and 
a derived mean $X_{\rm max}$, Figure~\ref{fig:xmax_no_bias}.


\section*{Acknowledgments}

The Telescope Array experiment is supported by the Japan Society for
the Promotion of Science(JSPS) through 
Grants-in-Aid
for Priority Area
431,
for Specially Promoted Research 
JP21000002, 
for Scientific  Research (S) 
JP19104006, 
for Specially Promoted Research 
JP15H05693, 
for Scientific  Research (S)
JP15H05741, for Science Research (A) JP18H03705, 
for Young Scientists (A)
JPH26707011, 
and for Fostering Joint International Research (B)
JP19KK0074,
by the joint research program of the Institute for Cosmic Ray Research (ICRR),
The University of Tokyo; 
by the Pioneering Program of RIKEN for Matter in the Universe (r-EMU);
by the U.S. National Science
Foundation awards PHY-1404495, PHY-1404502, PHY-1607727, PHY-1712517,
and PHY-1806797; 
by the National Research Foundation of Korea
(2017K1A4A3015188, 2020R1A2C1008230, \& 2020R1A2C2102800) ;
by the Russian Science Foundation grant 20-42-09010 (INR),
IISN project No. 4.4501.18,
and Belgian Science Policy under IUAP VII/37 (ULB).
The foundations of Dr. Ezekiel R. and Edna Wattis Dumke, Willard L. Eccles,
and George S. and Dolores Dor\'e Eccles all helped with generous donations.
The State of Utah supported the project through its Economic Development Board,
and the University of Utah through the Office of the Vice President for
Research. The experimental site became available through the cooperation of
the Utah School and Institutional Trust Lands Administration (SITLA),
U.S. Bureau of Land Management (BLM), and the U.S. Air Force.
We appreciate the assistance of the State of Utah and Fillmore offices of
the BLM in crafting the Plan of Development for the site.
Patrick A. Shea assisted the collaboration with valuable advice and supported
the collaboration’s efforts. The people and the officials of Millard County,
Utah have been a source of steadfast and warm support for our work which we
greatly appreciate. We are indebted to the Millard County Road Department
for their efforts to maintain and clear the roads which get us to our sites.
We gratefully acknowledge the contribution from the technical staffs of our
home institutions. An allocation of computer time from the Center for
High Performance Computing at the University of Utah is gratefully acknowledged.

\section*{Software}
\software{CONEX~\citep{Bergmann:2006yz}, CORSIKA~\citep{Heck:1998vt},
  CORSIKA/IACT~\citep{Bernlohr:2008kv}}

\section*{Appendix A}
\label{appendix:xmax}

Mean shower $X_{\rm max}$ for events included in the composition analysis
of TALE data are shown in Tables~\ref{table_xmax_epos} and ~\ref{table_xmax_qgs}.
The number of events in each energy bin, along with the $X_{\rm max}$ and
estimated errors are listed.  Table~\ref{table_xmax_no_bias} lists the
bias corrected $\langle X_{\rm max} \rangle$ of data events, 
estimated using the bias corrected fit fractions from EPOS-LHC based
analysis.

\startlongtable
\begin{deluxetable}{c|c|l}
  \tablecaption{$\langle X_{\rm max} \rangle$ of data events.  Missing energy correction based on EPOS-LHC hadronic model.
    \label{table_xmax_epos}
  }
  \tablehead{
    \colhead{energy-bin} & \colhead{Number} & \colhead{$\langle X_{\rm max} \rangle \pm \sigma_{stat.} +/- \sigma_{sys.}$}\\
\colhead{log$_{10}$ (E [eV])} & \colhead{of Events} & \colhead{g~cm$^{-2}$}
}
\startdata
15.30-15.40 &    39669 &   590.1 $\pm$   0.4 +  10.0 -  10.0 \\ 
15.40-15.50 &    71699 &   585.8 $\pm$   0.3 +  10.0 -  10.0 \\ 
15.50-15.60 &    98113 &   585.8 $\pm$   0.3 +  12.4 -  10.5 \\ 
15.60-15.70 &   109055 &   589.8 $\pm$   0.2 +  15.6 -  12.8 \\ 
15.70-15.80 &    97017 &   597.9 $\pm$   0.3 +  13.0 -  14.9 \\ 
15.80-15.90 &    91380 &   602.1 $\pm$   0.3 +  12.8 -  12.6 \\ 
15.90-16.00 &    78444 &   606.4 $\pm$   0.3 +  13.2 -  12.7 \\ 
16.00-16.10 &    64594 &   610.4 $\pm$   0.3 +  12.0 -  12.2 \\ 
16.10-16.20 &    51230 &   613.6 $\pm$   0.3 +  12.7 -  12.2 \\ 
16.20-16.30 &    40036 &   617.0 $\pm$   0.4 +  12.2 -  11.5 \\ 
16.30-16.40 &    30915 &   620.3 $\pm$   0.4 +  12.1 -  12.5 \\ 
16.40-16.50 &    23657 &   623.2 $\pm$   0.4 +  12.4 -  11.7 \\ 
16.50-16.60 &    17972 &   626.9 $\pm$   0.5 +  13.1 -  12.2 \\ 
16.60-16.70 &    13517 &   630.9 $\pm$   0.6 +  12.7 -  12.2 \\ 
16.70-16.80 &     9940 &   634.2 $\pm$   0.6 +  12.2 -  12.1 \\ 
16.80-16.90 &     7644 &   638.3 $\pm$   0.7 +  11.8 -  12.8 \\ 
16.90-17.00 &     5560 &   641.3 $\pm$   0.8 +  12.8 -  11.9 \\ 
17.00-17.10 &     4284 &   644.5 $\pm$   1.0 +  12.9 -  11.5 \\ 
17.10-17.20 &     3059 &   648.6 $\pm$   1.1 +  12.0 -  12.2 \\ 
17.20-17.30 &     1833 &   653.0 $\pm$   1.4 +  12.8 -  14.8 \\ 
17.30-17.40 &     1295 &   657.0 $\pm$   1.8 +  14.8 -  12.4 \\ 
17.40-17.50 &      807 &   665.5 $\pm$   2.2 +  15.1 -  15.9 \\ 
17.50-17.60 &      487 &   673.8 $\pm$   2.9 +  13.4 -  12.5 \\ 
17.60-17.70 &      300 &   674.9 $\pm$   3.6 +  10.6 -  11.9 \\ 
17.70-17.80 &      176 &   682.7 $\pm$   4.9 +  14.0 -  15.6 \\ 
17.80-17.90 &      112 &   685.4 $\pm$   5.7 +  33.4 -  12.0 \\ 
17.90-18.00 &       57 &   712.5 $\pm$   9.0 +  10.0 -  31.4 \\ 
18.00-18.10 &       25 &   712.8 $\pm$  13.4 +  10.0 -  14.8 \\ 
18.10-18.20 &       16 &   692.9 $\pm$  17.4 +  32.2 -  10.0 \\ 
18.20-18.30 &        5 &   701.3 $\pm$  21.8 +  10.0 -  10.0 \\ \hline
\enddata
\end{deluxetable}

\newpage

\startlongtable
\begin{deluxetable}{c|c|l}
  \tablecaption{$\langle X_{\rm max} \rangle$ of data events.  Missing energy correction based on QGSJetII-03 hadronic model.
    \label{table_xmax_qgs}
  }
  \tablehead{
    \colhead{energy-bin} & \colhead{Number } & \colhead{$\langle X_{\rm max} \rangle \pm \sigma_{stat.} +/- \sigma_{sys.}$}\\
    \colhead{log$_{10}$ (E [eV])} & \colhead{of Events} & \colhead{g~cm$^{-2}$}
  }
  \startdata
  15.30-15.40 &    43977 &   589.3 $\pm$   0.4 +  10.0 -  10.0 \\ 
  15.40-15.50 &    75692 &   585.6 $\pm$   0.3 +  10.0 -  10.0 \\ 
  15.50-15.60 &   100698 &   586.1 $\pm$   0.3 +  13.0 -  10.8 \\ 
  15.60-15.70 &   109617 &   590.8 $\pm$   0.2 +  15.3 -  13.3 \\ 
  15.70-15.80 &    96990 &   598.2 $\pm$   0.3 +  13.3 -  14.4 \\ 
  15.80-15.90 &    89810 &   602.8 $\pm$   0.3 +  12.8 -  12.6 \\ 
  15.90-16.00 &    76271 &   607.0 $\pm$   0.3 +  13.2 -  12.9 \\ 
  16.00-16.10 &    62828 &   610.8 $\pm$   0.3 +  12.1 -  11.8 \\ 
  16.10-16.20 &    49683 &   614.1 $\pm$   0.3 +  12.4 -  12.5 \\ 
  16.20-16.30 &    38534 &   617.4 $\pm$   0.4 +  12.1 -  11.4 \\ 
  16.30-16.40 &    29904 &   620.4 $\pm$   0.4 +  12.5 -  11.8 \\ 
  16.40-16.50 &    22732 &   623.8 $\pm$   0.4 +  12.1 -  12.1 \\ 
  16.50-16.60 &    17201 &   627.3 $\pm$   0.5 +  13.2 -  12.4 \\ 
  16.60-16.70 &    13166 &   631.3 $\pm$   0.6 +  12.5 -  11.7 \\ 
  16.70-16.80 &     9447 &   634.7 $\pm$   0.7 +  12.7 -  12.3 \\ 
  16.80-16.90 &     7338 &   638.6 $\pm$   0.8 +  11.8 -  12.4 \\ 
  16.90-17.00 &     5389 &   642.0 $\pm$   0.9 +  12.8 -  12.6 \\ 
  17.00-17.10 &     4075 &   645.3 $\pm$   1.0 +  11.4 -  12.4 \\ 
  17.10-17.20 &     2916 &   648.0 $\pm$   1.1 +  14.2 -  10.8 \\ 
  17.20-17.30 &     1719 &   653.9 $\pm$   1.5 +  13.0 -  14.3 \\ 
  17.30-17.40 &     1241 &   657.0 $\pm$   1.8 +  16.4 -  12.3 \\ 
  17.40-17.50 &      762 &   667.8 $\pm$   2.4 +  12.6 -  17.5 \\ 
  17.50-17.60 &      450 &   673.1 $\pm$   3.0 +  13.2 -  11.8 \\ 
  17.60-17.70 &      285 &   675.4 $\pm$   3.6 +  15.5 -  10.8 \\ 
  17.70-17.80 &      167 &   685.5 $\pm$   4.9 +  10.0 -  18.5 \\ 
  17.80-17.90 &      107 &   685.5 $\pm$   6.1 +  35.0 -  10.9 \\ 
  17.90-18.00 &       50 &   709.8 $\pm$   9.2 +  10.0 -  22.7 \\ 
  18.00-18.10 &       20 &   719.9 $\pm$  15.9 +  10.0 -  25.6 \\ 
  18.10-18.20 &       16 &   690.4 $\pm$  18.0 +  35.0 -  10.0 \\ 
  18.20-18.30 &        5 &   736.5 $\pm$  12.3 +  35.0 -  30.4 \\ \hline 
  \enddata
\end{deluxetable}

\startlongtable
\begin{deluxetable}{c|c|l}
  \tablecaption{Bias corrected $\langle X_{\rm max} \rangle$ of data events, 
    estimated using the bias corrected fit fractions from EPOS-LHC based
    analysis.  Statistical uncertainties are those on the mean of the
    corresponding measured data histogram bin.  Systematics are calculated
    based on estimated systematic uncertainties on primary fit fractions.
    \label{table_xmax_no_bias}
  }
  \tablehead{
    \colhead{energy-bin} & \colhead{Number} & \colhead{$\langle X_{\rm max} \rangle \pm \sigma_{stat.} +/- \sigma_{sys.}$} \\
    \colhead{log$_{10}$ (E [eV])} & \colhead{of Events} & \colhead{g~cm$^{-2}$}
  }
  \startdata
  15.338-15.434 &    40122 &   572.5 $\pm$   0.4 +  20.4 -  34.0 \\ 
  15.434-15.530 &    70340 &   579.6 $\pm$   0.3 +  19.2 -  22.6 \\ 
  15.530-15.626 &    94914 &   590.1 $\pm$   0.3 +  14.7 -  19.0 \\ 
  15.626-15.723 &   105576 &   599.8 $\pm$   0.2 +  11.0 -  17.7 \\ 
  15.723-15.819 &    95197 &   606.8 $\pm$   0.3 +   9.9 -  16.8 \\ 
  15.819-15.915 &    90591 &   614.7 $\pm$   0.3 +   8.0 -  16.6 \\ 
  15.915-16.012 &    78786 &   617.7 $\pm$   0.3 +  10.9 -  17.1 \\ 
  16.012-16.109 &    65331 &   621.2 $\pm$   0.3 +  13.3 -  16.3 \\ 
  16.109-16.205 &    52467 &   624.2 $\pm$   0.3 +  15.9 -  15.3 \\ 
  16.205-16.302 &    41031 &   626.5 $\pm$   0.3 +  16.2 -  14.1 \\ 
  16.302-16.399 &    31743 &   630.8 $\pm$   0.4 +  15.6 -  14.8 \\ 
  16.399-16.496 &    24609 &   632.7 $\pm$   0.4 +  15.3 -  13.9 \\ 
  16.496-16.594 &    18644 &   633.0 $\pm$   0.5 +  13.4 -  13.3 \\ 
  16.594-16.691 &    14051 &   635.5 $\pm$   0.6 +  12.2 -  12.6 \\ 
  16.691-16.788 &    10544 &   637.9 $\pm$   0.6 +  12.5 -  12.4 \\ 
  16.788-16.886 &     7920 &   639.4 $\pm$   0.7 +  12.0 -  11.0 \\ 
  16.886-16.984 &     5919 &   640.1 $\pm$   0.8 +  10.6 -  12.0 \\ 
  16.984-17.081 &     4485 &   644.0 $\pm$   0.9 +  11.9 -  11.0 \\ 
  17.081-17.179 &     3259 &   647.6 $\pm$   1.1 +  11.9 -  10.8 \\ 
  17.179-17.277 &     2031 &   651.0 $\pm$   1.3 +  10.1 -   7.0 \\ 
  17.277-17.375 &     1377 &   653.8 $\pm$   1.7 +  10.3 -   9.1 \\ 
  17.375-17.473 &      904 &   663.6 $\pm$   2.1 +   8.7 -  11.9 \\ 
  17.473-17.670 &      896 &   672.1 $\pm$   2.1 +   9.1 -   9.6 \\ 
  17.670-17.866 &      331 &   678.7 $\pm$   3.4 +  12.3 -   7.8 \\ 
  17.866-18.063 &      110 &   708.2 $\pm$   6.5 +  12.5 -  12.5 \\  \hline
  \enddata
\end{deluxetable}

\end{document}